\documentclass[pra,twocolumn,showpacs,amsmath,amssymb,superscriptaddress,floatfix,nofootinbib]{revtex4-1}
\usepackage{amsmath}
\usepackage{amssymb}
\usepackage{amsthm}
\usepackage{amsfonts}
\usepackage{mathtools}
\usepackage{dsfont}
\usepackage{listings}
\usepackage{enumerate}
\usepackage{latexsym}
\usepackage{mathdots}
\usepackage{hyperref}
\usepackage{ulem}
\usepackage{psfrag}
\usepackage[titletoc,toc,title]{appendix}
\usepackage{bm}
\usepackage{graphicx}
\usepackage{subfigure}
\usepackage{xcolor}

\newcommand{\spa}{\text{ }}

\newcommand{\beq}{\begin{equation}}
\newcommand{\eneq}{\end{equation}}

\newcommand{\bra}[1]{\left\langle#1\right|}
\newcommand{\ket}[1]{\left|#1\right\rangle}

\newcommand{\up}{\uparrow}

\newcommand{\mE}{\mathcal{E}}
\newcommand{\mP}{\mathcal{P}}
\newcommand{\hT}{\widehat{T}}
\newcommand{\hn}{\hat{n}}
\newcommand{\hC}{\widehat{C}}
\newcommand{\hV}{\widehat{V}}
\newcommand{\hO}{\widehat{O}}
\newcommand{\hF}{\widehat{F}}
\newcommand{\hB}{\widehat{B}}

\newcommand{\heff}{H_{\textrm{eff}}}

\newcommand{\twopartdef}[4]
{
	\left\{
		\begin{array}{ll}
			#1 & \mbox{if } #2 \\
			#3 & \mbox{if } #4
		\end{array}
	\right.
}

\newcommand{\cd}{c^\dagger}
\newcommand{\dd}{d^\dagger}

\newcommand{\sumal}[2]{\underset{#1}{\overset{#2}{\sum}}}

\input{epsf}

\newcommand{\nn}{\nonumber}

\newcommand{\eeq}{\end{equation}}

\begin{document}

\tolerance 10000

\newcommand{\vk}{{\bf k}}   
\title{Thermalization and its absence within Krylov subspaces of a constrained Hamiltonian}
\author{Sanjay Moudgalya}
\affiliation{Department of Physics, Princeton University, NJ 08544, USA}
\affiliation{Laboratoire de Physique de l'Ecole normale sup\'{e}rieure, ENS, Universit\'{e} PSL, CNRS, Sorbonne Universit\'{e}, Universit\'{e} Paris-Diderot, Sorbonne Paris Cit\'{e}, Paris, France}

\author{Abhinav Prem}
\affiliation{
Princeton Center for Theoretical Science, Princeton University, NJ 08544, USA}

\author{Rahul Nandkishore}
\affiliation{Department of Physics and Center for Theory of Quantum Matter, University of Colorado, Boulder, CO 80309, USA}

\author{Nicolas Regnault}
\affiliation{Laboratoire de Physique de l'Ecole normale sup\'{e}rieure, ENS, Universit\'{e} PSL, CNRS, Sorbonne Universit\'{e}, Universit\'{e} Paris-Diderot, Sorbonne Paris Cit\'{e}, Paris, France}
\affiliation{Department of Physics, Princeton University, NJ 08544, USA}

\author{B. Andrei Bernevig}
\affiliation{Department of Physics, Princeton University, NJ 08544, USA}

\begin{abstract}
We study the quantum dynamics of a simple translation invariant, center-of-mass (CoM) preserving model of interacting fermions in one dimension (1D), which arises in multiple experimentally realizable contexts. We show that this model naturally displays the phenomenology associated with fractonic systems, wherein single charges can only move by emitting dipoles.
This allows us to demonstrate the rich Krylov fractured structure of this model, whose Hilbert space shatters into exponentially many dynamically disconnected subspaces.
Focusing on \textit{exponentially large} Krylov subspaces, we show that these can be either be integrable or non-integrable, thereby establishing the notion of \textit{Krylov-restricted} thermalization.
We analytically find a \textit{tower} of integrable Krylov subspaces of this Hamiltonian, all of which map onto spin-1/2 XX models of various system sizes.
We also discuss the physics of the non-integrable subspaces, where we show evidence for weak Eigenstate Thermalization Hypothesis (ETH) restricted to each non-integrable Krylov subspace.
Further, we show that constraints in some of the thermal Krylov subspaces cause the long-time expectation values of local operators to deviate from behaviour typically expected from translation invariant systems.   
Finally, we show using a Schrieffer-Wolff transformation that such models naturally appear as effective Hamiltonians in the large electric field limit of the interacting Wannier-Stark problem, and comment on connections of our work with the phenomenon of Bloch many-body localization.
\end{abstract}
\maketitle
\date{\today}


\section{Introduction}
\label{intro}
Rapid advances in the coherent control and manipulation of cold atoms have enabled experiments to study the non-equilibrium dynamics of closed quantum many-body systems~\cite{weiss2006,gring2012,schreiber2015,smith2016,kaufman2016,kucsko2018}. Consequently, the question of how (and whether) an arbitrary quantum state evolving under closed system dynamics  achieves thermal equilibrium while evolving under unitary dynamics has moved to the forefront of contemporary research. An important theoretical development along these lines is the Eigenstate Thermalization Hypothesis (ETH)~\cite{deutsch1991quantum,srednicki1994chaos,rigol2008thermalization, polkovnikov2011colloquium}, which, in its strong form, states that, as far as expectation values of local observables are concerned, \textit{all} eigenstates of an ergodic system display thermal behaviour~\cite{d2016quantum,gogolin2016review,ueda2018review}. 
Although lacking a formal proof, it is widely held that generic interacting systems obey the strong version of ETH, as evinced by several numerical studies~\cite{rigol2008thermalization,d2016quantum,kim2014testing,garrison2018does}.
Notable exceptions are integrable models, which possess extensively many conserved quantities, and  many-body localized (MBL) systems~\cite{anderson1958,gornyi2005,baa2006}, where the \textit{emergence} of extensively many local integrals of motion prohibits the system from exploring all allowed configurations in Hilbert space~\cite{serbyn2013cons,huse2014fully}.
MBL systems thus evade ergodicity even at high energy densities and are able to retain a memory of their initial conditions in local observables for arbitrarily long times, leading to rich new physics which has been extensively studied numerically (see Refs.~\cite{rahul2015review,altman2015review,abanin2018review} for a review).
An important open question is whether similar phenomena, e.g. violation of ETH or memory of initial conditions at long times, can occur in translation invariant non-integrable systems~\cite{deroeck2014trans,deroeck2014AMBL,groverfisher, schiulaz2015,papic2015nodisorder,yao2016,smith2017a,smith2017b,michailidis2018,brenes2018}.
This has generated much interest in identifying non-integrable models which violate strong ETH but obey \textit{weak} ETH, where the latter consists of a measure zero set of non-thermal eigenstates and is sufficient for preventing complete thermalization of the system~\cite{biroli2010ETH,mori2016ETH}. 
One recent line of attack has been to identify \textit{exact} excited eigenstates~\cite{vafek2017entanglement, moudgalya2018exact} in the middle of the spectrum of non-integrable Hamiltonians that could shed significant light on ETH and its violation, given that the dynamics of a quantum system is governed by the properties of the full many-body spectrum and not only its low-lying features.
There has been promising progress in this direction--- Refs.~\cite{moudgalya2018exact,moudgalya2018nonint} identified and analyzed an infinite tower of exact eigenstates of the celebrated 1D Affleck-Kennedy-Lieb-Tasaki (AKLT) models~\cite{affleck1988valence, arovas1988some}, where some states of the tower are present in the bulk of the energy spectrum and are non-thermal, thus representing a novel type of strong ETH violation. 
Moreover, Refs.~\cite{lin2018exact, schecter2019weak, chattopadhyay2019quantum, iadecola2019quantum2} have recently found similar exact ETH-violating eigenstates in a variety of models.
In addition, Ref.~\cite{mori2017eth} proposed a general construction of ``embedding" ETH-violating eigenstates into a thermal spectrum, which has also been applied to construct systems with topological eigenstates in the middle of the spectrum~\cite{ok2019topological}.
Concurrently, an experiment on a 1D chain of Rydberg atoms observed persistent revivals upon quenching the system from certain initial conditions, while other initial conditions led to the system thermalizing rapidly~\cite{bernien2017probing}. This striking dependence on initial conditions was numerically demonstrated to be caused by a vanishing number of non-thermal states that co-exist with an otherwise thermal spectrum~\cite{turner2017quantum,schecter2018many, turner2018quantum, choi2018emergent,ho2018periodic}, dubbed ``quantum many-body scars".
Several explanations for the origin of quantum scars have been proposed: analogues to single-particle scarring~\cite{heller1984bound, turner2017quantum, ho2018periodic, michailidis2019slow}, proximity to integrability~\cite{khemani2019int}, existence of approximate quasiparticle towers of states~\cite{lin2018exact, surace2019lattice, iadecola2019quantum}, confinement~\cite{james2019nonthermal, robinson2019signatures}, and an emergent SU(2) symmetry~\cite{choi2018emergent}. 
Furthermore, recent works have constructed generalizations of the PXP model that show similar characteristics~\cite{schecter2018many,  2bull2019scar, moudgalya2019quantum}, studied the stability of the scars to perturbations~\cite{lin2019pxp}, and found quantum scars in Floquet settings~\cite{pai2019robust, mukherjee2019collapse}.  
These discoveries thus reveal new possibilities for quantum dynamics which may occur between the extremes of thermalization and the complete breaking of ergodicity. 
Although it remains largely unclear what the general desiderata are for the presence of scar states, systems with constrained dynamics, such as kinetically constrained models~\cite{olmos2009,vanH2015,lan2018slow} and the PXP model~\cite{sun2008numerical,olmos2009collective,olmos2012universal}, offer a promising platform for exploring ergodicity breaking.
\textit{Fractonic} systems, whose defining feature is the presence of excitations with restricted mobility, are natural candidates displaying constrained dynamics (see Ref.~\cite{fractonreview} for a review). Indeed, alongside their novel ground-state features, 3D gapped fracton models have garnered attention also for their slow quantum dynamics in the absence of spatial disorder~\cite{chamon2005,kimhaah2016,prem2017glass}. As first observed in Ref.~\cite{sub}, the conservation of higher (e.g., dipole or angular) moments in $U(1)$ symmetric systems places stringent constraints on the mobility of excitations, rendering isolated charges completely immobile. This insight has allowed the characteristic physics of fractons to be realized away from their initial conception in exactly solvable 3D lattice models, potentially even in 1D\footnote{For our purposes, fractonic behaviour refers to the strict immobility of isolated charges and (possibly) restricted mobility of bound charges (e.g., dipoles). It remains an open question whether the non-trivial topological features, such as a sub-extensive ground state degeneracy, associated with 3D gapped fracton models are possible in spatial dimension less than three.} (see e.g. Refs.~\cite{sous2019,pai2019confined}). This perspective was recently taken in Ref.~\cite{pai2018localization}, where random unitary dynamics in 1D with conserved dipole moment were shown to localize for reasons beyond usual locator-expansion techniques.
 
In this paper, we investigate the quantum dynamics of a translation invariant, non-integrable 1D fermionic chain with conserved center-of-mass (CoM). Rather than imposing constraints by hand, we show that the CoM conserving model we study has a natural origin in two distinct physical settings: in the thin-torus limit of the fractional quantum Hall effect and in the strong electric field limit of the interacting Wannier-Stark problem, a regime accessible to current cold-atom experiments\footnote{For example, tilting an optical lattice subjects the trapped ultracold atoms to a linear field.}~\cite{morschreview}. Focusing on systems close to half-filling, we define composite degrees of freedom in terms of which CoM conservation maps onto dipole moment conservation, revealing the underlying fractonic nature of the model. 
Once we resolve the Hamiltonian into its disparate symmetry sectors, we find that the Hilbert space further shatters into exponentially many dynamically disconnected sectors or \textit{Krylov subspaces}, which have previously been studied under various settings~\cite{znidaric2019coexistence, iadecola2018exact,sala2019ergodicity,khemani2019local,moudgalya2019quantum}. 
This shattering is a consequence of charge and center-of-mass conservation and, as discussed in Refs.~\cite{sala2019ergodicity,khemani2019local}, the presence of exponentially many small (finite size in the thermodynamic limit) closed Krylov subspaces can lead to effectively localized dynamics. 
Here, we instead focus on a new phenomenon within exponentially \textit{large} Krylov subspaces, which are of infinite size in the thermodynamic limit, and unveil a rich structure within these sectors, leading to new notions of \textit{Krylov-restricted} integrability and thermalization.
Specifically, we find that several such large Krylov subspaces are \textit{integrable}, thereby establishing the phenomenon of emergent integrability and further breaking of ergodicity \textit{within} closed Krylov sectors. Meanwhile, other large sectors remain non-integrable.
To bring this distinction into focus, we propose that a modified version of ETH applies to Krylov fractured systems, wherein conventional diagnostics of non-integrability, such as level statistics, are defined with respect to a symmetry sector \textit{and a Krylov subspace.} 
Using our modified definition, we conclude that the problem `thermalizes' within each non-integrable Krylov subspace, in that the long-time behaviour of a state belonging to a particular Krylov subspace coincides with the Gibbs ensemble \textit{restricted} to that subspace.
Remarkably, we find that this restricted thermalization within some of the Krylov subspaces leads to the `infinite temperature' state within the Krylov sectors showing atypical behaviour, in that the late-time charge density deviates from that expected from unconstrained translation invariant systems.
Violations of this modified or `Krylov-restricted ETH' require either integrability, conventional `disorder induced' many-body localization, or existence of further symmetries within the Krylov subspace. 
Armed with this understanding, we also revisit the problem of interacting Wannier-Stark localization~\cite{schulz2019stark, van2018bloch}, which we argue requires the ideas introduced in this paper for a more complete understanding.

This paper is organized as follows: we introduce the pair-hopping model, also studied in Ref.~\cite{moudgalya2019quantum}, in Sec.~\ref{sec:model} and show that it conserves center-of-mass. We then briefly discuss its origins in the thin torus limit of the fractional quantum Hall effect (FQHE) and in the limit of strong electric field in the interacting Wannier-Stark problem. In Sec.~\ref{sec:halffilling}, we introduce a convenient formalism to study this model at half-filling, and show that it exhibits fractonic phenomenology. In Sec.~\ref{sec:krylovfracture}, we discuss the notion of Krylov fracture i.e., the phenomenon where systems exhibit several closed subspaces that are dynamically disconnected with respect to product states. We show examples of integrable and non-integrable dynamically disconnected Krylov subspaces in Secs.~\ref{sec:integrablekrylov} and \ref{sec:nonintkrylov} respectively. The integrable subspaces we study exactly map onto XX models of various sizes, and the non-integrable subspaces show features that are typically not expected in non-integrable models, which we discuss in Sec.~\ref{sec:quasilocal}.
Finally, we make connections to Bloch MBL in Sec.~\ref{sec:stark} and conclude in Sec.~\ref{sec:conclusions}. Various details are relegated to appendices.

%
%

\section{Model and its symmetries}
\label{sec:model}
The ``pair-hopping model" we study is a one-dimensional chain of interacting spinless fermions with translation and inversion symmetry, with the Hamiltonian~\cite{seidel2005incompressible, moudgalya2019quantum}
\beq
\label{eq:pairhopping}
H = \sum_{j = 1}^{L_b}{H_j} = \sum_{j = 1}^{L_b}{\left(\cd_j \cd_{j+3} c_{j + 2} c_{j + 1} + h.c.\right)} \, ,
\eeq
where $L_b = L - 3$ for open boundary conditions (OBC), $L_b = L$ for periodic boundary conditions (PBC), and the subscripts are defined modulo $L$ for PBC. 
Note that we have set the overall energy scale equal to one for convenience.
Each term $H_j$ of Eq.~(\ref{eq:pairhopping}) vanishes on all spin configurations on sites $j$ to $j + 3$ except for
\begin{eqnarray}
    H_j \overset{j\;\;\;\;\;j+3}{\ket{0\spa 1 \spa 1 \spa 0}} &=& \overset{j\;\;\;\;\;j+3}{\ket{1 \spa 0 \spa 0 \spa 1}} \nn \, , \\
    H_j\overset{j\;\;\;\;\;j+3}{\ket{1 \spa 0 \spa 0 \spa 1}} &=& \overset{j\;\;\;\;\;j+3}{\ket{0 \spa 1 \spa 1 \spa 0}} \, ,
\label{eq:hamilnonvanish}
\end{eqnarray}
where $\ket{a \spa b \spa c \spa d}$ represents the occupation of sites $j$ to $j + 3$. In the rest of the paper, we will use the following shorthand notation
\begin{equation}
    \ket{1 \spa 0 \spa  0 \spa 1} \leftrightarrow \ket{0 \spa 1 \spa 1 \spa 0}
\label{eq:convenient}
\end{equation}
to represent Eq.~(\ref{eq:hamilnonvanish}) i.e., the action of individual terms of the Hamiltonian Eq.~\eqref{eq:pairhopping}. 
This pair-hopping model preserves the center-of-mass position i.e., the center-of-mass position operator~\cite{seidel2005incompressible}
\begin{equation}
    \widehat{C} \equiv \twopartdef{\sumal{j = 1}{L}{j \hat{n}_j}}{OBC}{\exp\left(\frac{2\pi i}{L}{\sumal{j = 1}{L}{j \hat{n}_j}}\right)}{PBC},
\label{eq:comoperator}
\end{equation}
where the number operator $\hat{n}_j \equiv \cd_j c_j$ commutes with the Hamiltonian of Eq.~(\ref{eq:hamilnonvanish}). 
Hamiltonians with such conservation laws, including the model given by Eq.~\eqref{eq:pairhopping}, were first discussed in Ref.~\cite{seidel2005incompressible} in the quest to build featureless Mott insulators. 
As emphasized by Ref.~[\onlinecite{seidel2005incompressible}], the spectra of center-of-mass preserving Hamiltonians have some unusual features.
For example, at a filling $\nu = p/q$ (with $p$ and $q$ coprime), the full spectrum is $q$-fold degenerate, which stems from the fact that the center-of-mass position operator $\widehat{C}$, and the translation operator $\widehat{T}$ do not commute. 
More precisely, consider a 1D chain of length $L$ with periodic boundary conditions. As shown in Ref.~[\onlinecite{seidel2005incompressible}],  
\begin{equation}
    \widehat{C}\widehat{T} = e^{2\pi i \nu} \widehat{T}\widehat{C} \, ,
\label{eq:comTcommute}
\end{equation}
where $\nu$ is the the filling fraction $\nu = p/q$. This results in a $q$-fold degeneracy of the spectrum with PBC. 
The pair-hopping model Eq.~(\ref{eq:pairhopping}), with even system size $L = 2N$ and with PBC, has an additional symmetry: sublattice particle number conservation. 
That is, the operators 
\begin{equation}
    \hat{n}_e = \sumal{j = 1}{N}{\hat{n}_{2j}},\;\;\; \hat{n}_o = \sumal{j = 1}{N-1}{\hat{n}_{2j + 1}} \, ,
\label{eq:sublatticecharge}
\end{equation}
both commute with Eq.~(\ref{eq:pairhopping}). 
This can be seen by writing the action of the terms of the pair-hopping Hamiltonian as
\begin{equation}
    \overset{e\ o\ e\ o}{\ket{1\ 0\ 0\ 1}} \leftrightarrow \overset{e\ o\ e\ o}{\ket{0\ 1\ 1\ 0}},\;\;\;\overset{o\ e\ o\ e}{\ket{1\ 0\ 0\ 1}} \leftrightarrow \overset{o\ e\ o\ e}{\ket{0\ 1\ 1\ 0}},
\label{eq:sublatticeaction}
\end{equation}
where the superscripts $o$ and $e$ label the parity of the sites. The actions of Eq.~(\ref{eq:sublatticeaction}) conserve the particle number on the odd and even sites separately. 
Sublattice number conservation of Eq.~(\ref{eq:sublatticecharge}) trivially implies the conservation of total particle number $\left(n_e + n_o\right)$. 
Note that the sublattice number conservation is a special property of the truncated Hamiltonian Eq.~(\ref{eq:pairhopping}), and does not hold in general for center-of-mass preserving Hamiltonians.
For example, the extended pair-hopping Hamiltonian $\sumal{j}{}{\left(\cd_j \cd_{j+3} c_{j+2} c_{j+1} + \cd_j\cd_{j+4}c_{j+3}c_{j+1} + \textrm{h.c.}\right)}$ preserves the center-of-mass position but does not conserve sublattice particle number.
\subsection*{Experimental Relevance}
An especially appealing feature of center-of-mass preserving terms, including the pair-hopping term Eq.~\eqref{eq:pairhopping}, is their natural appearance in multiple experimentally relevant systems. The first setting in which such models appear is in the quantum Hall effect, when translation invariant interactions are projected onto a single Landau level~\cite{bergholtz2006one, bergholtz2008quantum, moudgalya2019quantum}. We refer the reader to Ref.~\cite{moudgalya2019quantum} for a derivation, but summarize the general idea here: one works in the Landau gauge, such that the single particle orbitals in a Landau level can be written as eigenstates of the magnetic translation operators in the $\hat{y}$ direction, in which case the position in the $\hat{x}$ direction is the momentum quantum number in the $\hat{y}$ direction.  
The matrix elements of a translation invariant interaction between the single particle orbitals are hence momentum conserving in the $\hat{y}$ direction, which translates to center-of-mass conservation in the $\hat{x}$ direction of the effective one-dimensional model~\cite{bergholtz2006one}. 
A general interaction operator projected to a Landau level of an $L_x \times L_y$ quantum Hall system has the form
\begin{equation}
    H = \sumal{j = 1}{N_{\Phi}}{\ \sumal{k,m}{}{V_{km} \left(\cd_j \cd_{j + k + m} c_{j + k} c_{j + m} + h. c.\right)}} \, ,
\label{eq:gencomhamil}
\end{equation}
where $N_{\Phi} = L_x L_y/\left(2\pi\right)$ is the number of flux quanta and $V_{km} \sim \exp\left(-2\pi^2\left(k^2 + m^2\right)/L_y^2\right)$ with the magnetic length set to unity.  
Thus, in the ``thin-torus" limit ($L_y \rightarrow 0$), one of the dominant terms is the pair-hopping Hamiltonian Eq.~(\ref{eq:pairhopping}).
We note that such Hamiltonians also appear in the thin torus limit of the pseudopotential Hamitonians for several Fractional Quantum Hall states~\cite{lee2015geometric, papic2014solvable,rezayi1994,nakamura2012exactly, moudgalya2019quantum}.
A second origin of such center-of-mass preserving models is in the well-known Wannier-Stark problem~\cite{wannierrmp}: spinless fermions hopping on a finite one-dimensional lattice, subject to an electric field. While localization at the single-particle level has been long established~\cite{emin1987existence}, an interacting version of the problem has recently been studied and found to display behaviour associated with MBL systems at strong fields~\cite{van2018bloch,schulz2019stark}; this phenomenon goes under the name Bloch (or Stark) MBL. In Sec.~\ref{sec:stark}, we show that the dynamics of the Bloch MBL model in the limit of an infinitely strong electric field is governed by an effective center-of-mass preserving Hamiltonian, with the lowest order ``hopping" term given precisely by Eq.~\eqref{eq:pairhopping}. Specifically, the resulting Hamiltonian is again of the form Eq.~\eqref{eq:gencomhamil}, with $N_{\Phi}$ replaced by the system size\footnote{Note that for both the FQHE and the Bloch MBL case, the dominant center-of-mass conserving terms are nearest neighbor ($\hat{n}_j \hat{n}_{j + 1}$) and next nearest neighbor electrostatic terms ($\hat{n}_j \hat{n}_{j+2}$), but the lowest order ``hopping" is the pair-hopping Hamiltonian of Eq.~(\ref{eq:pairhopping}).}. This mapping hence allows us to present a new perspective on the phenomenon of  Bloch MBL (see Sec.~\ref{sec:stark}), in addition to providing a natural experimental setting, accessible to current cold-atom experiments, for realizing the model studied here.

%
%

\section{Hamiltonian at 1/2 filling}
\label{sec:halffilling}
We now proceed to study the spectrum of the pair hopping Hamiltonian Eq.~(\ref{eq:pairhopping}).
In this work, we will be focusing on systems at, or close to, half filling, and will restrict ourselves to even system sizes $L = 2N$. 
For the study of this Hamiltonian at other filling factors, see Refs.~\cite{wang2012spin, moudgalya2019quantum}.
\subsection{Composite degrees of freedom}\label{sec:compositedof}
To study this model, and to elucidate its relation to the physics of fractons, we define composite degrees of freedom formed by grouping neighboring sites of the original model. 
Assuming an even number of sites, we group sites $2j - 1$, $2j$ of the original lattice into a new site $j$ so as to form a new chain with $N = L/2$ sites.
We define new degrees of freedom for these composite sites as follows:
\begin{eqnarray}
&\ket{\uparrow} \equiv \ket{0 \spa 1}\, , \;\;\; \ket{\downarrow} \equiv \ket{1 \spa 0} \, , \nn \\ 
&\ket{+} \equiv \ket{1 \spa 1}\, , \;\;\; \ket{-} \equiv \ket{0 \spa 0}.
\label{eq:halffillingpart}
\end{eqnarray}
The choice of grouping is unambiguously defined for OBC, and we stick to it for most of this paper.
Writing the action of the Hamiltonian Eq.~\eqref{eq:hamilnonvanish} in terms of these composite degrees of freedom, we find
\begin{eqnarray}
    \ket{\ \fbox{01}\ \fbox{10}\ } &\leftrightarrow& \ket{\ \fbox{10}\ \fbox{01}\ }\nn \\
    \iff \ket{\uparrow \downarrow} &\leftrightarrow& \ket{\downarrow\uparrow} \, , \label{eq:spinscattering}\\
    \ket{\ \fbox{10}\ \fbox{11}\ \fbox{00}\ } &\leftrightarrow& \ket{\ \fbox{11}\ \fbox{00}\ \fbox{10}\ } \nn \\
    \iff \ket{\downarrow + -} &\leftrightarrow& \ket{+ - \downarrow} \, , \label{eq:dipolescattering1} \\
    \ket{\ \fbox{00}\ \fbox{11}\ \fbox{01}\ } &\leftrightarrow& \ket{\ \fbox{01}\ \fbox{00}\ \fbox{11}\ } \nn \\
    \iff \ket{- + \uparrow} &\leftrightarrow& \ket{\uparrow - +} \, , \label{eq:dipolescattering2} \\
    \ket{\ \fbox{10}\ \fbox{11}\ \fbox{01}\ } &\leftrightarrow& \ket{\ \fbox{11}\ \fbox{00}\ \fbox{11}\ } \nn \\
    \iff \ket{\downarrow + \uparrow} &\leftrightarrow& \ket{+ - +} \, , \label{eq:+fracton} \\
    \ket{\ \fbox{01}\ \fbox{00}\ \fbox{10}\ } &\leftrightarrow& \ket{\ \fbox{00}\ \fbox{11}\ \fbox{00}\ } \nn \\
    \iff \ket{\uparrow - \downarrow} &\leftrightarrow& \ket{- + -}, \label{eq:-fracton}
\end{eqnarray}
where $\fbox{$\cdots$}$ represents a grouping of some sites $2j - 1$ and $2j$, and $\ket{a} \leftrightarrow \ket{b}$ represents the action of a single term of the Hamiltonian on $\ket{a}$ resulting in $\ket{b}$ and vice versa (see Eqs.~(\ref{eq:hamilnonvanish}) and (\ref{eq:convenient})). 
For reasons that will become clear forthwith, we set the nomenclature of the composite degrees of freedom as follows:
\begin{center}
    \begin{tabular}{cc}
        $\ket{+}$, $\ket{-}$: & Fractons \\
        $\ket{+-}$, $\ket{-+}$: & Dipoles \\
        $\ket{\uparrow}$, $\ket{\downarrow}$: & Spins \\
    \end{tabular}
\end{center}
Here, Eqs.~(\ref{eq:dipolescattering1})-(\ref{eq:-fracton}) resemble the rules restricting the mobility of fractons, and are similar to those discussed in Ref.~\cite{pai2018localization} (see Ref.~\cite{fractonreview} for a review on fractons).

In particular, Eqs.~(\ref{eq:dipolescattering1}) and (\ref{eq:dipolescattering2}) represent the free propagation of dipoles when separated by spins, and Eqs.~(\ref{eq:+fracton}) and (\ref{eq:-fracton}) encode the characteristic movement of a fracton through the emission or absorption of a dipole, i.e. dipole assisted hopping.  
However, in contrast to usual fracton phenomenology, here the movement of fractons is also sensitive to the background spin configuration. For example, the fracton in the configuration $\ket{\cdots \downarrow + \uparrow \cdots}$ can move by emitting a dipole (see Eq.~\eqref{eq:+fracton}) while that in the configuration $\ket{\cdots \uparrow + \downarrow \cdots}$ cannot.
In our convention, the fractons $\ket{+}$ and $\ket{-}$ have spin $0$ and charges $+1$ and $-1$ respectively, while the spins $\ket{\uparrow}$ and $\ket{\downarrow}$ have charge $0$ and spins $+1$ and $-1$ respectively. Thus the unit cell charge and spin operators in terms of the original fermionic degrees of freedom read
\begin{equation}
    \widehat{Q}_j \equiv \hat{n}_{2j - 1} + \hat{n}_{2j} - 1,\;\;\; \widehat{S}^z_j \equiv - \hat{n}_{2j - 1} + \hat{n}_{2j},
\label{eq:chargespinops}
\end{equation}
where $j$ is the unit cell index, and $2j-1$, $2j$ are the site indices of the original configuration. 
We represent the total number of $+$, $-$, $\uparrow$, and $\downarrow$ by $N_+$, $N_-$, $N_\uparrow$, $N_\downarrow$ respectively. 
Thus, the \textit{total charge} is $N_+ - N_-$ and the \textit{total spin} is $N_\uparrow - N_\downarrow$.
\subsection{Symmetries in terms of the composite degrees}
\label{sec:compositesymmetry}
We now study the symmetries of the Hamiltonian whose terms act on the composite degrees of freedom through Eqs.~(\ref{eq:spinscattering})-(\ref{eq:-fracton}). 
As discussed in Sec.~\ref{sec:halffilling}, the pair-hopping model Eq.~(\ref{eq:pairhopping}) has several symmetries: sublattice charge conservation, center-of-mass conservation, inversion, and translation (for PBC). 
Using Eqs.~(\ref{eq:spinscattering})-(\ref{eq:-fracton}), we now interpret these symmetries in terms of the composite degrees of freedom defined in Eq.~(\ref{eq:halffillingpart}). 

The model in terms of the composite degrees of freedom conserves the total spin and the total charge, as is evident from Eqs.~(\ref{eq:spinscattering})-(\ref{eq:-fracton}). In other words, $N_\uparrow - N_\downarrow$ and $N_+ - N_-$ are separately conserved.  
Indeed, using the definitions of spin and charge in Eq.~(\ref{eq:chargespinops}), the total spin operator $\widehat{S}^z$ and total charge operator $\widehat{Q}$ can be expressed in terms of the operators in the original Hilbert space as follows:
\begin{equation}
    \widehat{Q} \equiv \sumal{j = 1}{N}{\widehat{Q}_j} = \hat{n}_e + \hat{n}_o - N,\;\;\; \widehat{S}^z \equiv \sumal{j = 1}{N}{\widehat{S}^z_j} = \hat{n}_o - \hat{n}_e,
\end{equation}
where $\hat{n}_e$ and $\hat{n}_o$ are the sublattice particle numbers defined in Eq.~(\ref{eq:sublatticecharge}).
Thus, the conservation of total charge and total spin in the fracton model is a direct consequence of the sublattice number conservation of the pair-hopping model.

Moreover, the fractonic behavior inherent in the rules specified by Eqs.~(\ref{eq:spinscattering})-(\ref{eq:-fracton}) suggests that the \textit{dipole moment} of the composite degrees of freedom is a conserved quantity~\cite{sub}. This operator is defined similarly to the center-of-mass operator Eq.~(\ref{eq:comoperator}) as:
\begin{equation}
    \widehat{D} \equiv \twopartdef{\sumal{j = 1}{N}{j \widehat{Q}_j}}{OBC}{\exp\left(i\frac{2 \pi}{N}\sumal{j = 1}{N}{j \widehat{Q}_j}\right)}{PBC}.
\label{eq:dipoleoperator}
\end{equation}
To explicitly show that $\widehat{D}$ is in fact a conserved quantity of the composite fractonic model, we observe that
\begin{align}
    \sumal{j = 1}{N}{j \widehat{Q}_j} &= \sumal{j = 1}{N}{j \left(\hat{n}_{2j - 1} + \hat{n}_{2j} - 1\right)} \nn \\
    &= \sumal{j = 1}{N}{\frac{(2 j  - 1) \hat{n}_{2j - 1} + 2j \hat{n}_{2j}}{2}} + \sumal{j = 1}{N}{\frac{\hat{n}_{2j - 1}}{2}} - \sum_{j=1}^N j \nn \\
    &= \frac{1}{2}\sumal{j = 1}{L}{j \hat{n}_{j}} + \frac{\hat{n}_o}{2} - \frac{N(N+1)}{2}.
\label{eq:dipolesimp}
\end{align}
Then, using Eqs.~(\ref{eq:comoperator}), (\ref{eq:sublatticecharge}), and (\ref{eq:dipolesimp}), in terms of the original operators in the pair-hopping model, the operator $\widehat{D}$ can be expressed as
\begin{equation}
    \widehat{D} = \twopartdef{\frac{1}{2}\widehat{C} + \frac{1}{2}\hat{n}_o - \frac{N(N+1)}{2}}{OBC}{\widehat{C}^{\frac{1}{2}}e^{i\frac{\pi }{L}\hat{n}_o}e^{-i\frac{\pi N(N + 1)}{L}}}{PBC}.
\label{eq:dipolecomrelation}
\end{equation}
Since $\widehat{C}$ and $\hat{n}_o$ are conserved operators of the pair-hopping Hamiltonian, as discussed in Sec.~\ref{sec:model}, it follows from Eq.~(\ref{eq:dipolecomrelation}) that $\widehat{D}$ is conserved in the composite model. %
To complete our discussion, we note that the composite model also preserves inversion as well as translation symmetry (with PBC), neither of which commute with $\widehat{D}$. Details of the symmetries are relegated to App.~\ref{sec:symcomp}. 

%
%

\section{Krylov Fracture}
\label{sec:krylovfracture}
We now study the dynamics of $H$, and show that it exhibits exponentially many dynamically disconnected subspaces.
More precisely, we construct \textit{Krylov subspaces} of the form
\begin{equation}
    \mathcal{K}\left(H, \ket{\psi_0}\right) \equiv \textrm{span}\{\ket{\psi_0}, H\ket{\psi_0}, H^2\ket{\psi_0}, \cdots\}
\label{eq:krylov}
\end{equation}
that are by definition closed under the action of the Hamiltonian $H$.
While $\ket{\psi_0}$ in Eq.~(\ref{eq:krylov}) can in principle be an arbitrary state, we are interested in the dynamics of initial product states, which are more easily accessible to experiments. Hence, we focus on Krylov subspaces generated by product states $\ket{\psi_0}$, which we dub \textit{root states} of the Krylov subspace $\mathcal{K}\left(H, \ket{\psi_0}\right)$.
For a generic non-integrable Hamiltonian $H$ without any symmetries, one expects that $\mathcal{K}\left(H, \ket{\psi_0}\right)$ for \textit{any} initial product state $\ket{\psi_0}$ is the \textit{full} Hilbert space of the system. 
For a non-integrable Hamiltonian with some symmetry, and with $\ket{\psi_0}$ an eigenstate of the symmetry, one typically expects that $\mathcal{K}\left(H, \ket{\psi_0}\right)$ spans \textit{all} states with the same symmetry quantum number as $\ket{\psi_0}$. 
Surprisingly, however, we show that the pair-hopping Hamiltonian~\eqref{eq:pairhopping} exhibits \textit{Krylov fracture} i.e., even after resolving the charge and center-of-mass symmetries, we find generically that $\mathcal{K}\left(H, \ket{\psi_0}\right)$ does \textit{not} span all states with the same symmetry quantum numbers as $\ket{\psi_0}$.
Thus the full Hilbert space of the system $\mathcal{H}$ is of the form
\beq
\mathcal{H} = \bigoplus_{\bf{s}} \mathcal{H}^{(\bf{s})}, \quad \mathcal{H}^{(\bf{s})} = \bigoplus_{i = 1}^{K^{(\bf{s})}} {\mathcal{K}\left(H, \ket{\psi_i^{(\bf{s})}}\right)} \, ,
\label{eq:fullhilbert}
\eeq
where $\bf{s}$ labels the distinct symmetry quantum numbers, such as charge and center-of-mass, $K^{(\bf{s})}$ denotes the number of disjoint Krylov subspaces generated from product states with the same symmetry quantum numbers, and $\ket{\psi_i^{(\bf{s})}}$ are the root states generating the Krylov subspaces. Note that the root states in Eq.~\eqref{eq:fullhilbert} are chosen such that they generate distinct disconnected Krylov subspaces, since the same subspace can be generated by different root states. Stated symbolically, 
\begin{equation}
    \mathcal{K}\left(H, \ket{\psi^{(\bf{s})}_i}\right) \cap \mathcal{K}\left(H, \ket{\psi^{(\bf{s'})}_{i'}}\right) 
    = \delta_{\bf{s}, \bf{s'}} \delta_{i, i'} \mathcal{K}\left(H, \ket{\psi^{(\bf{s})}_i}\right). 
\end{equation}

Fracture of the form Eq.~(\ref{eq:fullhilbert}), where the total number of Krylov subspaces $K^{(\bf{s})}$ is exponentially large in the system size, was recently shown to \textit{always} exist in Hamiltonians and random-circuit-models with center-of-mass conservation~\cite{sala2019ergodicity, khemani2019local} (alternatively referred to as ``dipole moment" conservation). 
While the presence of these symmetries guarantees fracture, one can distinguish between ``strong" and ``weak" fracture~\cite{sala2019ergodicity,khemani2019local}, depending respectively on whether or not the ratio of the largest Krylov subspace to the Hilbert space within a given global symmetry sector vanishes in the thermodynamic limit. 
Strong (resp. weak) fracture is associated with the violation of weak (resp. strong) ETH with respect to the full Hilbert space. 
The pair-hopping model Eq.~\eqref{eq:pairhopping} (which is equivalent to the Hamiltonian $H_4$ in Ref.~\cite{sala2019ergodicity} with spin-$1/2$) numerically appears to exhibit \textit{strong} fracture within several symmetry sectors.
However, the addition of longer-range CoM preserving terms numerically appears to cause the Hilbert space to fracture only weakly~\cite{sala2019ergodicity}, with the fracture disappearing with the addition of infinite-range CoM preserving terms, even if the interaction strength decays exponentially with range~\cite{fremling2018dynamics}.
By definition, distinct Krylov subspaces are \textit{dynamically disconnected} i.e., no state initialized completely within one of the Krylov subspaces can evolve out to a different Krylov subspace. 
Indeed, exponentially many of these Krylov subspaces are one-dimensional \textit{static} configurations---product states that are eigenstates of $H$. 
For instance, the Hamiltonian vanishes on any product state that does not contain the patterns $``\cdots 0110 \cdots"$ or $``\cdots 1001\cdots"$, since those are the only configurations on which terms of $H$ act non-trivially (see Eq.~(\ref{eq:hamilnonvanish})). The charge-density-wave (CDW) state 
\begin{equation*}
    \ket{1111000011110000\dots\dots1111000011110000}
\end{equation*}
is one example of a static configuration that is an eigenstate.
In terms of the composite degrees of freedom we can equivalently consider configurations with only $+$, $-$, and no spins, such as $$\ket{\cdots + + - - + + - - \cdots },$$ with a pattern that alternates between $+$ and $-$ with `domain walls' that are at least 2  sites apart. According to  Eqs.~(\ref{eq:spinscattering})-(\ref{eq:-fracton}), all terms of the Hamiltonian vanish on these configurations: since there are \textit{exponentially} many such patterns, there are equally many one-dimensional Krylov subspaces.
We can also construct small Krylov subspaces by embedding finite non-trivial blocks, on which the Hamiltonian acts non-trivially, into the static configurations, thereby leading to exponentially many Krylov subspaces of every size~\cite{sala2019ergodicity,khemani2019local}.
For example, the following configurations $\ket{\psi_\pm}$
\begin{eqnarray}
    &\ket{\psi_\pm} = \frac{1}{\sqrt{2}}\left(\ket{++--\cdots++--\uparrow\downarrow++--\cdots++--}\right. \nn \\
    &\left.\pm \ket{++--\cdots++--\downarrow\uparrow++--\cdots++--}\right)  
\end{eqnarray}
are composed of one non-trivial block $\uparrow\downarrow$ sandwiched within a frozen configuration, and they thus have energies $E_\pm = \pm 1$. Exponentially many configurations with energies $E = \pm 1$ can be constructed by changing the frozen configuration around the non-trivial block. 

The presence of exponentially many static states (within each symmetry sector) in the the Hilbert space leaves an imprint on the dynamical behaviour of such systems. Specifically, time-evolution starting from randomly chosen product states looks highly non-generic from the perspective of the full Hilbert space.
For example, in the absence of Krylov fracture one typically expects that the bipartite entanglement entropy evolves to the Page value~\cite{page1993}, the average bipartite entanglement entropy of states in the Hilbert space.
For a system of Hilbert space dimension $D\left[L\right] = 2^L$, the Page value is $\log D\left[L/2\right] \approx L/2 \log 2$. 
However, in the presence of Krylov fracture, we expect that the late-time bipartite entanglement entropy of product states $\ket{\psi_0}$ is smaller and typically $\sim \log {D_{\mathcal{K}}\left[L/2\right]}$, where $D_{\mathcal{K}}\left[L\right]$ is the dimension of the Krylov subspace $\mathcal{K}\left(H, \ket{\psi_0}\right)$ for a system size $L$. 
The phenomenon of Krylov fracture can thus be regarded as a breaking of ergodicity with respect to the full Hilbert space, resulting in (at the very least) violation of strong ETH. 
However, what remains unclear is whether, for systems exhibiting Krylov fracture, thermalization occurs \textit{within} each of the Krylov subspaces. 
Of course, thermalization or ETH-violation are only well-posed concepts for \textit{large} Krylov subspaces $\mathcal{K}$ (with dimension $\mathcal{D}_{\mathcal{K}}[L] \rightarrow \infty$ as $L \rightarrow \infty$)\footnote{Note that the dimension of the Krylov subspace $D_{\mathcal{K}}[L]$ could in principle scale polynomially with $L$; however, we are not aware of any such example in the pair-hopping model Eq.~(\ref{eq:pairhopping}).} and do not have a clear meaning when the Krylov subspace has a finite dimension in the thermodynamic limit, as is the case for the exponentially many static configurations discussed above. 
Indeed, there exist exponentially large Krylov subspaces of the Hamiltonian Eq.~\eqref{eq:pairhopping} at filling $\nu = p/(2p + 1)$ for which Krylov-restricted thermalization appears to hold for most initial states, as recently demonstrated by some of the present authors ~\cite{moudgalya2019quantum}. 
There, we demonstrated the existence of Krylov subspaces with Wigner-Dyson level statistics, despite such Krylov subspaces hosting \textit{quantum scars} i.e., evenly spaced towers of anomalous states in the spectrum that lead to revivals in the fidelity of time evolution from particular initial states.
Those Krylov subspaces are examples of ones that violated Krylov-restricted \textit{strong} ETH, although Krylov-restricted \textit{weak} ETH is satisfied. 
However, it has not yet been established if Krylov-restricted \textit{weak} ETH is \textit{necessarily} satisfied for large dimensional Krylov subspaces, or if there are examples of \textit{semi-integrable} systems with both integrable and non-integrable Krylov subspaces, opening the door to further violations of ergodicity within Krylov sectors. 

\begin{table*}[t]
    \centering
    \hspace{-6mm}
    \begin{tabular}{|c|c|c|c|}
        \hline
        {\bf Krylov Subspace} & {\bf Root Configuration} & {\bf Quantum Numbers}  & {\bf Restricted Hamiltonian}\\
        \hline
        Spin & $\ket{\uparrow \downarrow \cdots \downarrow \uparrow}$  & $N_\uparrow$ & $H_{XX}\left[N\right]$\\
        \hline
        Single $+-$ dipole & $\ket{\uparrow \cdots \downarrow + - \uparrow \cdots \downarrow}$ & $N^{(1)}_\uparrow, N^{(2)}_{\uparrow}$ & $H_{XX}\left[N-1\right]$\\
        \hline
        Two separated $+-$ dipoles & $\ket{\uparrow \cdots \downarrow +-\uparrow\cdots\downarrow+- \uparrow \cdots \downarrow}$ & $N^{(1)}_\uparrow, N^{(2)}_\uparrow \geq 1, N^{(3)}_\uparrow$ & $H_{XX}\left[N-2\right]$\\
        \hline 
        Two adjacent $+-$ dipoles  & $\ket{\uparrow \cdots \downarrow +-+- \uparrow \cdots \downarrow}$ & $N^{(1)}_\uparrow, N^{(2)}_\uparrow = 0, N^{(3)}_\uparrow$  & $H_{XX}\left[N-1\right]$\\
        \hline
        $X$ separated $+-$ dipoles & $\ket{\uparrow \cdots \downarrow +- \uparrow \cdots +- \cdots \downarrow+- \uparrow \cdots \downarrow}$ & $N^{(1)}_\uparrow, \{N^{(2)}_\uparrow, \cdots, N^{(X-1)}_\uparrow\} \geq 1, N^{(X)}_\uparrow$ & $H_{XX}\left[N-(X-1)\right]$ \\
        \hline
        $X$ adjacent $+-$ dipoles & $\ket{\uparrow \cdots \downarrow +-+-\cdots+- \uparrow \cdots \downarrow}$ & $N^{(1)}_\uparrow, \{N^{(2)}_\uparrow, \cdots, N^{(X-1)}_\uparrow\}  = 0, N^{(X)}_\uparrow$ & $H_{XX}\left[N-1\right]$ \\
        \hline
    \end{tabular}
    \caption{Table of integrable Krylov subspaces (by no means an exhaustive list) of the pair-hopping model for system size $L = 2N$, with OBC at half-filling. For each type of Krylov subspace, we provide the root configuration generating it, the associated quantum numbers, and the Hamiltonian restricted to that subspace. Dipole subspaces for the oppositely oriented $-+$ dipoles can be constructed analogously (see main text for discussion).}
    \label{tab:krylov}
\end{table*}
Thus, in what follows we will focus on high dimensional \textit{irreducible} Krylov subspaces $\mathcal{K}\left(H, \ket{\psi}\right)$, defined as those with exponentially large dimension $\mathcal{D}_{\mathcal{K}}\left[L\right] \sim \alpha^L$ as $L \to \infty$ ($\alpha > 1$), and which satisfy 
\begin{equation}
    \mathcal{K}\left(H, \ket{\psi}\right) \neq  \mathcal{K}\left(H, \ket{\psi_1}\right) \oplus  \mathcal{K}\left(H, \ket{\psi_2}\right)
\label{eq:irreduciblekrylov}
\end{equation}
for any product states $\ket{\psi_1}$ and $\ket{\psi_2}$, after resolving charge and center-of-mass symmetries.
Remarkably, we find several examples of both integrable and non-integrable subspaces in the model Eq.~(\ref{eq:pairhopping}), demonstrating the rich dynamical structure inherent in systems with fractured Hilbert spaces. Studying the dynamics of root states that generate large irreducible Krylov subspaces thus allows us to establish that integrability or non-integrability of a system is correctly defined only \textit{within} each Krylov subspace. 

%
%

\section{Integrable subspaces}
\label{sec:integrablekrylov}
In this section, we illustrate several \textit{integrable} irreducible Krylov subspaces with exponentially large dimension present in the pair-hopping model Eq.~(\ref{eq:pairhopping}). 

\subsection{Spin subspace}
\label{sec:spinsubspace}
The simplest example of a large integrable Krylov subspace can be generated by a root state $\ket{\psi_0}$ (see Eq.~(\ref{eq:krylov})) which is any product state of only spin degrees of freedom: $\uparrow$ and $\downarrow$ as defined in Eq.~(\ref{eq:halffillingpart}). 
From Eq.~(\ref{eq:spinscattering}), we find that the Hamiltonian restricted to this subspace can be written as a nearest neighbor Hamiltonian with actions:
\begin{equation}
    \ket{\uparrow \uparrow} \rightarrow 0,\;\;\;\ket{\downarrow \downarrow} \rightarrow 0,\;\;\;\ket{\uparrow \downarrow} \leftrightarrow \ket{\downarrow \uparrow},
\label{eq:spinscatteringkryl}
\end{equation}
where $\ket{a} \rightarrow 0$ and $\ket{a} \leftrightarrow \ket{b}$ represent the action of a single term of the Hamiltonian. 
Thus, starting from a root state with $N_\uparrow$ spin $\uparrow$'s (and hence $(N - N_\up)$ spin $\downarrow$'s), such as
\begin{equation*}
    \ket{\uparrow\downarrow\uparrow\uparrow\downarrow}, \quad (N, N_\uparrow) = (5, 3),
\end{equation*}
the action of the Hamiltonian only rearranges the spins.
In particular, note that: (i) The number of $\uparrow$'s and $\downarrow$'s in the root state $N_\uparrow$ and $N - N_\uparrow$ respectively are preserved upon the action of the Hamiltonian, (ii) no fractons (i.e. $+$'s or $-$'s) are created, and (iii) \textit{all} product configurations with $N$ spins and a fixed value of $N_\uparrow$ are part of the Krylov subspace $\mathcal{K}\left(H, \ket{\psi_0}\right)$ associated with the root state $\ket{\psi_0}$.
Furthermore, since the Hamiltonian restricted to this subspace only interchanges the spins (see Eq.~(\ref{eq:spinscatteringkryl})), it maps \textit{exactly} onto that of the spin-1/2 XX model:
\begin{equation}
    H_{XX}\left[N\right] \equiv \sum_{j = 1}^{N}{\left(\sigma^+_j \sigma^-_{j+1} + \sigma^-_j \sigma^+_{j+1}\right)} \, ,
\label{eq:hamilXX}
\end{equation}
where $\{\sigma^+_j\}$ and $\{\sigma^-_j\}$ are onsite Pauli matrices. 
This mapping was first noted in earlier works on half-filled Landau levels~\cite{bergholtz2005half, bergholtz2006one, bergholtz2008quantum}, and is formally illustrated in App.~\ref{sec:XXformal}.
As is well known, the Hamiltonian Eq.~(\ref{eq:hamilXX}) can be solved using a Jordan-Wigner transformation~\cite{lieb1963exact}, upon which it maps onto a non-interacting problem. 
We numerically observe that the full ground state of the Hamiltonian Eq.~(\ref{eq:pairhopping}) belongs this Krylov subspace with $\left(N, N_\uparrow\right) = \left(N, \left\lfloor{\frac{N}{2}}\right\rfloor\right)$.
We refer to App.~\ref{sec:energies} for a complete discussion of the structure of the eigenstates within this Krylov subspace.
An important note regarding symmetries: each Krylov subspace generated from a root state with only spins and with a fixed $N_\uparrow$ (dubbed the spin Krylov subspace) only generates one symmetry sector of the XX model with a fixed $S_z$.  
All symmetry sectors of the XX model can be generated by starting from root states with different $N_\uparrow$, so that the full spectrum of the XX model of $N$ sites is embedded within the spectrum of the pair-hopping Hamiltonian $H$~(\ref{eq:pairhopping}), both for OBC and PBC. 

With respect to the symmetries of $H$, these Krylov subspaces lie within the sector $(Q, D, S^z) = (0, 0, 2 N_\uparrow - N)$, where $Q$, $D$, and $S^z$ are the total charge, dipole moment, and spin respectively, discussed in Sec.~\ref{sec:compositesymmetry}. 
However, these are \textit{not} the only states within that $(Q,D, S^z)$  symmetry sector, providing evidence for the Krylov fracture in the pair-hopping Hamiltonian $H$. 
For example, the product state
\begin{equation}
    \ket{\ast \cdots \ast + - - + \ast \cdots \ast},
\end{equation}
where $\ast =\ \uparrow, \downarrow$ and with $\left(N_\uparrow - 1\right)$ $\uparrow$'s (and hence $(N - N_\uparrow - 1)$ $\downarrow$'s) lies within the symmetry sector $(Q, D, S^z) = (0, 0, 2 N_\uparrow - N)$ but outside the spin Krylov subspace constructed above.
\subsection{Single dipole subspace}
\label{sec:dipolesubspace}
Restricting our attention to OBC, we now demonstrate the existence of another set of integrable Krylov subspaces $\mathcal{K}\left(H, \ket{\psi_0}\right)$, which are generated from root states containing only a single dipole. Such root states are of the form
\begin{equation}
    \ket{\psi_0} = \ket{\ast \cdots\ast + -\ast \cdots\ast },\;\;\;\ket{\psi_0} = \ket{\ast \cdots \ast - +\ast \cdots\ast },
\label{eq:singledipolerootstate}
\end{equation}
where $\ast =\ \uparrow, \downarrow$. 
The action of the Hamiltonian Eq.~(\ref{eq:pairhopping}) on configurations of the form Eq.~(\ref{eq:singledipolerootstate}) is given by
\begin{eqnarray}
    &\ket{\downarrow + -} \leftrightarrow \ket{+ - \downarrow}, \;\;\; \ket{\uparrow - +} \leftrightarrow \ket{- + \uparrow} \nn \\
    &\ket{\uparrow + -} \rightarrow 0,\;\;\; \ket{+ - \uparrow} \rightarrow 0\nn \\
    &\ket{\downarrow - +} \rightarrow 0,\;\;\;\ket{- + \downarrow} \rightarrow 0.
\label{eq:dipolescattering}
\end{eqnarray}
Since dipole moment is conserved, the dipole does not ``disintegrate" under the action of the Hamiltonian Eq.~(\ref{eq:dipolescattering}), i.e. the dipole does not separate into its constituent $+$ and $-$ fractons. 
As it turns out, Krylov subspaces generated by root states of the form~(\ref{eq:singledipolerootstate}) with $N$ sites are \textit{isomorphic} to Hilbert spaces of $(N - 1)$ spin-1/2's, with the effective Hamiltonians within these Krylov subspaces given by XX models of $(N - 1)$ sites.
In the following, we focus on the Krylov subspace corresponding to a $+-$ dipole.
As we discuss later, the generalization to $-+$ dipoles follows similarly. 
To show this, we first observe that as a consequence of Eq.~(\ref{eq:dipolescattering}),  a dipole $+-$  in the root state can \textit{never} cross an $\uparrow$ spin to its left or to its right. In other words, the dipole $+-$ can only hop left (right) if there is a $\downarrow$ spin immediately to its left (right).
Hence, all product states in the Krylov subspace generated by a root state $\ket{\psi_0}$ with one dipole $+-$ preserve the number of $\uparrow$ spins to the left and right of the dipole separately.
Denoting these conserved quantities by $N^{(1)}_\uparrow$ and $N^{(2)}_{\uparrow}$ respectively, we see that product states in the Krylov subspace $\mathcal{K}\left(H, \ket{\psi_0}\right)$ always have the form 
\begin{equation}
	\underset{\underbrace{\hspace{10mm}}_{N^{(1)}_\uparrow}\hspace{6mm}\underbrace{\hspace{10mm}}_{N^{(2)}_\uparrow}}{\ket{\ast \cdots \ast + - \ast \cdots \ast}},
\label{eq:singledipolequants}
\end{equation}
where $\ast =\ \uparrow, \downarrow$. 
This Krylov subspace can thus be uniquely labelled by the tuple $(N, N^{(1)}_{\uparrow}, N^{(2)}_\uparrow)$. 
For example, the Krylov subspace $\mathcal{K}\left(H, \ket{\psi_0}\right)$ generated by the configuration $\ket{\psi_0} = \ket{\uparrow\downarrow +-\uparrow \downarrow}$ with OBC consists of the following basis states:
\begin{eqnarray}
    &\ket{\uparrow\downarrow +-\uparrow \downarrow}, \ket{\downarrow\uparrow +-\uparrow \downarrow}, \ket{\uparrow\downarrow +-\downarrow \uparrow}, \ket{\downarrow\uparrow +- \downarrow\uparrow}\nn \\
    &\ket{\uparrow +-\downarrow \uparrow \downarrow}, \ket{\uparrow +-\uparrow \downarrow \downarrow}, \ket{\uparrow +-\downarrow \downarrow\uparrow } \nn \\
    &\ket{\uparrow \downarrow\downarrow+-\uparrow}, \ket{\downarrow\uparrow \downarrow +-\uparrow}, \ket{\downarrow\downarrow \uparrow+-\uparrow}.
\end{eqnarray}
Note that all the states in $\mathcal{K}\left(H, \ket{\psi_0}\right)$ are labelled by $(N, N^{(1)}_\uparrow, N^{(2)}_{\uparrow}) = (6, 1, 1)$. 
In order to map configurations of the form Eq.~(\ref{eq:singledipolequants}) onto an effective spin-1/2 Hilbert space, note that the rules of Eq.~(\ref{eq:dipolescattering}) are \textit{identical} to those of Eq.~(\ref{eq:spinscattering}) when the dipole $+-$ is replaced by an $\uparrow$ spin.
This observation allows us to establish two crucial results on the single-dipole Krylov subspace $\mathcal{K}\left(H, \ket{\psi_0}\right)$.
Firstly, product states in the single dipole Krylov subspace consisting of a $+ -$ dipole can be \textit{uniquely} mapped onto product states of $(N - 1)$ spin-1/2's with $(N^{(1)}_\uparrow + N^{(2)}_\uparrow + 1)$ $\uparrow$'s by replacing the $+ -$ dipole with an $\uparrow$.
For example, the following holds:
\begin{equation}
    \underset{(A)}{\ket{\uparrow \uparrow \downarrow + - \uparrow \downarrow \uparrow \uparrow \uparrow}} \iff  \underset{(B)}{\ket{\uparrow \uparrow \downarrow \uparrow \uparrow \downarrow \uparrow \uparrow\uparrow}},
\label{eq:singledipolemapping}
\end{equation}
where configuration (A) in the Krylov subspace with $(N, N^{(1)}_\uparrow, N^{(2)}_{\uparrow}) = (10, 2, 4)$ maps onto the configuration (B) in the spin subspace with $(N, N_\uparrow) = (9, 6)$ by replacing the $+ -$ dipole with an $\uparrow$. 
The inverse mapping from the spin-1/2 Hilbert space of $(N - 1)$ sites and $(N^{(1)}_\uparrow + N^{(2)}_\uparrow + 1)$ $\uparrow$'s to the single dipole Krylov subspace $(N, N^{(1)}_\uparrow, N^{(2)}_\uparrow)$ proceeds by identifying one $\uparrow$ to be the $+ -$ dipole such that the resulting configuration has the correct $N^{(1)}_\uparrow$ and $N^{(2)}_\uparrow$.  
For instance in Eq.~(\ref{eq:singledipolemapping}), given $(N, N^{(1)}_\uparrow, N^{(2)}_\uparrow) = (10, 2, 4)$, the mapping from (B) to (A) is possible only if the third $\uparrow$ in the configuration (B) is replaced by a $+ -$ dipole.
The mapping for the single $-+$ dipole subspace follows analogously, with $\uparrow$ replaced by $\downarrow$ i.e., by identifying $-+$'s with $\downarrow$'s instead. In that case, the quantities $N^{(1)}_\downarrow$ and $N^{(2)}_\downarrow$, defined as
\begin{equation}
\underset{\underbrace{\hspace{10mm}}_{N^{(1)}_\downarrow}\hspace{6mm}\underbrace{\hspace{10mm}}_{N^{(2)}_\downarrow}}{\ket{\ast \cdots \ast - + \ast \cdots \ast}} \, ,
\end{equation}
are preserved within the Krylov subspace.
Thus, the single dipole Krylov subspace with OBC and a fixed $(N, N^{(1)}_\uparrow, N^{(2)}_{\uparrow})$ (resp. $(N, N^{(1)}_\downarrow, N^{(2)}_{\downarrow})$) is \textit{isomorphic} to the Hilbert space of $(N - 1)$ spin-1/2's with $(N^{(1)}_\uparrow + N^{(2)}_\uparrow + 1)$ $\uparrow$'s (resp. $(N^{(1)}_\downarrow + N^{(2)}_\downarrow + 1)$ $\downarrow$'s).
Secondly, since Eq.~(\ref{eq:dipolescattering}) is identical to Eq.~(\ref{eq:spinscatteringkryl}) when the dipole $+ - $ (resp. $-+$) is replaced with an $\uparrow$ (resp. $\downarrow$), the effective Hamiltonian within each such Krylov subspace is the XX model of  $(N - 1)$ sites with OBC.\footnote{Once an $\uparrow$ spin is identified, note that the action of the XX Hamiltonian also preserves $N^{(1)}_\uparrow$ and $N^{(2)}_\uparrow$, the number of $\uparrow$ spins to the left and to the right of the identified $\uparrow$ spin respectively.}
In particular, the spectrum of $H$ in Eq.~(\ref{eq:hamilXX}) restricted to the single Krylov subspace labelled by $(N, N^{(1)}_\uparrow, N^{(2)}_\uparrow)$ (resp. $(N, N^{(1)}_\uparrow, N^{(2)}_\uparrow)$) is precisely the spectrum of the quantum number sector $S_z = (2(N^{(1)}_\uparrow + N^{(2)}_\uparrow) + 3 - N)$ (resp. $S_z = -(2(N^{(1)}_\downarrow + N^{(2)}_\downarrow) + 3 - N)$) of the XX model.
Note that with PBC this Krylov subspace is no longer isomorphic to the spin-1/2 Hilbert space of the XX model, since the inverse mapping from the spin-1/2 Hilbert space to the dipole subspace is not unique.
Thus, the effective Hamiltonian within this Krylov subspace cannot map exactly onto the XX model of Eq.~(\ref{eq:hamilXX}) with PBC, and it remains unclear whether or not the resulting Hamiltonian is integrable for any finite system size.
\subsection{Multidipole subspaces}
\label{sec:multidipolesubspace}
We now consider Krylov subspaces generated by root configurations containing multiple identically oriented dipoles.
All such subspaces turn out to be integrable and governed by effective XX Hamiltonians of various sizes. 
As with a single dipole discussed in the previous section, spins and dipoles interact according to Eq.~(\ref{eq:dipolescattering}). 
A crucial property of these rules, which we will make use of throughout this section, is that the $+ -$ (resp. $- +$) dipole cannot cross any $\uparrow$ (resp. $\downarrow$) spins under the action of the Hamiltonian $H$.  
We first illustrate the case where the root state contains two $+-$ dipoles before discussing the general setting.
Since the dipoles $+ -$ cannot cross $\uparrow$'s, the Krylov subspace generated from a root state with two identically oriented dipoles preserves three quantities of the root state: $(N^{(1)}_{\uparrow}, N^{(2)}_{\uparrow}, N^{(3)}_{\uparrow})$, depicted schematically by the following configurations:
\begin{eqnarray}
	&\underset{\underbrace{\hspace{10mm}}_{N^{(1)}_\uparrow}\hspace{6mm}\underbrace{\hspace{10mm}}_{N^{(2)}_\uparrow}\hspace{6mm}\underbrace{\hspace{10mm}}_{N^{(3)}_\uparrow}}{\ket{\ast \cdots \ast + - \ast \cdots \ast + - \ast \cdots \ast}}, 
\label{eq:multidipoleroot}
\end{eqnarray}
where $\ast =\ \uparrow, \downarrow$. 
That is, for a Krylov subspace generated by root states with two $+ - $ dipoles, the number of $\uparrow$ spins to the left of the left dipole, in between the two dipoles, and to the right of the right dipole are each separately conserved. 
Thus, the quantities $(N, N^{(1)}_{\uparrow}, N^{(2)}_{\uparrow}, N^{(3)}_{\uparrow})$ uniquely label the Krylov subspace.  
We now restrict our discussion to the Krylov subspace containing two $+ -$ dipoles, with the generalization to the two $- +$ dipole subspace being straightforward.
Provided $N^{(2)}_{\uparrow} \geq 1$ in the root state $\ket{\psi_0}$, the two dipoles are always separated by an $\uparrow$ spin and can never be adjacent to each other; the action of the Hamiltonian is therefore entirely specified by Eq.~(\ref{eq:dipolescattering}).
Product states in the Krylov subspace can be mapped onto configurations of $(N-2)$ spin-1/2's with $\left(N^{(1)}_\uparrow + N^{(2)}_\uparrow + N^{(3)}_\uparrow + 2\right)$ $\uparrow$'s by replacing the $+ -$ dipoles by $\uparrow$'s.
For example,
\begin{equation}
    \underset{(A)}{\ket{\uparrow \downarrow \uparrow +- \uparrow \downarrow\uparrow +- \uparrow \downarrow}} \iff \underset{(B)}{\ket{\uparrow\downarrow\uparrow\uparrow\uparrow\downarrow\uparrow\uparrow\uparrow\downarrow}},
\label{eq:twodipolemapping}
\end{equation}
where the configuration (A) in the two-dipole Krylov subspace labeled by $\left(N, N^{(1)}_\uparrow, N^{(2)}_\uparrow, N^{(3)}_\uparrow\right) = (12, 2, 2, 1)$, maps onto configuration (B).
Similar to the single dipole case, the inverse mapping is unique once $(N, N^{(1)}_\uparrow, N^{(2)}_\uparrow, N^{(3)}_\uparrow)$ are specified. 
This inverse mapping proceeds by identifying two of the $\uparrow$'s to be $+ -$ dipoles such that the resulting configuration has the required values of $N^{(1)}_\uparrow$, $N^{(2)}_\uparrow$, and $N^{(3)}_\uparrow$. 
For example, given that $(N, N^{(1)}_\uparrow, N^{(2)}_\uparrow, N^{(3)}_\uparrow) = (12, 2, 2, 1)$, the two-dipole configuration (A) in Eq.~(\ref{eq:twodipolemapping}) is the unique two-dipole configuration corresponding to spin configuration (B). 
The mapping for the two-dipole subspace with $-+$ dipoles follows analogously, with $\uparrow$ replaced by $\downarrow$ i.e., by identifying $-+$'s with $\downarrow$'s instead.
The action of the Hamiltonian is completely specified by Eq.~\eqref{eq:dipolescattering} when the dipoles are not allowed to be adjacent each other; as discussed in Sec.~\ref{sec:dipolesubspace}, Eq.~(\ref{eq:dipolescattering}) is identical to Eq.~(\ref{eq:spinscatteringkryl}) when the $+ -$ (resp. $- +$) dipole is identified with $\uparrow$ (resp. $\downarrow$) spin.
Thus, the Hamiltonian restricted to the two $+-$ (resp. $-+$) dipole Krylov subspace is identical to the XX model of $(N - 2)$ sites within the $S_z = (2 (N^{(1)}_\uparrow + N^{(2)}_\uparrow + N^{(3)}_\uparrow) + 6 - N)$ (resp. $S_z = -(2 (N^{(1)}_\downarrow + N^{(2)}_\downarrow + N^{(3)}_\downarrow) + 6 - N)$) sector.
We emphasize that the two-dipole Krylov subspace of $N$ is isomorphic to the spin-1/2 Hilbert space of $(N - 2)$ sites \textit{only} when the two $+ -$ (resp. $- +$) dipoles have at least one $\uparrow$ (resp. $\downarrow$) spin between them i.e., only if $N^{(2)}_\uparrow \geq 1$ (resp. $N^{(2)}_\downarrow \geq 1$).
When the two dipoles are adjacent to each other, using Eqs.~(\ref{eq:+fracton}) and (\ref{eq:-fracton}) we find that the action of the Hamiltonian $H$ reads
\begin{eqnarray}
    &&\ket{+ - + -} \leftrightarrow \ket{\downarrow + \uparrow -},\;\;\;\ket{+ - + -} \leftrightarrow \ket{+ \uparrow - \downarrow}, \nn \\
    &&\ket{- + - +} \leftrightarrow \ket{\uparrow - \downarrow +},\;\;\;\ket{- + - +} \leftrightarrow \ket{- \downarrow + \uparrow}. \nn \\
\label{eq:multidipolescattering}
\end{eqnarray}
As a consequence, the action of the Hamiltonian on root states of the form $\ket{\cdots + - + -\cdots}$ result in the ``disintegration" of dipoles, resulting in configurations of the form:
$$\ket{\cdots \downarrow + \uparrow -  \cdots}, \quad \ket{\cdots + \uparrow - \downarrow \cdots},$$
which cannot be mapped onto a configuration of $(N - 2)$ spin-1/2's through the map described earlier in this section. 
Nevertheless, we find that such Krylov subspaces \textit{does} map onto the XX model, albeit one with $(N - 1)$ spin-1/2's; we discuss this mapping in App.~\ref{sec:dipoletouch}. 
The preceding discussion straightforward generalizes to three or more dipoles. 
For a Krylov subspace generated by a root state containing $n$ identically oriented dipoles, with OBC the system can be partitioned into $(n+1)$ segments separated by the dipoles. 
We introduce the quantities $N^{(1)}_\uparrow, N^{(2)}_\uparrow, \cdots, N^{(n + 1)}_{\uparrow}$, where 
$N^{(j)}_\uparrow$ (resp. $N^{(j)}_\downarrow$) represents the number of $\uparrow$ (resp. $\downarrow$) spins in the $j$-th segment of the chain in the root state:
\begin{eqnarray}
    &\overset{1\;\;\;\;\;\;\;\;\;\;\;\;2\;\;\;\;\;\;\;\;\;\;\;\;n-1\;\;\;\;\;\;\;\;\;\;\;\;n}{\underset{\;\;\;\;\underbrace{\hspace{4mm}}_{N^{(1)}_\uparrow}\;\;\;\;\;\;\underbrace{\hspace{4mm}}_{N^{(2)}_\uparrow}\;\;\;\;\;\;\;\;\;\;\;\;\;\;\;\;\;\;\;\;\;\underbrace{\hspace{4mm}}_{N^{(n)}_\uparrow}\;\;\;\;\;\;\underbrace{\hspace{4mm}}_{N^{(n+1)}_\uparrow}}{\ket{\cdots + - \cdots + - \cdots + - \cdots + -\cdots}}},
\label{eq:ndipolekrylov}
\end{eqnarray}
with the superscripts $1, 2, \cdots, n$ indexing the dipoles. 
Since a $+ -$ dipole is not allowed to cross an $\uparrow$ spin under the action of the Hamiltonian, the quantities $\{N^{(j)}_\uparrow \geq 1\}$ are invariant under the dynamics i.e., these quantities are identical for all product states within the Krylov subspace generated by the root state of the form Eq.~(\ref{eq:ndipolekrylov}). As with two dipoles, this is true provided no dipoles are adjacent in the root state, which corresponds to the constraint $N^{(j)}_\uparrow \geq 1$ for any $j$. 

In this case ($N^{(j)} \neq 0 \, \forall \, j$), the $n$ dipole Krylov subspace exactly maps onto a spin-1/2 Hilbert space with $(N - n)$ sites and $(n + \sumal{j = 1}{n + 1}{N^{(j)}_\uparrow})$ $\uparrow$'s by identifying each $+ -$ dipole with an $\uparrow$ spin.
For example,
\begin{equation}
    \underset{(A)}{\ket{\uparrow\downarrow + - \downarrow \uparrow \uparrow + - \downarrow \uparrow + - \uparrow \uparrow}} \iff \underset{(B)}{\ket{\uparrow \downarrow \uparrow \downarrow \uparrow \uparrow \uparrow \downarrow \uparrow \uparrow \uparrow \uparrow}},
\label{eq:threedipolemapping}
\end{equation}
where $n=3$ and where the three dipole configuration (A) with $(N^{(1)}_\uparrow, N^{(2)}_\uparrow, N^{(3)}_\uparrow, N^{(4)}_\uparrow) = (1, 2, 1, 2)$ maps onto the spin configuration (B).
This mapping onto the spin-1/2 Hilbert space is invertible provided the tuple $(N^{(1)}_\uparrow, N^{(2)}_\uparrow, \cdots, N^{(n+1)}_\uparrow)$ is known, and it proceeds by identifying $n$ $\uparrow$ spins in each product configuration with $+ -$ dipoles such that the resulting configuration has the requisite $(N^{(1)}_\uparrow, N^{(2)}_\uparrow, \cdots, N^{(n+1)}_\uparrow)$ values.
For example, given $(N^{(1)}_\uparrow, N^{(2)}_\uparrow, N^{(3)}_\uparrow, N^{(4)}_\uparrow) = (1, 2, 1, 2)$, configuration (B) in Eq.~(\ref{eq:threedipolemapping}) uniquely maps onto (A) by identifying the appropriate $\uparrow$ spins with $+-$ dipoles.
The mapping with $-+$ dipoles proceeds in a similar way by replacing the $-+$ dipole by $\downarrow$. 
The quantities $\{N^{(j)}_\downarrow\}$ are thus preserved within the Krylov subspaces, where
\begin{equation}
    \overset{1\;\;\;\;\;\;\;\;\;\;\;\;2\;\;\;\;\;\;\;\;\;\;\;\;n-1\;\;\;\;\;\;\;\;\;\;\;\;n}{\underset{\;\;\;\;\underbrace{\hspace{4mm}}_{N^{(1)}_\downarrow}\;\;\;\;\;\;\underbrace{\hspace{4mm}}_{N^{(2)}_\downarrow}\;\;\;\;\;\;\;\;\;\;\;\;\;\;\;\;\;\;\;\;\;\underbrace{\hspace{4mm}}_{N^{(n)}_\downarrow}\;\;\;\;\;\;\underbrace{\hspace{4mm}}_{N^{(n+1)}_\downarrow}}{\ket{\cdots - + \cdots - + \cdots - + \cdots - + \cdots}}}.
\end{equation}
Since the Hamiltonian Eq.~(\ref{eq:dipolescattering}) is identical to Eq.~(\ref{eq:spinscattering}) upon the identification of dipoles with spins, the Hamiltonian restricted to the Krylov subspace for the $n$ $+-$ (resp. $- +$) dipole case is the XX model with $(N - n)$ sites within the quantum number sector $S_z = (3n + \sumal{j = 1}{n+1}{N^{(j)}_\uparrow} - N)$ (resp. $S_z = -(3n + \sumal{j = 1}{n+1}{N^{(j)}_\downarrow} - N)$).
When $N^{(j)}_\uparrow = 0$ or $N^{(j)}_\downarrow = 0$ for some $j$ in the root state Eq.~(\ref{eq:ndipolekrylov}), the mapping prescribed above fails because the action of the Hamiltonian causes the adjacent dipoles to disintegrate, as shown in Eq.~(\ref{eq:multidipolescattering}).   
Nevertheless, as we show in App.~\ref{sec:dipoletouch}, we find that the Krylov subspace remains integrable even if some dipoles in the root state are adjacent.
Specifically, we find that the Hamiltonian restricted to a Krylov subspace with \textit{only} $n$ $+ -$ (resp. $- +$) dipoles is the XX model of $(N - n + X)$ sites, where $X$ is the number of segments $j$ containing no spins, such that $N^{(j)}_\uparrow = 0$ (resp. $N^{(j)}_\downarrow = 0$). 
For example, the effective Hamiltonians restricted to the Krylov subspaces generated by the root states $$\ket{\ast \cdots \ast + - + - + - \ast \cdots \ast}$$ and $$\ket{\ast \cdots \ast + - + - \ast \cdots \ast + - \ast\cdots \ast},$$ where $\ast =\ \uparrow, \downarrow$ are the XX models acting on $(N - 1)$ and $(N - 2)$ spin-1/2's respectively.
As was the case for a single dipole, the mapping onto XX models does not work with PBC.
However, it is not clear if the effective Hamiltonian restricted to this sector with PBC is solvable for a finite system size, although integrability of this sector should be restored in the thermodynamic limit and the energy spectrum should display Poisson level statistics for a large enough system size.  
Finally, we note that upon the addition of electrostatic terms or disorder (discussed in App.~\ref{sec:electroaction}), the spin subspace described in Sec.~\ref{sec:spinsubspace} maps onto the XXZ model or disordered XX model, and thus remains integrable. 
However, the dipole subspaces are no longer integrable, and they show all the signs of usual non-integrability, including GOE level statistics~\cite{poilblanc1993poisson}. 
\subsection{Systematic construction of integrable subspaces}
Having illustrated the existence of several integrable Krylov subspaces of the pair-hopping model Eq.~(\ref{eq:pairhopping}), we briefly discuss a general prescription for constructing additional irreducible integrable subspaces by using the integrable subspaces of Secs.~\ref{sec:spinsubspace}-\ref{sec:multidipolesubspace} as building blocks.
As also emphasized in Refs.~\cite{sala2019ergodicity, khemani2019local}, one can introduce \textit{blockades} i.e., regions of the chain on which terms of the Hamiltonian vanish.
For example, consider the following root state with a configuration of the form:
\begin{equation}
    \underset{\underbrace{\hspace{10mm}}_A\hspace{20mm}\underbrace{\hspace{10mm}}_B}{\ket{\ast \cdots \ast ++\cdots++ \ast \cdots \ast}}, 
\label{eq:blockade}
\end{equation}
where $\ast =\ \uparrow, \downarrow$, with $N_+ \geq 2$ and $N_- = 0$.
Following the rules Eqs.~(\ref{eq:spinscattering})-(\ref{eq:-fracton}), the Hamiltonian can act non-trivially only on sites contained within regions $A$ and $B$ of the root state Eq.~(\ref{eq:blockade}).

Due to this, all basis states of the Krylov subspace generated from the root state Eq.~(\ref{eq:blockade}) retain the same schematic form, with $++ \cdots ++$ ($N_+ \geq 2$) acting as a blockade that spatially disconnects two parts of the Krylov subspace. 

Thus, one can show that the effective Hamiltonian restricted to such blockaded Krylov subspaces is simply given by the sum of two independent XX models acting on distinct degrees of freedom lying in regions $A$ and $B$. 
Note that blockades can also be constructed using exponentially many  other ``static" patterns~\cite{sala2019ergodicity, khemani2019local},  such as $--\cdots--$, $++--\cdots++--$, or $++\uparrow\cdots\uparrow++$,   which in turn lead to exponentially many integrable subspaces.
Similarly, we can also introduce blockades for the dipole Krylov subspaces considered in Secs.~\ref{sec:dipolesubspace}-\ref{sec:multidipolesubspace}, as long as the dipoles do not interact with the blockade. 
For example, consider the root configuration of the form of Eq.~(\ref{eq:blockade}) where region $A$ is a root configuration for an integrable subspace with one or more $-+$ dipoles, and region $B$ is a root configuration for an integrable subspace with $+-$ dipoles:
\begin{equation}
    \underset{\underbrace{\hspace{25mm}}_A\hspace{20mm}\underbrace{\hspace{25mm}}_B}{\ket{\ast \cdots \ast - + \ast \cdots \ast ++\cdots++ \ast \cdots \ast + - \ast \cdots \ast}}, 
\label{eq:blockade2}
\end{equation}
where $\ast = \uparrow, \downarrow$.
Upon successive applications of the Hamiltonian on the root state of Eq.~(\ref{eq:blockade2}), the dipoles in regions $A$ and $B$ do not interact with the string of $+$'s in between the regions.
Thus, the string of $+$'s acts as a blockade, and the Krylov subspaces generated by such root configurations are integrable, since the restricted Hamiltonian is a sum of XX models on regions $A$ and $B$. 
While we have only illustrated the simplest cases where blockades are introduced between regions $A$ and $B$, each of which are integrable regions that do not interact with the blockade, we can of course generalise by introducing $n$ blockades separating $n+1$ regions, each of which contain the integrable subspaces that do not interact with the neighboring blockades.
In such a case, the Hamiltonian restricted to the Krylov subspace is a sum of $n+1$ independent XX models.

A detailed study delineating \textit{all} integrable subspaces of the pair-hopping model Eq.~\eqref{eq:pairhopping} is beyond the scope of this work. 
Nevertheless, the above examples suffice to illustrate the existence of \textit{exponentially many} integrable Krylov subspaces, clearly establishing the possibility of emergent, Krylov-restricted integrability in systems exhibiting Krylov fracture.

%
%

\section{Non-integrable subspaces and Krylov-Restricted ETH}
\label{sec:nonintkrylov}
\begin{figure*}[t!]
    \centering
    \begin{tabular}{cc}
    \includegraphics[scale = 0.45]{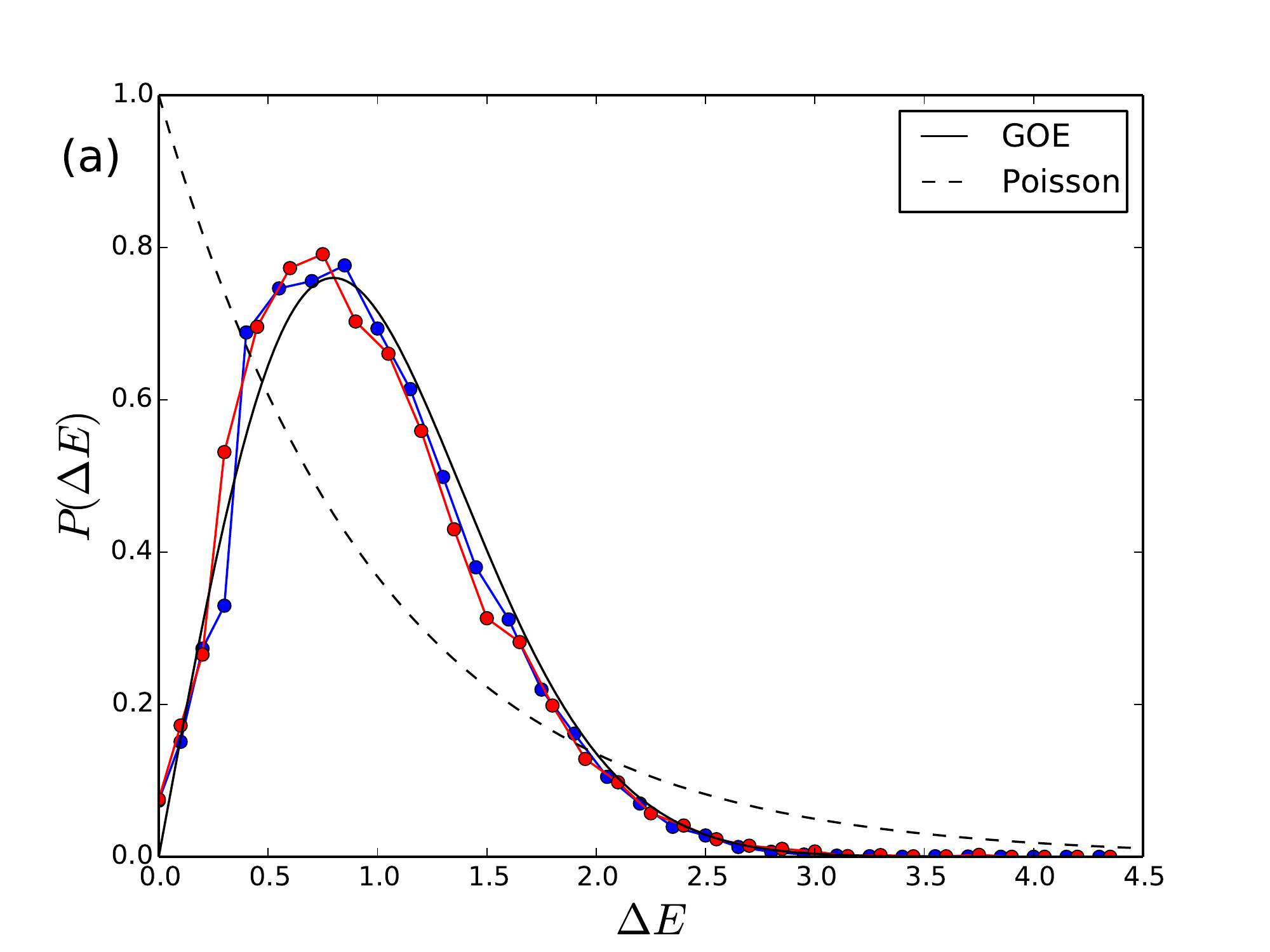}&\includegraphics[scale = 0.45]{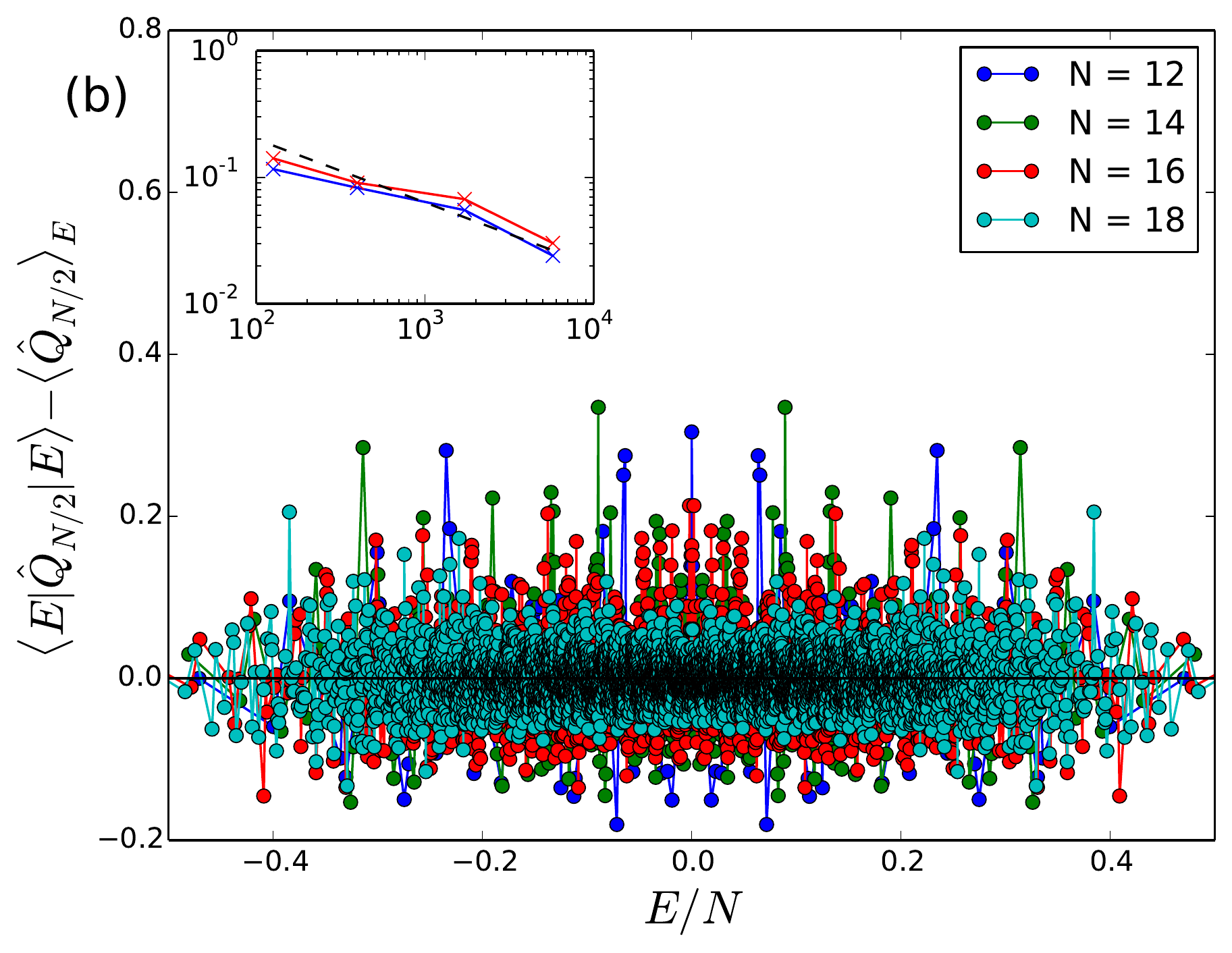}
    \end{tabular}
    \caption{(Color online) (a) Level statistics within the Krylov subspace $\mathcal{K}\left(H, \ket{\psi_0}\right)$ generated by various root states $\ket{\psi_0}$ with OBC, where $H$ is the pair-hopping Hamiltonian Eq.~(\ref{eq:pairhopping}). Red: $\ket{\psi_0} = \ket{\downarrow \uparrow \uparrow \uparrow \downarrow \downarrow \downarrow - + + - \downarrow \downarrow \downarrow \uparrow \uparrow \downarrow \downarrow}$ (Krylov subspace dimension $\mathcal{D}_{\mathcal{K}}$ = 18849), Blue: $\ket{\psi_0} = \ket{\uparrow \downarrow \uparrow \downarrow \uparrow + \uparrow \uparrow \uparrow \uparrow \downarrow \downarrow \downarrow - \downarrow \uparrow \downarrow \uparrow \downarrow}$ (Krylov subspace dimension $\mathcal{D}_{\mathcal{K}}$ = 21660). The configurations of spins in the root configurations have been chosen to ensure that the Krylov subspace does not have any symmetries. The standard $\langle r \rangle$ parameter~\cite{atas2013distribution} in these subspaces is $0.5331$ and $0.5276$ respectively, close to the GOE value of $0.53$. 
    (b) Evidence for the Eigenstate Thermalization Hypothesis (ETH) in the non-integrable Krylov subspace $\mathcal{K}\left(H, \ket{\psi_0}\right)$ generated by the root state shown in Eq.~(\ref{eq:ethrootstate}), which for $N = 18$ reads $\ket{\psi_0} = \ket{\uparrow\downarrow\uparrow\downarrow\uparrow\downarrow\uparrow-++-\downarrow\uparrow\downarrow\uparrow\downarrow\uparrow\downarrow}$. In order to break the symmetries within this Krylov subspace, the couplings $\{J_j\}$ of the terms of the pair-hopping Hamiltonian (see Eq.~(\ref{eq:disoderedpairhopping})) are chosen from a uniform distribution $[1-W, 1 + W]$, with $W = 0.1$. This disorder preserves the Krylov fractured structure of the Hilbert space. Main: The difference between $\bra{E}\widehat{Q}_{N/2}\ket{E}$, the expectation value of the charge operator in an eigenstate at energy $E$ and $\langle \widehat{Q}_{N/2}\rangle_E$, the thermal expectation value at that energy determined by averaging $\bra{E}\widehat{Q}_{N/2}\ket{E}$ in an energy window of $\Delta E = 0.05$~\cite{beugeling2014finite}. Inset: The standard deviations of that difference as a function of the Hilbert dimension $\mathcal{D}_{\mathcal{K}}$ scales $\sim 1/\sqrt{\mathcal{D}_{\mathcal{K}}}$ (dotted line) for two operators: the charge operator $\widehat{Q}_{N/2}$ (blue) and the spin operator $\widehat{S}^z_{N/2}$ (red), consistent with ETH within the Krylov subspace $\mathcal{K}\left(H, \ket{\psi_0}\right)$.}
    \label{fig:nonint}
\end{figure*}
Given that large swaths of the spectrum of the pair-hopping Hamiltonian are solvable, it is natural to ask whether this model is \textit{completely} integrable.
The standard diagnostic for probing non-integrability of some Hamiltonian is the appearance of random matrix behavior \textit{within} a sector resolved by symmetries of that Hamiltonian. For example, the energy level statistics~\cite{poilblanc1993poisson, rahul2015review} and the matrix elements of local operators in the energy eigenbasis (according to ETH)~\cite{srednicki1994chaos} are expected to follow random matrix behavior for non-integrable systems. 

Generally, in unconstrained models, symmetry sectors are themselves examples of well-defined dynamically disconnected Krylov subspaces.
In other words, a root-state which is an eigenstate of the symmetry typically generates a Krylov subspace which spans all states within that symmetry sector. However, for systems exhibiting Krylov fracture, there exist several dynamically disconnected Krylov subspaces \textit{within} each symmetry sector.
As was also emphasized by Refs.~\cite{sala2019ergodicity,khemani2019local}, resolving eigenstates by symmetries alone may hence be insufficient for identifying ergodicity, given the possibility of Krylov fracture.
Thus, we pose the crucial question that motivates the title of the paper: Whether symmetries are only a subset of the more general phenomena of Krylov fracture, and if ergodicity or its absence should correspondingly be defined within dynamically disconnected irreducible Krylov subspaces. 
In the previous section, we encountered examples of Krylov subspaces within symmetry sectors which display the characteristic trademarks of integrable systems e.g., Poisson level statistics. Now, we wish to ask whether Krylov subspaces that are \textit{not} integrable exhibit conventional diagonostics of ergodic systems, such as Wigner-Dyson level statistics and ETH~\cite{srednicki1994chaos}.
Of course, random matrix theory is a statement about ``large" matrices i.e., in the limit that the size of the matrix goes to infinity; consequently, the question of thermalization within Krylov subspaces is only well-posed for ``large" Krylov subspaces, whose size tends to infinity in the thermodynamic limit. Thus, we explore some simple non-integrable Krylov subspaces of the pair-hopping model Eq.~(\ref{eq:pairhopping}) and, in the process, establish the notion of Krylov-restricted ETH.
Indeed, there exist Krylov subspaces of the pair-hopping Hamiltonian which are not integrable. 
Consider for instance the Krylov subspace generated by the root state containing both $+ -$ \textit{and} $-+$ dipoles:
\begin{equation}
    \ket{\psi_0} = \ket{\ast \cdots \ast - + + - \ast \cdots \ast},
\label{eq:psi01}
\end{equation}
where $\ast =\ \uparrow, \downarrow$. 
Since the dipoles are of opposite orientation, the mapping of the $+ -$ and $- +$ dipoles to $\uparrow$ and $\downarrow$ spins would only be justified if $\ket{- + + -} \leftrightarrow \ket{+ - - +}$ under the action of the Hamiltonian, which is strictly prohibited by the rules given in Eqs.~(\ref{eq:spinscattering})-(\ref{eq:-fracton}). As a result, the Hamiltonian restricted to this Krylov subspace does not need to map onto an integrable model.
Another example is the Krylov subspace generated by the root state containing two separated fractons:
\begin{equation}
    \ket{\psi_0} = \ket{\ast \cdots \ast + \ast \cdots \ast - \ast \cdots \ast}, 
\label{eq:psi02}
\end{equation}
where $\ast =\ \uparrow, \downarrow$, and the $\ast \cdots \ast$ in between the $+$ and $-$ contains both $\uparrow$ and $\downarrow$ spins. 
The latter condition is required to ensure that $\ket{\psi_0}$ does not belong to any of the integrable multidipole Krylov subspaces discussed in App.~\ref{sec:dipoletouch}.

We have numerically studied the behaviour of Krylov subspaces generated by root states such as those given in Eqs.~\eqref{eq:psi01} and~\eqref{eq:psi02}. 
As shown in Fig.~\ref{fig:nonint}(a), we find that eigenstates of the Hamiltonian within these Krylov subspaces $\mathcal{K}\left(H, \ket{\psi_0}\right)$ exhibit GOE level statistics, providing evidence for the non-integrability of the Krylov subspace.
We further {\bf conjecture} that such non-integrable Krylov subspaces satisfy the Eigenstate Thermalization Hypothesis (ETH)~\cite{deutsch1991quantum, srednicki1994chaos, rigol2008thermalization, polkovnikov2011colloquium, d2016quantum}. 
ETH states that the matrix elements of local operators in the energy eigenstates of a non-integrable model take the form~\cite{d2016quantum}
\begin{equation}
    \bra{E_m}\widehat{O}\ket{E_n} = \bar{O}\left(E\right)\delta_{m, n} + R_{m, n} e^{-S\left(E\right)/2} f_O\left(E, \omega \right),
\label{eq:ethmatrixel}
\end{equation}
where $\widehat{O}$ is a local operator that is invariant under the symmetries of the Hamiltonian, $\ket{E_m}$ and $\ket{E_n}$ are the energy eigenstates with energies $E_m$ and $E_n$ with the same symmetry quantum numbers, $E = \left(E_m + E_n\right)/2$, $\omega = E_m - E_n$, $R_{m,n}$ is a random variable with zero mean and unit variance, $\bar{O}\left(E\right)$ is a smooth function of $E$ and represents the thermal expectation value of $\widehat{O}$ at energy $E$,\footnote{The thermal value here is determined by averaging the eigenstate expectation values $\bra{E} \widehat{O} \ket{E}$ over a small energy window $\Delta E$, where $\ket{E}$ is an eigenstate with energy $E$.~\cite{beugeling2014finite}} $f_{O}\left(E, \omega\right)$ is a smooth function of $E$ and $\omega$ which do not scale with the system size~\cite{d2016quantum}, and $S\left(E\right)$ is the thermodynamic entropy at energy $E$.
In Eq.~(\ref{eq:ethmatrixel}), since $S(E) \sim \log \mathcal{D}$ for states in the middle of the spectrum, where $\mathcal{D}$ is the Hilbert space dimension, the standard deviation of expectation values of operators in the eigenstates is expected to scale as $\sim 1/\sqrt{\mathcal{D}}$ for eigenstates in the middle of the spectrum~\cite{beugeling2014finite}.
Here, we want to test whether Eq.~(\ref{eq:ethmatrixel}) holds within a non-integrable Krylov subspace. 
We focus on the Krylov subspace with the root states (with OBC):
\begin{equation}
    \ket{\psi_0} = \twopartdef{\ket{\uparrow\downarrow \cdots \uparrow \downarrow - + + - \uparrow \downarrow \cdots \uparrow \downarrow}}{N = 4p}{\ket{\uparrow\downarrow \cdots \uparrow \downarrow \uparrow  - + + - \downarrow \uparrow \downarrow \cdots \uparrow \downarrow}}{N = 4p + 2},
\label{eq:ethrootstate}
\end{equation}
with two dipoles $-+$ and $+-$ placed at the center of the chain.
Furthermore, to probe the validity of Eq.~(\ref{eq:ethmatrixel}), we need to choose an operator $\widehat{O}$ that preserves the Krylov subspaces. 
Hence we choose the charge operator on the $N/2$-th site $\widehat{O} = \widehat{Q}_{N/2}$, which is diagonal in the basis of product states. 
Since the Krylov subspace $\mathcal{K}\left(H, \ket{\psi_0}\right)$ has symmetries (e.g. inversion symmetry), we add disorder to the couplings of the pair-hopping Hamiltonian (see Eq.~(\ref{eq:disoderedpairhopping})), which does not affect the structure of the Krylov subspaces of the Hamiltonian, and focus on testing the ergodicity within the Krylov subspace. 
To probe the validity of Eq.~(\ref{eq:ethmatrixel}) within non-integrable Krylov subspaces, in Fig.~\ref{fig:nonint}(b) we plot the quantity $\left(\langle E | \widehat{O} | E\rangle - \bar{O}\left(E\right)\right)$, as a function of $E$, where $\ket{E}$ is the eigenstate with energy $E$.
The inset show the variance of the difference as a function of the Krylov subspace dimension $\mathcal{D}_\mathcal{K}$.
Two observations in Fig.~\ref{fig:nonint}(b) suggest the validity of ETH within the Krylov subspace.  
Firstly, the quantity $\langle E | \widehat{O} | E\rangle - \bar{O}\left(E\right)$ is centered about $0$, which shows that eigenstate expectation values approach the thermal expectation value.
Secondly, the standard deviation of the difference (shown in the inset) scales as $\sim 1/\sqrt{\mathcal{D}_\mathcal{K}}$, the dimension of the Krylov subspace.
Hence these observations provide evidence for ``diagonal ETH" \textit{within} non-integrable Krylov subspaces, supporting the existence of Krylov-restricted ETH in systems exhibiting Krylov fracture.
%
%
%

\section{Quasilocalization from Thermalization}
\label{sec:quasilocal}
\begin{figure*}[t!]
    \centering
    \begin{tabular}{cc}
    \includegraphics[scale = 0.45]{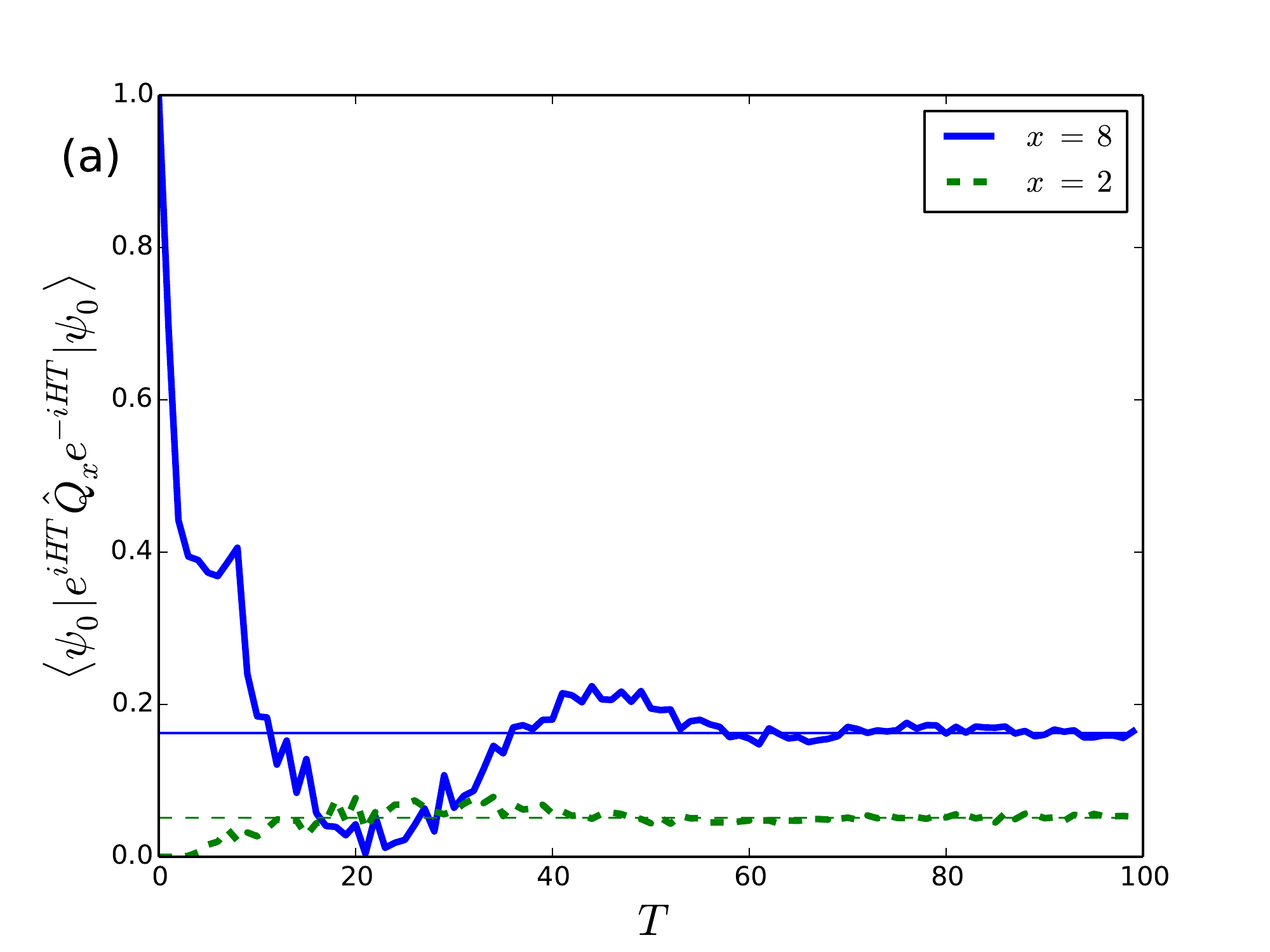}&\includegraphics[scale = 0.45]{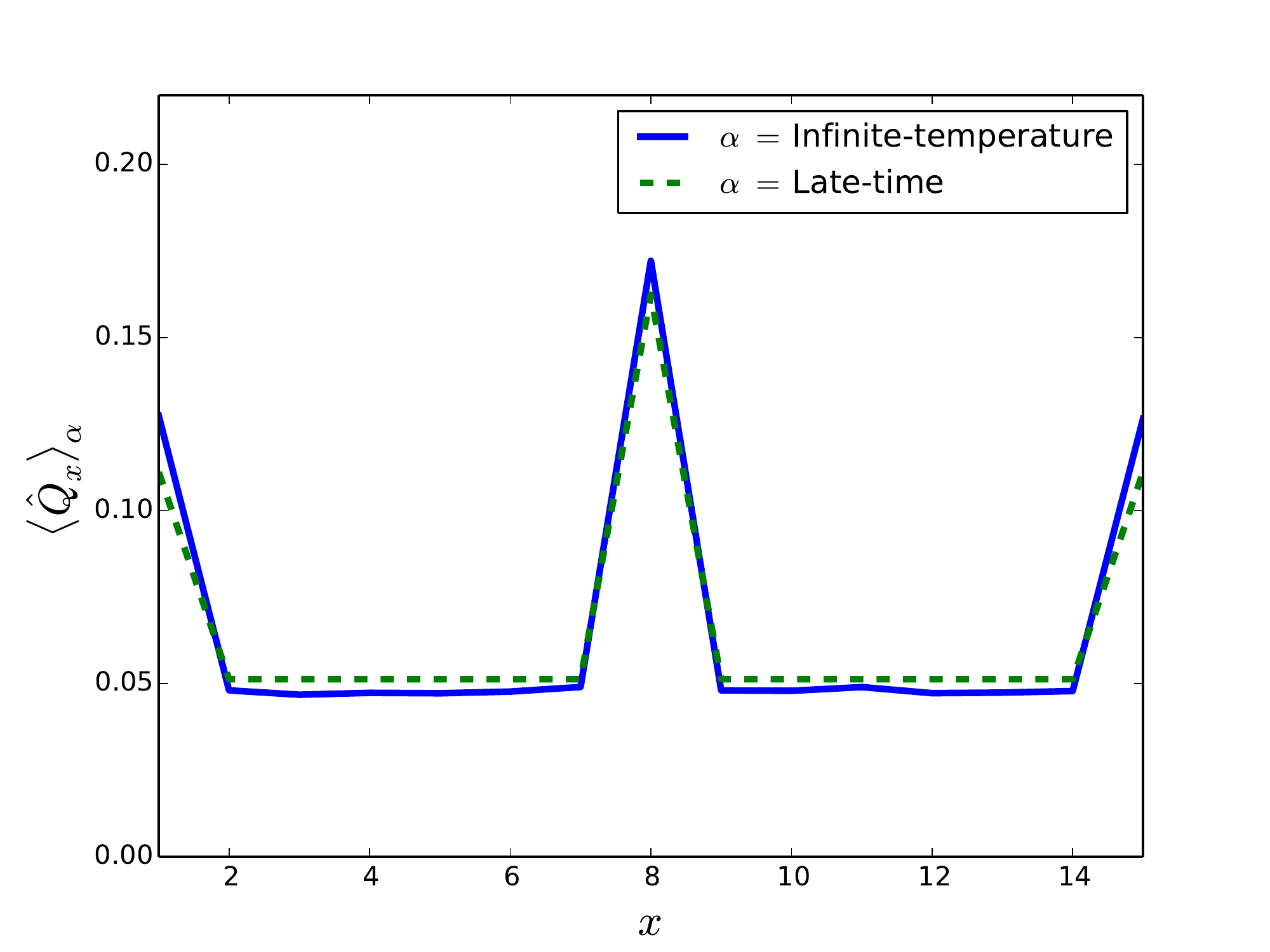}
    \end{tabular}
    \caption{(Color online) (a) Time-evolution of the expectation value of on-site charge operators for the middle site and a site away from the middle under the pair hopping Hamiltonian with PBC, starting from an initial state of the form $\ket{\psi_0} = \ket{\ast \cdots \ast + \ast \cdots \ast}$ where $\ast =\ \uparrow, \downarrow$ for $N = 15$ with total spin $S^z = 0$. The horizontal lines show the infinite-temperature expectation values of the same charge operators. Data averaged over 10 configurations of the $\ast$'s such that $S^z = 0$. (b) Late-time charge profile on sites of the chain matches the infinite temperature value \textit{within} the Krylov subspace $\mathcal{K}\left(H, \ket{\psi_0}\right)$. They both show a peak on the middle site, providing an example of  \textit{quasilocalization from thermalization}.
    }
    \label{fig:localization}
\end{figure*}
Based on the results of the previous section, which established the phenomenon of Krylov-restricted ETH, we expect that the long-time behaviour of typical states within a particular non-integrable (resp. integrable) Krylov subspace coincides with the Gibbs ensemble (resp. generalized Gibbs ensemble) restricted to that subspace. 
Such Krylov-restricted thermalization can lead to surprising behaviour within some Krylov subspaces.
For example, in the following we show that the \textit{thermal} expectation value of charge density on the chain within a particular Krylov subspace is spatially non-uniform for any finite system size. 
To illustrate this behaviour, we consider the dynamics of a single fracton immersed in a spin background i.e., we study the Krylov subspace generated by the root state with PBC: 
\begin{equation}
    \ket{\psi_0} = \ket{\ast \cdots \ast + \ast \cdots \ast} \, ,
\label{eq:fractonrootstate}
\end{equation}
where $\ast = \uparrow, \downarrow$ such that $N_\uparrow = N_\downarrow$. 
The configuration $\ket{\psi_0}$ thus belongs to the quantum number sector $Q = 1, D = \exp\left(i \pi (N+1)/N \right), S^z = 0$, where $N$ is the length of the chain. 
Since we impose PBC here, all configurations of $\ast$'s in the root state generate the same Krylov subspace as the spins can rearrange amongst themselves under the action of the Hamiltonian (see Eq.~(\ref{eq:spinscattering})). 
Hence, in the following, we only explicitly describe the action of the Hamiltonian on the fractons, given that all possible spin configurations (with $N_\uparrow = N_\downarrow$) are generated within this subspace. 
An explicit example of the complete list of product configurations in the Krylov subspace generated by $\ket{\psi_0} = \ket{\uparrow\uparrow\uparrow+\downarrow\downarrow\downarrow}$ for $N=7$ is given in App.~\ref{sec:fractonkrylov}.
There are two possibilities for how the state $\ket{\psi_0}$ of Eq.~(\ref{eq:fractonrootstate}) evolves under one application of the Hamiltonian $H$: either the spins can rearrange amongst themselves or the fracton moves by emitting a dipole, according to Eq.~(\ref{eq:+fracton}).  
Since we are only focusing on the fracton, in the latter case, the new basis state reads
\begin{equation}
    \ket{\psi_1} = \ket{\ast\ \cdots\ \ast + - + \ast \cdots \ast} \,, 
\label{eq:psi1}
\end{equation}
where $\ast =\ \uparrow, \downarrow$.
Upon further actions of the Hamiltonian, the emitted $+-$ or $-+$ dipole in Eq.~(\ref{eq:psi1}) can propagate in the spin background to the left or to the right, leaving behind a free $+$ fracton and resulting in one of the following two configurations:
\begin{eqnarray}
    \ket{\psi_2} = \left\{
		\begin{array}{ll}
			\ket{\ast\ \cdots\ \ast + - \downarrow \cdots \downarrow + \ast \cdots \ast}\\
			\ket{\ast\ \cdots\ \ast +  \uparrow \cdots \uparrow - + \ast \cdots \ast} 
		\end{array}
	\right.,
\label{eq:psi2}
\end{eqnarray}
where $\ast = \uparrow, \downarrow$ such that $N_\uparrow = N_\downarrow$.
With either a string of $\downarrow$'s or $\uparrow$'s (upper and lower situation in Eq.~(\ref{eq:psi2}) respectively), further actions of the Hamiltonian enable the isolated fracton in Eq.~\eqref{eq:psi2} to move through the emission of an additional dipole, which can then propagate in the spin background. This results in configurations of the form:
\begin{eqnarray}
    \ket{\psi_3} = \left\{
		\begin{array}{ll}
			\ket{\ast\ \cdots\ \ast + - + - \downarrow \cdots \downarrow + \ast \cdots \ast}\\
			\ket{\ast\ \cdots\ \ast +  \uparrow \cdots \uparrow - + - + \ast \cdots \ast}, 
		\end{array}
	\right.
\label{eq:psi3}
\end{eqnarray}
where $\ast = \uparrow, \downarrow$ such that $N_\uparrow = N_\downarrow$.
Once configurations of the form Eq.~\eqref{eq:psi3} are generated, a fracton can absorb a dipole when acted upon by the Hamiltonian, as allowed by Eq.~\eqref{eq:-fracton}. The resulting configurations are of the form:
\begin{eqnarray}
    \ket{\psi_4} = \left\{
		\begin{array}{ll}
			\ket{\ast\ \cdots\ \ast + \uparrow -  \downarrow \cdots \downarrow + \ast \cdots \ast}\\
			\ket{\ast\ \cdots\ \ast +  \uparrow \cdots \uparrow - \downarrow + \ast \cdots \ast}, 
		\end{array}
	\right.
\label{eq:psi4}
\end{eqnarray}
where $\ast = \uparrow, \downarrow$ such that $N_\uparrow = N_\downarrow$.
Following the above discussion, one can show that the repeated emission and absorption of multiple dipoles generates product states within the Krylov subspace that are necessarily of the form:
\begin{equation}
    \ket{\cdots \ast + \uparrow \cdots \uparrow - \downarrow \cdots \downarrow + \uparrow \cdots \uparrow - \downarrow \cdots \downarrow + \ast \cdots}, 
\label{eq:emergentconst}
\end{equation}
i.e. with strings of only $\uparrow$'s or $\downarrow$'s between consecutive fractons.
Given the symmetries of the Hamiltonian, only strings of the form Eq.~(\ref{eq:emergentconst}), that have the same $(Q,D,S^z)$ quantum numbers as the root state $\ket{\psi_0}$, are allowed in the Krylov subspace. Hence, this subspace is characterized by the presence of an emergent string-order (equivalently, it is non-locally constrained).
To illustrate the novel features of this Krylov subspace, we compare the time evolution of the charge density on the middle site (the site on which the fracton resides initially) with that on a different site, which initially hosts a spin.
The results are shown in Fig.~\ref{fig:localization}(a), which compares the charge density at the middle site (in blue) to that at a different site (in green) as a function of time.
Irrespective of the spin configuration in the initial state, we consistently find that the middle site exhibits a higher charge density as compared to any other site.  
Moreover, as shown in Fig.~\ref{fig:localization}(b), we find that this late-time charge density \textit{matches} that predicted by ETH, assuming the initial state lies in the middle of the spectrum of the Krylov subspace. 
The charge density at an inverse temperature $\beta$ \textit{restricted} to the Krylov subspace $\mathcal{K}$ is then given by
\begin{equation}
    \langle \widehat{Q}_{\textrm{mid}} \rangle_{\beta} = \frac{\textrm{Tr}\left(\widehat{Q}_{\textrm{mid}} e^{-\beta \mathcal{H}_{|\mathcal{K}}}\right)}{\textrm{Tr}\left(e^{-\beta \mathcal{H}_{|\mathcal{K}}}\right)},
\label{eq:chargemidbeta}
\end{equation}
where $\mathcal{H}_{|\mathcal{K}}$ is the restriction of the Hamiltonian  $H$ to the Krylov subspace $\mathcal{K}$, and $\widehat{Q}_{\textrm{mid}}$ is the charge operator of the middle site, using the established convention: spins are charge neutral, whereas $+$ and $-$ fractons have charges $+1$ and $-1$ respectively.

Assuming infinite temperature ($\beta = 0$) in Eq.~(\ref{eq:chargemidbeta}), we obtain 
\begin{equation}
    \langle \widehat{Q}_{\textrm{mid}} \rangle_{\beta} = \frac{\textrm{Tr}\left(\widehat{Q}_{\textrm{mid}}\right)}{\textrm{Tr}\left(\mathds{1}|_{\mathcal{K}}\right)} \equiv \frac{\mathcal{Q}_N}{\mathcal{D}_N} = \frac{3}{N}, 
\label{eq:quasilocalization}
\end{equation}
where $\mathds{1}|_{\mathcal{K}}$ is the identity restricted to the Krylov subspace $\mathcal{K}$, and thus $\textrm{Tr}\left(\mathds{1}|_{\mathcal{K}}\right) = \mathcal{D}_N$, the Hilbert space dimension of $\mathcal{K}\left(H, \ket{\psi_0}\right)$ of the chain of $N$ sites.
We provide analytical and numerical arguments for the result of $3/N$ in Eq.~(\ref{eq:quasilocalization}) in App.~\ref{sec:fractonkrylov} (see Eq.~(\ref{eq:quasilocapp})). On the other hand, the late time expectation value of the charge density on any other site in the middle of the chain is $1/N$.
We dub this phenomenon as \textit{quasi-localization} of the fracton, since it is localized for any finite system size although the  localization vanishes in the thermodynamic ($N \rightarrow \infty$) limit. 
We emphasize that unlike usual mechanisms for localization, which rely on the existence of localized eigenstates~\cite{rahul2015review, sala2019ergodicity, khemani2019local}, the phenomenon here is \textit{quasi-localization from thermalization}, which is a consequence of ergodicity, albeit ergodicity \textit{within} a constrained Krylov subspace.
%

%
%

\section{Connections with Bloch MBL}
\label{sec:stark}
Having established some consequences of Krylov fracture, we now discuss the relationship between our model and the Bloch (or Stark) MBL problem~\cite{van2018bloch,schulz2019stark}. The latter is an interacting extension of the well-known single particle Wannier-Stark localization~\cite{emin1987existence}, with the Hamiltonian given by
\begin{align}
    H_{\textrm{Bloch}} =\ & t \sumal{j=1}{L-1}{\left(\cd_j c_{j+1} + h.c.\right)} + E\,\sumal{j=1}{L}{j \, \hat{n}_j} \nonumber \\
    & + V_0 \sum_{j=1}^{L}{w_j\hat{n}_j} + V_1 \sum_{j=1}^{L-1} \hat{n}_j \hat{n}_{j+1},
\label{eq:starkhamil}
\end{align}
where $\hat{n}_j = \cd_j c_j$ is the fermionic number operator, $t$ is the hopping strength, $w_j$ is an on-site disorder ($w_j$ random) or curvature ($w_j \sim j^2$) whose strength is set by $V_0$, and $V_1$ is the nearest-neighbour repulsion strength. 
Here, the model is defined on a chain with $L$ sites and with open boundary conditions.
Observe that the term $\sum_j{j \hat{n}_j}$, representing the uniform electric field, is \textit{precisely} the center-of-mass operator $\widehat{C}$ for OBC, defined in Eq.~\eqref{eq:comoperator}. 
As detailed in App.~\ref{sec:blochSW}, we can perform a Schrieffer-Wolff transformation~\cite{bravyi2011schrieffer} perturbatively at large $E/t$ for an infinite chain to derive the effective CoM preserving Hamiltonian (see Eq.~(\ref{eq:hefffinal})):
\begin{align}
H_{\text{eff}} &=\ V_0\sumal{j}{}{\widetilde{w}_j\hn_j} + \widetilde{V}_1\sum_j \hat{n}_j \hat{n}_{j+1} + \widetilde{V}_2\sum_j \hat{n}_j \hat{n}_{j+2} \nn \\
&- \frac{t^2 V_1}{E^2} \sumal{j}{}{\left(\cd_j \cd_{j + 3} c_{j + 2} c_{j + 1} + h. c.\right)} + \mathcal{O}\left(\frac{t^3}{E^3}\right),
\label{eq:effhamiltext}
\end{align}
where $\widetilde{w}_j$, $\widetilde{V}_1$, $\widetilde{V}_2$ are defined in Eq.~(\ref{eq:tildedefns}).
$\widetilde{w}_j$ and $\widetilde{V}_1$ are the disorder and nearest neighbor interaction strengths respectively ``renormalized" by corrections of $\mathcal{O}\left(t^2/E^2\right)$, and $\widetilde{V}_2$ is the effective next-nearest neighbor interaction of $\mathcal{O}\left(t^2/E^2\right)$.  
Hence, the leading order hopping term in the effective Hamiltonian governing the Wannier-Stark model is the pair-hopping term studied in this paper, given by Eq.~\eqref{eq:pairhopping}. 
Longer range center-of-mass preserving terms, including $n$-body terms for $n > 2$ appear at higher orders in perturbation theory, and are therefore suppressed by higher powers of $t/E$; we thus expect their strength to drop off exponentially with range as $\sim t^n/E^{n}$, for terms which have support over $\sim n$ sites.

Given this mapping, we now comment briefly on the phenomenon of Bloch MBL, as discussed in Refs.~\cite{van2018bloch,schulz2019stark}.
We begin by noting that the electric field in itself is \textit{not} sufficient to give MBL, since while the electric field `switches off' single particle hopping, it leaves in place the correlated center-of-mass preserving hopping processes discussed above.
As we have discussed in the preceding sections, eigenstates of such processes are by no means guaranteed to be localized.
Thus, different physics must underlie the numerical observation of MBL in the Bloch MBL problem.
Strictly in the $E/t \rightarrow \infty$ limit, the effective Hamiltonian consists only of the nearest-neighbor electrostatic term $V_1 \sum_j{\hn_j \hn_{j+1}}$ and the onsite potential term $V_0 \sum_j{w_j \hn_j}$.
When $w_j = 0$, i.e. without disorder or curvature, the eigenstates are clearly not localized since the spectrum of $V_1 \sum_j{\hn_j \hn_{j+1}}$ is highly degenerate. 
However, that degeneracy is lifted by small disorder or curvature; thus, when $w_j$ is random or $w_j \sim j^2$, all the eigenstates of Eq.~(\ref{eq:effhamiltext}) have low entanglement.
This is consistent with the fact that Refs.~\cite{schulz2019stark, van2018bloch} do not observe MBL without curvature or disorder respectively.
Moving away from the $E/t \rightarrow \infty$ limit, we obtain the effective Hamiltonian of Eq.~(\ref{eq:effhamiltext}) for large but finite $E/t$, which exhibits Krylov fracture.
The fracture is said to be `strong'~\cite{sala2019ergodicity,khemani2019local} if the dimension of the largest Krylov subspace is a vanishing fraction of the full Hilbert space dimension in the thermodynamic limit.
This leads to the non-thermalization of generic initial product states with respect to the entire Hilbert space~\cite{sala2019ergodicity,khemani2019local}, for example the entanglement entropy does not saturate to the maximum value allowed by the full Hilbert space.
For a `minimal' center-of-mass preserving Hamiltonian, such as the pair-hopping model of Eq.~(\ref{eq:pairhopping}), obtained by retaining only the leading order hopping terms in the effective Hamiltonian, strong fracture indeed occurs.\footnote{The pair-hopping Hamiltonian Eq.~(\ref{eq:pairhopping}) is equivalent to a $S = 1/2$ spin Hamiltonian for which evidence of strong fracture was found in Ref.~\cite{sala2019ergodicity}. We have also verified numerically up to $L = 24$ that the size of the largest Krylov subspace $\sim 2^L$ while the Hilbert space dimension $\sim 4^L$, consistent with strong fracture.}
A simple example of such non-thermalization is,the CDW state $\ket{0101\cdots01}$ used as a diagnostic of localization in Ref.~\cite{schulz2019stark}.
This state forms a one-dimensional Krylov subspace under the pair-hopping Hamiltonian of Eq.~(\ref{eq:pairhopping}): it maps onto the state $\ket{\uparrow \uparrow \dots \uparrow \uparrow}$ under the mapping defined in Sec.~\ref{sec:halffilling}.
Clearly, once initialized with this state, the system will forever retain memory of its initial condition under time evolution with the minimal pair-hopping Hamiltonian. 
\begin{figure}[t!]
    \centering
    \includegraphics[scale = 0.45]{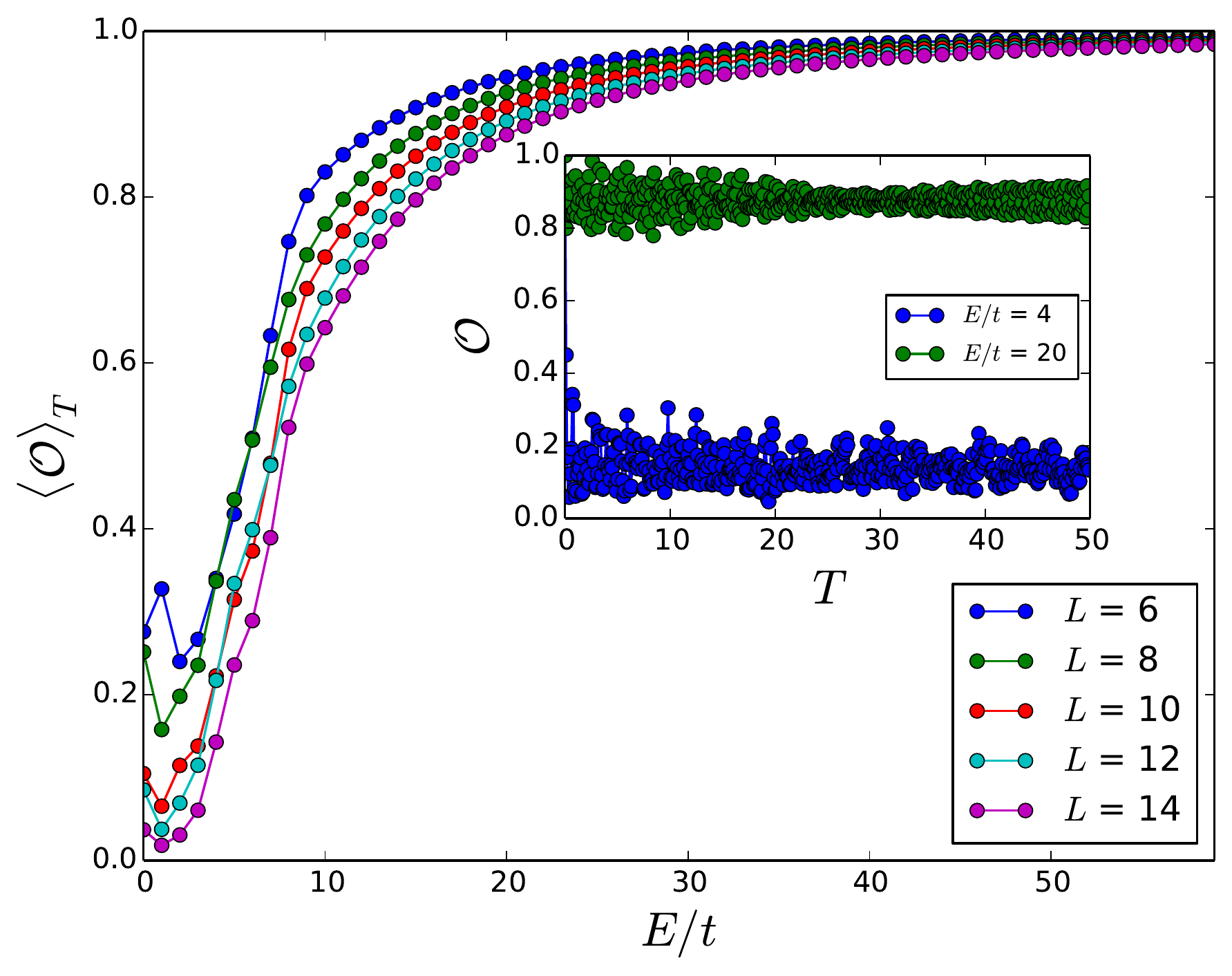}
    \caption{(Color online) Inset: Weight of the state $e^{-i H_{\textrm{Bloch}} T}\ket{\psi_0}$ within the Krylov subspace $\mathcal{K}\left(H_{\textrm{eff}}, \ket{\psi_0}\right)$, captured by the quantity $\mathcal{O}(T)$ (defined in  Eq.~(\ref{eq:OTdefn})) for two values of the electric field $E$. Main: Time-average of  $\mathcal{O}\left(T\right)$, denoted by $\langle \mathcal{O}\rangle_T$ as a function of $E/t$. $\mathcal{O}$ is close to $1$ for larger values of $E$, justifying that the pair-hopping Hamiltonian $H$ is a good approximation for $H_{\textrm{Bloch}}$. Data is shown for $V_0/t = 0$, $V_1/t = 1$, and $\ket{\psi_0} = \ket{\uparrow\downarrow \uparrow \downarrow \cdots \cdots} = \ket{01100110 \cdots \cdots}$.}
\label{fig:blochmbl}
\end{figure}
However, we note that the effective Hamiltonian $H_{\textrm{eff}}$ of Eq.~(\ref{eq:effhamiltext}) is a good approximation to the Bloch MBL Hamiltonian $H_{\textrm{Bloch}}$, given by Eq.~(\ref{eq:starkhamil}), only for large values of $E/t$.
To test the effectiveness of $H_{\textrm{eff}}$, we study the quantity
\begin{equation}
	\mathcal{O}\left(T\right) = \sumal{\ket{\phi_n} \in \mathcal{K}\left(H_{\textrm{eff}}, \ket{\psi_0}\right)}{}{|\bra{\phi_n} e^{-i H_{\textrm{Bloch}} T} \ket{\psi_0}|^2},
\label{eq:OTdefn}
\end{equation}
which is the weight of the state $e^{-i H_{\textrm{Bloch}} T} \ket{\psi_0}$ within the Krylov subspace $\mathcal{K}\left(H_{\textrm{eff}}, \ket{\psi_0}\right)$.\footnote{Note that since $H_{\textrm{eff}}$ of Eq.~(\ref{eq:effhamiltext}) and $H$ of Eq.~(\ref{eq:pairhopping}) only differ by diagonal terms, $\mathcal{K}\left(H_{\textrm{eff}}, \ket{\psi_0}\right) = \mathcal{K}\left(H, \ket{\psi_0}\right)$.}
We expect $H_{\textrm{eff}}$ to correctly capture the dynamics of $H_{\textrm{Bloch}}$ only for values of $E/t$ when 
\begin{equation}
    \langle \mathcal{O}\rangle_T \equiv \lim_{\tau\rightarrow\infty}{\frac{1}{\tau}\int{\mathrm{d}\tau\  \mathcal{O}\left(\tau\right)}} \approx 1.
\end{equation}
In Fig.~\ref{fig:blochmbl}, we show the behavior of $\langle \mathcal{O}\rangle_T$ for the initial state $\ket{\psi_0} = \ket{\uparrow \downarrow \uparrow \downarrow \cdots \cdots} = \ket{01100110\cdots\cdots}$.
Thus, we find that $H_{\textrm{eff}}$ is a good approximation for $H_{\textrm{Bloch}}$ only for $E/t \gtrsim 50$ when $V_0, V_1 \sim \mathcal{O}\left(1\right)$ and for system sizes up to $L = 14$. 
In Fig.~\ref{fig:blochmbl}, we also find that for a fixed value of $E/t$, $H_{\textrm{eff}}$ becomes a worse approximation for $H_{\textrm{Bloch}}$ with increasing system size.
Thus, it is not clear whether Krylov fracture of the pair-hopping model of Eq.~(\ref{eq:pairhopping}) plays a significant role in the observations of Refs.~\cite{schulz2019stark, van2018bloch}, which focus on the regimes where $E/t \sim \mathcal{O}\left(10\right)$.
To conclude this section, we speculate on two mechanisms that give rise to localized eigenstates at the smaller values of $E/t$ with disorder, which could provide a partial explanation for the Bloch MBL phenomenon in Refs.~\cite{van2018bloch,schulz2019stark}:
(i) At smaller values of $E/t$, terms at higher order in perturbation theory cannot be neglected in the effective Hamiltonian.
However, since terms generated at all orders in perturbation theory are necessarily center-of-mass preserving, the hopping term of the effective Hamiltonian at any finite order exhibits exponentially many frozen eigenstates~\cite{sala2019ergodicity, khemani2019local}.
The addition of disorder breaks the exponentially large degeneracy of these frozen states under the effective Hamiltonian, which results in exponentially many product eigenstates of the effective Hamiltonian at any finite order.  
(ii) When disorder is added in Eq.~\eqref{eq:starkhamil} (as is done in Ref.~\cite{van2018bloch}), then this can give rise to conventional `disorder-induced' MBL~\cite{rahul2015review} within Krylov subspaces of the effective Hamiltonian.
This can happen even when the disorder is weak compared to the bare single particle hopping $t$, because the disorder may be strong compared to the largest hopping term: from Eq.~\eqref{eq:effhamiltext}, we see that the hopping term is of $\mathcal{O}\left(t^2 V_1/E^2\right)$, while the disorder is an $\mathcal{O}\left(V_0\right)$ term, which suggests the possibility of conventional MBL in the effective Hamiltonian. 
%
%
%
%
\section{Conclusions and Open Questions}
\label{sec:conclusions}
In this paper, we have studied a simple translation invariant model which conserves both charge and center-of-mass, and which provides a natural platform for realising the physics of fractonic systems. Specifically, we find that the pair-hopping model Eq.~\eqref{eq:pairhopping} exhibits the phenomenon of Krylov fracture, wherein various regions of Hilbert space are dynamically disconnected even if they belong to the same global symmetry sectors. In addition to exponentially many product eigenstates, whose effect on quantum dynamics was studied in Refs.~\cite{sala2019ergodicity,khemani2019local}, the pair-hopping model also hosts several \textit{large} closed Krylov subspaces with dimensions that grow exponentially in the system size at half-filling.

We find that exponentially many of such large Krylov subspaces admit a mapping onto spin-$1/2$ XX models of various sizes and hence, constitute examples of \textit{integrable} Krylov subspaces.
However, not all large Krylov subspaces show signs of integrability; instead, the model also possesses exponentially many non-integrable subspaces, many of which show level-repulsion and behaviour consistent with ETH. 
Moreover, some of these Krylov subspaces are highly constrained, which leads to atypical dynamical behaviour even within a \textit{thermal} Krylov subspace, an effect we dub ``quasilocalization due to thermalization".
By this, we specifically mean that the late-time expectation values of local operators within such subspaces deviate from the expected behaviour in generic translation invariant systems.
Finally, since the pair-hopping model appears as the leading order hopping term in the strong-field limit of the interacting Wannier-Stark problem, we make contact between our work and Bloch MBL.
Besides shedding new light on Bloch MBL, our work hence also provides an experimentally relevant setting for studying the dynamics of center-of-mass preserving systems.
Our results, which illustrate the rich structure that can arise as a consequence of Krylov fracture, harbour several implications for the dynamics of isolated quantum systems.
Firstly, in the presence of Krylov fracture, we have demonstrated that notions of ergodicity and its violation are well-defined once restricted to large Krylov subspaces.
Moreover, we showed that usual diagnostics, such as energy level-statistics, accurately capture whether such Krylov subspaces are integrable or not. These results thus suggest that a modified version of ETH, restricted to large Krylov subspaces, holds for systems with fractured Hilbert spaces.

Secondly, our results provide a clear example of a ``semi-integrable" model i.e., one where integrable as well as non-integrable exponentially large Krylov subspaces co-exist~\cite{znidaric2019coexistence, iadecola2018exact}.
When viewed from the perspective of the entire Hilbert space (within a particular symmetry sector), the integrable Krylov subspaces are examples of quantum many-body scars, since they are ETH-violating states embedded within the entire many-body spectrum.
Unlike the exponentially many static configurations (one-dimensional Krylov subspaces) which necessarily exist for any center-of-mass (dipole moment) conserving Hamiltonian~\cite{sala2019ergodicity, khemani2019local}, these integrable subspaces have an exponentially large dimension, which can lead to non-trivial dynamics in an otherwise non-integrable model. For the cognoscenti, we note that such subspaces are qualitatively distinct from subspaces generated from states containing a blockaded region~\cite{tomasi2019dynamics}.  
Thus, the existence of such integrable Krylov subspaces of dimension much smaller than that of the full Hilbert space, even if only approximately closed, might be related to quantum many-body scars which by now have been observed in several constrained systems, including the PXP model~\cite{turner2017quantum, 2bull2019scar}.

Additionally, even large non-integrable subspaces show ergodicity breaking with respect to the entire Hilbert space~\cite{sala2019ergodicity} and instead obey ETH only once restricted to the Krylov subspace, resulting in highly non-general \textit{thermal} expectation values of local operators within such Krylov subspaces. 
Note also that we have only focused on Krylov subspaces generated by root states that are product states (see Eq.~(\ref{eq:fullhilbert})), but one could also study closed Krylov subspaces generated by other low-entanglement states; whether this leads to further fracturing within the Krylov subspaces of the pair-hopping model is a question for future work. 
On a different note, we described the emergent fractonic behaviour of composite degrees of freedom in a simple model, one which can be realised by subjecting fermions hopping on a chain to a strong electric field. It would be interesting to study whether similar emergent behaviour appears in higher dimensions. 
For instance, one can impose the conservation of \textit{quadrupole moment} in two-dimensions, which could be arranged e.g., by adding strong field-gradients. 
Such a system could allow one to study the relation, if any, between the dynamics of fracton models~\cite{chamon2005,kimhaah2016,prem2017glass} and Krylov fracture.

\textit{Note added}: During the completion of this work, there appeared Refs.~\cite{khemani2019localization, taylor2019experimental} which also discuss connections between center-of-mass preserving models and the Bloch MBL phenomenon, and Ref.~\cite{rakovszky2019statistical} which discusses labelling the Krylov subspaces of a related model by non-local symmetries. Our results agree wherever there is overlap.
\section*{Acknowledgements}
We thank Dan Arovas, Vedika Khemani, Alan Morningstar, Frank Pollmann, Gil Refael, Max Schultz, Shivaji Sondhi, Ruben Verresen, and particularly David Huse for useful discussions. S.M. acknowledges the hospitality of the Laboratoire de Physique de l'Ecole Normale Sup\'{e}rieure, where parts of the manuscript were completed. A.P. acknowledges the hospitality of the Aspen Center of Physics, where part of this work was completed during a visit to the program ``Realizations and Applications of Quantum Coherence in Non-Equlibrium Systems." The Aspen Center for Physics is supported by National Science Foundation grant PHY-1607611. A.P. is supported by a PCTS fellowship at Princeton University. This material is based in part (R.M.N.) upon work supported by Air Force Office of Sponsored Research under grant no. FA9550-17-1-0183. R.M.N. also acknowledges the hospitality of the KITP, where part of this work was done, during a visit to the program ``Dynamics of Quantum Information." The KITP is supported in part by the National Science Foundation under grant PHY-1748958. B.A.B. and N.R. were supported by the Department of Energy Grant No. DE-SC0016239, the National
Science Foundation EAGER Grant No. DMR 1643312,
Simons Investigator Grant No. 404513, ONR Grant No.
N00014-14-1-0330, the Packard Foundation, the Schmidt
Fund for Innovative Research, and a Guggenheim Fellowship from the John Simon Guggenheim Memorial Foundation. 
%
%
\appendix
%
%
%
\section{Symmetries of the pair-hopping Hamiltonian in terms of composite degrees of freedom}\label{sec:symcomp}
In this appendix, we discuss some of the symmetries of the pair-hopping Hamiltonian Eq.~\eqref{eq:pairhopping} in terms of the composite degrees of freedom defined in Eq.~(\ref{eq:halffillingpart}).
Similarly to the center-of-mass operator Eq.~(\ref{eq:comoperator}), for PBC the dipole moment operator $\widehat{D}$ in Eq.~(\ref{eq:dipoleoperator}) does not commute with translation by one unit cell (which corresponds to translation by two sites in the original degrees of freedom). 
To see this, note that under translation $j \rightarrow j + 1$, 
\beq
\sumal{j = 1}{N}{j \widehat{Q}_j} \mapsto \sumal{j = 1}{N}{j \widehat{Q}_j} +  \widehat{Q},
\label{eq:dipoleoperatortrans}
\eeq
with $\widehat{Q}$ the total charge operator.
This operator obeys non-trivial commutation relations with translations along the chain, since 
\beq
\widehat{T} \widehat{D} \widehat{T}^{-1} = \widehat{D} \exp\left(\frac{2 \pi i}{L} \widehat{Q} \right) =  \widehat{D} \exp\left(2 \pi i \, \frac{p}{q} \right),
\eeq
where $\widehat{T}$ is the operator for translation by one unit cell (two sites of the original system), and we are focusing on states with a fixed charge $Q$ such that $Q/N = p/q$. 
Thus,  
\begin{equation}
    \left[\widehat{T}^q, \widehat{D}\right] = 0,
\end{equation}
within the charge $Q$ sector. 
Further, as discussed in Sec.~\ref{sec:model}, the pair-hopping Hamiltonian $H$ is inversion symmetric (i.e., under the exchange of sites $j$ and $L - j + 1$).
After grouping sites using Eq.~(\ref{eq:halffillingpart}), the inversion symmetry of $H$ also flips the composite spin degrees of freedom $\ket{\uparrow} \leftrightarrow \ket{\downarrow}$ in addition to interchanging the sites $j$ and $N - j + 1$. 
For example, when $L = 10$ ($N = 5$), under inversion about the center bond in the third unit cell, the configuration 
\begin{equation*}
    \ket{\ \fbox{01}\ \fbox{10}\ \fbox{11}\ \fbox{01}\ \fbox{00}\ } \to \ket{\ \fbox{00}\ \fbox{10}\ \fbox{11}\ \fbox{01}\ \fbox{10}\ } \, .
\end{equation*}
In terms of composite degrees of freedom, this corresponds to the transformation 
\begin{equation*}
\ket{\uparrow\ \downarrow + \uparrow - } \to \ket{-\downarrow+\uparrow\ \downarrow} \, ,
\end{equation*}
which is the usual inversion about the center site followed by a spin flip. 
However, inversion also does not commute with translation symmetry (with PBC). Under inversion, a momentum eigenstate with momentum $k$ goes to a state with momentum $-k$. 
Similarly, under inversion symmetry, note that
\begin{equation}
    \sumal{j = 1}{N}{j \widehat{Q}_j} \rightarrow \sumal{j = 1}{N}{\left(N + 1 - j\right) \widehat{Q}_j} = \left(N + 1\right)\widehat{Q} - \sumal{j = 1}{N}{j \widehat{Q}_j} \, ,
\label{eq:inversiondipole}
\end{equation}
such that the dipole moment operator transforms as
\begin{eqnarray}
    \widehat{D} \rightarrow \twopartdef{\left(N + 1\right)\widehat{Q} - \widehat{D}}{OBC}{\exp\left(\frac{2\pi i Q}{N}\right)\widehat{D}^{-1}}{PBC}.
\label{eq:inversionDtransform}
\end{eqnarray}
Thus, for PBC, the inversion symmetry can be diagonalized only in sectors with dipole moment $D$ that satisfies: $$D^2 = \exp{\left(\frac{2\pi i Q}{N}\right)}.$$
%
%
\section{Formal mapping of the spin Krylov subspace to the XX model}
\label{sec:XXformal}

In this Appendix, we show the formal mapping from the spin Krylov subspace in the pair-hopping Hamiltonian Eq.~(\ref{eq:pairhopping}) at half-filling to the XX model.
We define spin-1/2 raising and lowering operators using the fermionic operators $c_j$ and $\cd_j$,
\begin{eqnarray}
    \sigma_j^+ \equiv \cd_{2j - 1} c_{2j} \, ,\nn \\
    \sigma_j^- \equiv \cd_{2j} c_{2j - 1} \, .
\label{eq:spsmdefn}
\end{eqnarray}
Using Eq.~(\ref{eq:spsmdefn}), we obtain
\begin{eqnarray}
    \{\sigma^+_j, \sigma^-_j\} &=& \{\cd_{2j-1} c_{2j},\ \cd_{2j} c_{2j-1} \} \nn \\
    &=& \cd_{2j-1} c_{2j-1} c_{2j} \cd_{2j} + \cd_{2j} c_{2j} c_{2j-1} \cd_{2j-1} \nn \\
    &=& \hat{n}_{2j-1}\left(1 - \hat{n}_{2j}\right)  + \hat{n}_{2j}\left(1 - \hat{n}_{2j-1}\right) \nn \\
    &=& \hat{n}_{2j-1} + \hat{n}_{2j} - 2\hat{n}_{2j-1}\hat{n}_{2j}.
\end{eqnarray}
Since $n_{2j-1}, n_{2j} \in \{0,1\}$, $\sigma^+_j$ and $\sigma^-_j$ are valid Pauli operators only within the subspace of configurations that satisfy
\begin{eqnarray}
    &n_{2j-1} + n_{2j} = 1, \;\;\; n_{2j-1} n_{2j} = 0 \label{eq:validconditions}\\
    &\implies \{\sigma^+_j, \sigma^-_j\} = 1.
\label{eq:validmap}
\end{eqnarray}
The conditions in Eq.~(\ref{eq:validmap}) are only satisfied if the composite degrees of freedom on unit cells $j$ and $j+1$ are $\ket{\uparrow}$ or $\ket{\downarrow}$ (see Eq.~(\ref{eq:halffillingpart})), and hence the mapping from fermions to effective spin degrees of freedom is restricted only to the spin Krylov subspace.
First, we re-write the pair-hopping Hamiltonian Eq.~(\ref{eq:pairhopping}) (with PBC) as 
\begin{eqnarray}
    H &=& \sumal{j = 1}{N}{\left(\cd_{2j - 1} \cd_{2j+2} c_{2j + 1} c_{2j} + \cd_{2j} \cd_{2j+3} c_{2j + 2} c_{2j + 1} +  h.c\right)} \nn \\ 
    &=& \sumal{j = 1}{N}{\left(\cd_{2j - 1} c_{2j} \cd_{2j + 2} c_{2j + 1} + \cd_{2j} \cd_{2j+3} c_{2j + 2}c_{2j + 1}   + h.c\right)}. \nn \\
\label{eq:intermediate}
\end{eqnarray}
Given the conditions in Eq.~(\ref{eq:validconditions}), either $n_{2j+1} = 0$, $n_{2j+2} = 1$  or $n_{2j+1} = 1$, $n_{2j+2} = 0$ for every $j$ for configurations within the spin Krylov subspace. 
Since the second term of Eq.~(\ref{eq:intermediate}) contains $c_{2j + 2} c_{2j + 1}$, it and its Hermitian conjugate always vanish on states within the spin Krylov subspace. 
Thus, we obtain
\begin{eqnarray}
    H_{XX}\left[N\right] &=& \sumal{j = 1}{N}{\left(\sigma^+_j \sigma^-_{j+1} + h.c\right)} \nn \\
    &=& \frac{1}{2}\sumal{j = 1}{N}{\left(\sigma^x_j \sigma^x_{j+1}  + \sigma^y_j \sigma^y_{j+1}\right)},
\end{eqnarray}
which is the familiar XX model. 

The XX model is solved via the Jordan-Wigner transformation~\cite{lieb1963exact}, which proceeds by defining the operators
\begin{eqnarray}
\sigma^+_j &=& (-1)^{\sum_{l<j}{\dd_l d_l}} \dd_j \nn \\
\sigma^-_j &=& (-1)^{\sum_{l<j}{\dd_l d_l}} d_j \nn \\
\sigma^z_j &=& 2 \dd_j d_j - 1,
\label{eq:jwtransform}
\end{eqnarray}
where $d_j$'s and $\dd_j$'s are fermionic operators.
Using Eq.~(\ref{eq:jwtransform}), the Hamiltonian $H_{XX}\left[N\right]$ is mapped onto a non-interacting fermionic hopping Hamiltonian:
\begin{equation}
    H_{d} = \sum_{j = 1}^{N}{\left(\dd_j d_{j+1} + h.c.\right)}.
\label{eq:halffillingeffective}
\end{equation}
Thus, the many-body ground state is a Fermi sea of the $d$ fermions, with the Fermi momentum $k_F = \pm \pi/2$:
\begin{equation}
    \ket{G} = \prod_{k < k_F}{\dd_k}\ket{0} \, ,
\end{equation}
where the vacuum $\ket{0}$ is defined by
\begin{equation}
    d_j \ket{0} = 0, \;\;\; 1 \leq j \leq N. 
\end{equation}

%
%
\section{Energies of the Integrable Krylov subspaces}
\label{sec:energies}
We now discuss the energies of the various integrable Krylov subspaces discussed in Sec.~\ref{sec:integrablekrylov}, which map onto XX models of various sizes. 
The ground state energies of $H_{XX}\left[N\right]$ with PBC (and approximately for OBC) can be written as (see Ref.~\cite{de2008xx}) 
\begin{equation}
    E_N = \twopartdef{-2\sin\left(\frac{p\pi}{2p+1}\right)\csc\left(\frac{\pi}{2p+1}\right)}{N = 2p + 1}{-2\csc\left(\frac{\pi}{2p}\right)}{N = 2p}.
\label{eq:XXenergy}
\end{equation}
As described in Sec.~\ref{sec:integrablekrylov}, starting with a root state with a single dipole in the spin background (with restrictions discussed in Sec.~\ref{sec:multidipolesubspace}) results in a Krylov subspace for which the Hamiltonian maps onto an XX model with $(N - 1)$ sites. 
Interestingly, this state is separated by a finite gap from the ground state of the full pair-hopping model, which we numerically observe to be in the spin subspace discussed in Sec.~\ref{sec:spinsubspace}.

Since this finite gap corresponds to the insertion of a dipole, we associate it with the energy of creating a single dipole.
Using Eq.~(\ref{eq:XXenergy}), this \textit{dipole gap} in the thermodynamic limit (where the OBC and PBC spectra are the same) is
\begin{align}
    \Delta E_{d} =& -2 \lim_{p \rightarrow \infty} \left(\sin\left(\frac{p \pi}{2p+1}\right)\csc\left(\frac{\pi}{2p + 1}\right) - \csc\left(\frac{\pi}{2 p}\right)\right) \nn \\
    =& \frac{2}{\pi} \approx 0.64.
\label{eq:dipolegap}
\end{align}
The dipole gap also corresponds to the gap between the ground states of the single-dipole and two-dipole Krylov subspaces illustrated in Secs.~\ref{sec:dipolesubspace} and \ref{sec:multidipolesubspace}, and more generally, between the ground states of the $n$ dipole and the $(n + 1)$ dipole Krylov subspaces illustrated in Sec.~\ref{sec:multidipolesubspace}. 
%
%
Thus, the pair-hopping model Eq.~(\ref{eq:pairhopping}) exhibits an equally spaced tower of integrable Krylov subspaces.

%
%

\section{Mapping to the XX model when dipoles are adjacent to each other}
\label{sec:dipoletouch}
Here, we study the dipole Krylov subspaces generated by root states containing adjacent, identically oriented dipoles. The discussion will focus on OBC throughout this appendix.
\subsection{Multidipole subspace}
\label{sec:multidipoleapp}
We first consider the two-dipole Krylov subspace $\mathcal{K}\left(H, \ket{\psi_0}\right)$ generated by the root state
\begin{equation}
    \ket{\psi_0} = \underset{\underbrace{\hspace{10mm}}_{N^{(1)}_\uparrow}\hspace{15mm}\underbrace{\hspace{10mm}}_{N^{(3)}_\uparrow}}{\ket{\ast \cdots \ast + - + - \ast \cdots \ast}},
\label{eq:dipoletouchrootstate}
\end{equation}
where $\ast =\ \uparrow, \downarrow$ and where $N^{(1)}_\uparrow$ and $N^{(3)}_\uparrow$ represent the number of $\uparrow$'s to the left and right of the dipoles respectively. 
Here $N^{(2)}_\uparrow = 0$, where $N^{(2)}_\uparrow$ is the number of $\uparrow$'s between the two dipoles (see Eq.~(\ref{eq:multidipoleroot})).
Under the action of the Hamiltonian, which results in one of the dipoles moving via repeated applications of Eq.~(\ref{eq:dipolescattering1}), we obtain product states of the form
\begin{equation}
    \ket{\psi_1} = \underset{\underbrace{\hspace{10mm}}_{N^{(1)}_\uparrow}\;\;\;\;\;\;\;\;\underbrace{\hspace{10mm}}_{d}\;\;\;\;\;\;\;\;\;\underbrace{\hspace{10mm}}_{N^{(3)}_\uparrow}}{\ket{\ast \cdots \ast + - \downarrow \cdots \downarrow + - \ast \cdots \ast}},
\label{eq:dipoletouch1}
\end{equation}
within the Krylov subspace, where $\ast =\ \uparrow,\ \downarrow$, $d$ is the number of $\downarrow$ spins in between the dipoles, and the total number of $\downarrow$ spins is conserved. 

Similarly, applying the Hamiltonian on $\ket{\psi_0}$ and using the rules Eqs.~(\ref{eq:+fracton}) and~(\ref{eq:-fracton}), which results in fractons absorbing a dipole, product states of the form 
\begin{equation}
    \ket{\psi_2} = \underset{\underbrace{\hspace{10mm}}_{N^{(1)}_\uparrow}\;\;\;\;\;\;\;\;\;\;\;\;\;\;\;\underbrace{\hspace{10mm}}_{N^{(3)}_\uparrow}}{\ket{\ast \cdots \ast + \uparrow -  \ast \cdots \ast}} 
\label{eq:dipoletouch2}
\end{equation}
are generated within the Krylov subspace.
Due to Eqs.~(\ref{eq:+fracton}) and (\ref{eq:-fracton}), the $+$ and $-$ fractons in Eq.~(\ref{eq:dipoletouch2}) cannot move without the emission of a dipole (which results in Eq.~(\ref{eq:dipoletouchrootstate})), \textit{all} product states in $\mathcal{K}\left(H, \ket{\psi_0}\right)$ are of the form specified by Eqs.~(\ref{eq:dipoletouchrootstate}), (\ref{eq:dipoletouch1}), or~(\ref{eq:dipoletouch2}).
We now map each product state of the form $\ket{\psi_0}$, $\ket{\psi_1}$, or $\ket{\psi_2}$ onto product states in a spin-1/2 Hilbert space of $(N - 1)$ sites.
The mapping proceeds similarly to the multi-dipole case discussed in Sec.~\ref{sec:multidipolesubspace} i.e., by identifying the $+ -$ dipoles as $\uparrow$'s.
However, to account for the action of the Hamiltonian that results in product states of the form $\ket{\psi_2}$ in the Krylov subspace, we introduce a $\downarrow$ spin between the dipoles and map $\ket{\psi_0}$, $\ket{\psi_1}$, and $\ket{\psi_2}$ according to
\begin{eqnarray}
\underset{\underbrace{\hspace{10mm}}_{N^{(1)}_\uparrow}\hspace{15mm}\underbrace{\hspace{10mm}}_{N^{(3)}_\uparrow}}{\ket{\ast \cdots \ast + - + - \ast \cdots \ast}} &\iff& \underset{\underbrace{\hspace{9mm}}_{N^{(1)}_\uparrow}\hspace{7mm}\underbrace{\hspace{9mm}}_{N^{(3)}_\uparrow}}{\ket{\ast \cdots \ast \uparrow \downarrow \uparrow \ast \cdots \ast}} \nn \\
\underset{\underbrace{\hspace{10mm}}_{N^{(1)}_\uparrow}\;\;\;\;\;\;\;\;\underbrace{\hspace{10mm}}_{d}\;\;\;\;\;\;\;\;\;\underbrace{\hspace{10mm}}_{N^{(3)}_\uparrow}}{\ket{\ast \cdots \ast + - \downarrow \cdots \downarrow + - \ast \cdots \ast}} &\iff& \underset{\underbrace{\hspace{8mm}}_{N^{(1)}_\uparrow}\;\;\underbrace{\hspace{10mm}}_{d+1}\;\;\underbrace{\hspace{8mm}}_{N^{(3)}_\uparrow}}{\ket{\ast \cdots \ast \uparrow \downarrow \cdots \downarrow \uparrow \ast \cdots \ast}} \nn \\
\underset{\underbrace{\hspace{8mm}}_{N^{(1)}_\uparrow}\;\;\;\;\;\;\;\;\;\;\;\;\;\;\;\underbrace{\hspace{8mm}}_{N^{(3)}_\uparrow}}{\ket{\ast \cdots \ast + \uparrow -  \ast \cdots \ast}} &\iff& \underset{\underbrace{\hspace{10mm}}_{N^{(1)}_\uparrow}\hspace{4mm}\underbrace{\hspace{10mm}}_{N^{(3)}_\uparrow}}{\ket{\ast \cdots \ast \uparrow \uparrow \ast \cdots \ast}},
\label{eq:dipoletouchmapping}
\end{eqnarray}
where the $\ast$'s remain in the same configurations as in the original configurations.  

The reverse mapping from the spin-1/2 Hilbert space is unique provided the quantities $(N^{(1)}_\uparrow, N^{(3)}_\uparrow)$ are fixed, and it proceeds as follows.
In the spin-1/2 configuration, we identify two $\uparrow$ spins such that there are $N^{(1)}_\uparrow$ and $N^{(3)}_\uparrow$ $\uparrow$ spins to the left and right of them respectively. 
Since the spin-1/2 configuration has $(N^{(1)}_\uparrow + N^{(3)}_\uparrow + 2)$ $\uparrow$ spins, we are guaranteed that the two chosen $\uparrow$'s only have $\downarrow$'s between them. 
Depending on the number of $\downarrow$'s between the chosen $\uparrow$'s, we then use Eq.~(\ref{eq:dipoletouchmapping}) to obtain the corresponding configuration in the Krylov subspace $\mathcal{K}\left(\ket{\psi_0}, H\right)$.
Through the mapping Eq.~(\ref{eq:dipoletouchmapping}), the action of the Hamiltonian restricted to this Krylov subspace is equivalent to the XX model of $(N - 1)$ sites. 
To see this, note that the action of the terms of the Hamiltonian on the states in Eq.~(\ref{eq:dipoletouchmapping}) can be of three kinds, which can be mapped onto the action on spin degrees of freedom through Eq.~(\ref{eq:dipoletouchmapping}):
\begin{eqnarray}
    \ket{\cdots + - \downarrow \cdots} &\leftrightarrow& \ket{\cdots \downarrow + - \cdots} \nn \\
    \iff \ket{\cdots \uparrow \downarrow \cdots} &\leftrightarrow& \ket{\cdots \downarrow \uparrow \cdots} \, ,\label{eq:hamilactions1} \\
    \ket{\cdots + - + - \cdots} &\leftrightarrow& \ket{\cdots \downarrow + \uparrow - \cdots} \nn \\
    \iff \ket{\cdots \uparrow \downarrow \uparrow \cdots} &\leftrightarrow& \ket{\cdots \downarrow \uparrow \uparrow \cdots} \, ,\label{eq:hamilactions2} \\
    \ket{\cdots + - + - \cdots} &\leftrightarrow& \ket{\cdots + \uparrow - \downarrow \cdots} \nn \\
    \iff \ket{\cdots \uparrow \downarrow \uparrow \cdots} &\leftrightarrow& \ket{\cdots \uparrow \uparrow \downarrow \cdots},
\label{eq:hamilactions3}
\end{eqnarray}
which are precisely the actions of the XX model on spin-$1/2$'s. 
The mapping for two adjacent $-+$ dipoles proceeds analogously, with $\downarrow$ and $\uparrow$ interchanged in Eqs.~(\ref{eq:dipoletouchmapping}) and (\ref{eq:hamilactions1})-(\ref{eq:hamilactions3}). 
The preceding discussion for the two-dipole subspace can be extended to the Krylov subspace generated by a root state containing $n$ dipoles with no spins between them ($\ast$'s are $\uparrow$ or $\downarrow$):
\begin{equation}
    \ket{\psi_0} = \underset{\underbrace{\hspace{10mm}}_{N^{(1)}_\uparrow}\underbrace{\hspace{35mm}}_{n\ \textrm{dipoles}}\underbrace{\hspace{10mm}}_{N^{(n+1)}_\uparrow}}{\ket{\ast \cdots \ast \fbox{$+ - + - \cdots + - + -$} \ast \cdots \ast}}, 
\label{eq:multidipolerootstate}
\end{equation}
and $N^{(1)}_\uparrow$ and $N^{(n+1)}_\uparrow$ denote the number of $\uparrow$'s to the left and right of the string of dipoles respectively.  
We map the root state Eq.~(\ref{eq:multidipolerootstate}) to a spin configuration by identifying each $+ -$ dipole by an $\uparrow$, and inserting a $\downarrow$ spin between the dipoles. 
Thus, we find that $\ket{\psi_0}$ maps onto a spin-1/2 configuration by replacing the $n$ consecutive dipoles by a ``N\'{e}el state" of $(2n - 1)$ spins: 
\begin{equation}
    \ket{\psi_0} \iff \underset{\underbrace{\hspace{10mm}}_{N^{(1)}_\uparrow}\underbrace{\hspace{20mm}}_{(2n-1)\ \textrm{spins}}\underbrace{\hspace{10mm}}_{N^{(n+1)}_\uparrow}}{\ket{\ast \cdots \ast \fbox{$\uparrow \downarrow \uparrow \cdots \uparrow \downarrow \uparrow$} \ast \cdots \ast}}
\label{eq:neelstate}
\end{equation}
We do not attempt to rigorously prove the mapping for arbitrary $n$, but instead illustrate the mapping for the case when $n = 3$ and provide a conjecture for arbitrary $n$.
When the dipoles interact among themselves according to Eqs.~(\ref{eq:+fracton}) and (\ref{eq:-fracton}), the mappings read ($\ast$'s are $\uparrow$ or $\downarrow$):
\begin{eqnarray}
    \ket{\ast \cdots \ast + - + - + - \ast \cdots \ast} & \iff & \ket{\ast \cdots \ast \uparrow \downarrow \uparrow \downarrow \uparrow \ast \cdots \ast} \nn \\
    \ket{\ast \cdots \ast \downarrow + \uparrow - + - \ast \cdots \ast} & \iff & \ket{\ast \cdots \ast \downarrow \uparrow \uparrow \downarrow \uparrow \ast \cdots \ast} \nn \\
    \ket{\ast \cdots \ast + \uparrow - \downarrow + - \ast \cdots \ast} & \iff & \ket{\ast \cdots \ast \uparrow \uparrow \downarrow \downarrow \uparrow \ast \cdots \ast} \nn \\
    \ket{\ast \cdots \ast + - \downarrow + \uparrow - \ast \cdots \ast} & \iff & \ket{\ast \cdots \ast \uparrow \downarrow \downarrow\uparrow  \uparrow \ast \cdots \ast} \nn \\
    \ket{\ast \cdots \ast + - + \uparrow - \downarrow \ast \cdots \ast} & \iff & \ket{\ast \cdots \ast \uparrow \downarrow \uparrow \uparrow \downarrow \ast \cdots \ast} \nn \\
    \ket{\ast \cdots \ast \downarrow + \uparrow \uparrow - \downarrow \ast \cdots \ast} & \iff & \ket{\ast \cdots \ast \downarrow \uparrow \uparrow \uparrow \downarrow \ast \cdots \ast} \nn \\
    \ket{\ast \cdots \ast + \uparrow - + - \downarrow \ast \cdots \ast} & \iff & \ket{\ast \cdots \ast \uparrow \uparrow \downarrow \uparrow \downarrow \ast \cdots \ast} \nn \\
    \ket{\ast \cdots \ast + \uparrow \uparrow - \downarrow \downarrow \ast \cdots \ast} & \iff & \ket{\ast \cdots \ast \uparrow \uparrow \uparrow \downarrow \downarrow \ast \cdots \ast} \nn \\
    \ket{\ast \cdots \ast \downarrow + - + \uparrow - \ast \cdots \ast} & \iff & \ket{\ast \cdots \ast \downarrow \uparrow \downarrow\uparrow  \uparrow \ast \cdots \ast} \nn \\
    \ket{\ast \cdots \ast \downarrow \downarrow + \uparrow \uparrow - \ast \cdots \ast} & \iff & \ket{\ast \cdots \ast \downarrow \downarrow \uparrow \uparrow  \uparrow \ast \cdots \ast}, \nn \\
\label{eq:threedipoletouchall}
\end{eqnarray}
where the quantities $N^{(1)}_\uparrow$ and $N^{(4)}_\uparrow$ (shown in Eq.~(\ref{eq:neelstate})) are conserved in each of the above configurations. 
Apart from these, applying the Hamiltonian to configurations in Eq.~(\ref{eq:threedipoletouchall}) where dipoles move according to Eq.~(\ref{eq:hamilactions1}), we derive the following maps ($\ast$'s are $\uparrow$ or $\downarrow$): 
\begin{widetext}
\begin{eqnarray}
    \underset{\;\underbrace{\hspace{8mm}}_{d'}\;\;\;\hspace{6mm}\underbrace{\hspace{8mm}}_{d}\;\;}{\ket{\ast \cdots \ast + - \downarrow \cdots \downarrow + - \downarrow  \cdots \downarrow + - \ast \cdots \ast}}  &\iff&  \underset{\;\;\underbrace{\hspace{8mm}}_{d'+1}\;\;\;\;\underbrace{\hspace{8mm}}_{d+1}\;\;}{\ket{\ast \cdots \ast \uparrow \downarrow \cdots \downarrow \uparrow \downarrow \cdots \downarrow \uparrow \ast \cdots \ast}} \nn \\
    \underset{\;\;\;\;\underbrace{\hspace{8mm}}_d}{\ket{\ast \cdots \ast + \uparrow - \downarrow \cdots \downarrow + - \ast \cdots \ast}} & \iff & \underset{\;\;\underbrace{\hspace{8mm}}_{d+1}}{\ket{\ast \cdots \ast \uparrow \uparrow \downarrow \cdots \downarrow \uparrow \ast \cdots \ast}} \nn \\
    \underset{\underbrace{\hspace{8mm}}_d\;\;\;\;\;}{\ket{\ast \cdots \ast + - \downarrow \cdots \downarrow + \uparrow - \ast \cdots \ast}} & \iff & \underset{\underbrace{\hspace{8mm}}_{d+1}\;\;}{\ket{\ast \cdots \ast \uparrow \downarrow \cdots \downarrow\uparrow  \uparrow \ast \cdots \ast}}. \nn \\
\label{eq:threedipoletouchalldipmove}
\end{eqnarray}
\end{widetext}
Note that in order to derive the mapping for a particular configuration within this Krylov subspace, one should start from the mapping of the root state in Eq.~(\ref{eq:neelstate}) and follow the actions of the Hamiltonian in Eqs.~(\ref{eq:hamilactions1})-(\ref{eq:hamilactions3}).
Note that the mappings of Eqs.~(\ref{eq:multidipolerootstate}), (\ref{eq:threedipoletouchall}), and (\ref{eq:threedipoletouchalldipmove}) are only valid if there are no other dipoles or fractons other than the ones shown.  
We discuss the case of multiple dipole blocks in the next subsection. 
\subsection{Systematic construction of integrable Krylov subspaces}
\label{sec:systematicapp}
%
%
In the previous section, we conjectured that the Krylov subspace generated by a root state with $n$ contiguous dipoles is integrable and maps onto a particular quantum number sector of an XX model with $(N - 1)$ sites, and we showed an example for $n = 3$.
This mapping can be extended to Krylov subspaces generated by root states containing configurations with $m$ blocks of contiguous $+ -$ dipoles, with at least one $\uparrow$ separating the blocks.
We start by illustrating the case when $m = 2$. The root state with two blocks of dipoles reads
\begin{equation}
    \ket{\psi_0} = \overset{\overbrace{\hspace{18mm}}^{n_1\  \textrm{dipoles}}\hspace{10mm}\overbrace{\hspace{18mm}}^{n_2\  \textrm{dipoles}}}{\underset{\hspace{5mm}\underbrace{\hspace{9mm}}_{N^{(1)}_\uparrow}\hspace{18mm}\underbrace{\hspace{9mm}}_{N^{(n_1 +1)}_\uparrow}\hspace{17mm}\underbrace{\hspace{9mm}}_{N^{(n_1 + n_2 + 2)}_\uparrow}}{\ket{\ast \cdots \ast + - \cdots + - \ast \cdots \ast  + - \cdots + - \ast \cdots \ast}}}, 
\label{eq:dipolesetroot}
\end{equation}
where $\ast =\ \uparrow, \downarrow$, $N^{(j)}_\uparrow$ denotes the number of $\uparrow$ spins in the $j$-th segment of the chain, which is the part of the chain between the $j$-th and $(j+1)$-th $+-$ dipole.
In Eq.~(\ref{eq:dipolesetroot}), $N^{(j)}_\uparrow = 0$ if $2\leq j \leq n_1$ or $n_1 + 2 \leq j \leq n_1 + n_2 + 1$, 
and we are considering the case where $N^{(n_1+1)}_\uparrow \geq 1$.
The mapping from $\ket{\psi_0}$ onto a configuration of spin-$1/2$'s proceeds as follows.
The two groups of $n_1$ and $n_2$ dipoles are mapped onto N\'{e}el states of $(2 n_1 - 1)$ and $(2 n_2 - 1)$ spins respectively.
Since the Hamiltonian acts on the state $\ket{\psi_0}$ in Eq.~(\ref{eq:dipolesetroot}) according to Eqs.~(\ref{eq:hamilactions1})-(\ref{eq:hamilactions3}), the Hamiltonian $H$ restricted to this Krylov subspace is the XX model of size $(N - 2)$.
The full dictionary of mappings to the spin-1/2 Hilbert space can be derived by starting from the mapping for the root configuration and following the actions of the Hamiltonian in Eqs.~(\ref{eq:hamilactions1})-(\ref{eq:hamilactions3}).
This mapping directly generalizes to a root state with $m$ dipole groups with the form
\begin{equation}
    \ket{\psi_0} = \underset{\hspace{4mm}\underbrace{\hspace{9mm}}_{N^{(1)}_\uparrow}\hspace{50mm}\underbrace{\hspace{9mm}}_{N^{(\sumal{k = 1}{m}{n_k} + m)}_\uparrow}}{\overset{\overbrace{\hspace{18mm}}^{n_1\  \textrm{dipoles}}\hspace{16mm}\overbrace{\hspace{18mm}}^{n_m\  \textrm{dipoles}}}{\ket{\ast \cdots \ast + - \cdots + - \diamond \cdots \cdots \diamond  + - \cdots + - \ast \cdots \ast}}}, 
\label{eq:dipolesetrootgenm}
\end{equation}
where $\ast =\ \uparrow, \downarrow$, and $\diamond \cdots \diamond$ consists of spins and $(m -2)$ blocks of $+-\cdots+-$, with two adjacent blocks separated by at least one $\uparrow$.  
The mapping to the spin-1/2 chain proceeds by mapping each sequence of $n_l$ adjacent dipoles in Eq.~(\ref{eq:dipolesetrootgenm}) onto a N\'{e}el state of $(2 n_l -1)$ spins, as depicted in Eq.~(\ref{eq:neelstate}).
Since the Hamiltonian acts on this subspace according to Eqs.~(\ref{eq:hamilactions1})-(\ref{eq:hamilactions3}), the Hamiltonian restricted to this Krylov subspace is the XX model with $\left(N - m\right)$ sites.
In general, the Krylov subspace consisting of $n$ $+-$ (resp. $-+$) dipoles with $m$ values of $j$ such that $N^{(j)}_\uparrow = 0$ (resp. $N^{(j)}_\downarrow = 0$), precisely maps onto the XX model with $(N - n + m)$ sites.
%
%
%
%
%

\section{Effect of electrostatic terms and disorder}
\label{sec:electroaction}
\subsection{Electrostatic terms}
We now briefly discuss the effect of adding electrostatic terms to the analysis of the integrable Krylov subspaces discussed in Sec.~\ref{sec:integrablekrylov}.
In particular, we consider two simple perturbations to the Hamiltonian
\begin{equation}
    \delta H_1 = V_1\sumal{j = 1}{L_b}{\hat{n}_j \hat{n}_{j+1}},\;\;\; \delta H_2 = V_2\sumal{j = 1}{L'_b}{\hat{n}_j \hat{n}_{j+2}},
\label{eq:electro}
\end{equation}
where $L_b = L -1$ (resp. $L_b = L$) and $L'_b = L-2$ (resp. $L'_b = L$) for OBC (resp. PBC).
The terms in Eq.~(\ref{eq:electro}) are the simplest two electrostatic terms. In experimentally relevant settings, these terms typically have strengths greater than or comparable to that of the pair-hopping Hamiltonian $H$; see Eq.~(\ref{eq:gencomhamil}) and Sec.~\ref{sec:model} for a discussion of their sizes.  

The electrostatic terms are nearest neighbor terms that are diagonal in the basis of product states of the composite degrees of freedom i.e., of the spins and fractons defined in Eq.~(\ref{eq:halffillingpart}). 
Since composite degrees of freedom are formed by grouping pairs of neighboring sites, in terms of composite degrees of freedom the electrostatic Hamiltonians $\delta H_1$ and $\delta H_2$ in Eq.~(\ref{eq:electro}) map onto nearest-neighbor Hamiltonians $\delta \mathcal{H}_1$ and $\delta \mathcal{H}_2$, which have the forms: 
\begin{equation}
    \delta\mathcal{H}_1 = \sumal{j = 1}{N_b}{{\left(\delta\mathcal{H}_1\right)}_{j, j+1}} + \sumal{j = 1}{N}{{\left(\delta\mathcal{H}_1\right)}_{j}},\;\;\delta\mathcal{H}_2 = \sumal{j = 1}{N_b}{{\left(\delta\mathcal{H}_2\right)}_{j, j+1}},
\end{equation}
where $N_b = N$ (resp. $N_b = N -1$) for PBC (resp. OBC), and $\{\left(\delta\mathcal{H}_\alpha\right)_{j,j+1}\}$ and $\{\left(\delta\mathcal{H}_\alpha\right)_j\}$ are nearest-neighbor and onsite terms respectively.
The actions of each of the nearest-neighbor terms follows directly by using Eq.~(\ref{eq:electro}) and the definitions Eq.~(\ref{eq:halffillingpart}), and can be tabulated as:
\begin{equation}
    \begin{tabular}{c|c|c}
        Config. & $\delta\mathcal{H}_1$ & $\delta\mathcal{H}_2$ \\
        \hline
        $\ket{++}$     & $V_1$ & $2V_2$ \\
        $\ket{+\uparrow}$   & 0 & $V_2$ \\
        $\ket{+\downarrow}$ & $V_1$ & $V_2$\\
        $\ket{+-}$   & 0 & 0 \\
        $\ket{\uparrow +}$  & $V_1$ & $V_2$ \\
        $\ket{\uparrow \uparrow}$  & 0 & $V_2$ \\
        $\ket{\uparrow\downarrow}$  & $V_1$ & 0\\
        $\ket{\uparrow-}$ & 0 & 0 
    \end{tabular}\;\;\;\;\;\;
    \begin{tabular}{c|c|c}
        Config. & $\delta\mathcal{H}_1$ & $\delta\mathcal{H}_2$ \\
        \hline
        $\ket{\downarrow +}$ & 0 & $V_2$ \\
        $\ket{\downarrow\uparrow}$ & 0 & 0\\
        $\ket{\downarrow \downarrow}$ & 0 & $V_2$ \\
        $\ket{\downarrow -}$ & 0 & 0\\
        $\ket{-+}$ & 0 & 0\\
        $\ket{-\uparrow}$   & 0 & 0 \\
        $\ket{-\downarrow}$ & 0 & 0\\
        $\ket{--}$ & 0 & 0 
    \end{tabular},
\label{eq:electroterms}
\end{equation}
whereas the onsite terms read:
\begin{equation}
    \begin{tabular}{c|c}
        Config. & $\delta\mathcal{H}_1$ \\
        \hline
        $\ket{+}$ & $V_1$ \\
        $\ket{-}$ & 0
    \end{tabular}\;\;\;
    \begin{tabular}{c|c}
        Config. & $\delta\mathcal{H}_1$ \\
        \hline
        $\ket{\uparrow}$ & 0 \\
        $\ket{\downarrow}$ & 0
    \end{tabular}.
\label{eq:onsite}
\end{equation}
Importantly, since these terms are diagonal in the product basis, they do not change the structure of Krylov subspaces generated from product states. In other words, the full Hilbert space is still expressed in the same form as Eq.~(\ref{eq:fullhilbert}) irrespective of whether $H$ contains electrostatic terms or not. 
Within the integrable spin subspace discussed in Sec.~\ref{sec:spinsubspace}, the onsite terms always vanish according to Eq.~(\ref{eq:onsite}). Further, according to Eq.~(\ref{eq:electroterms}), the actions of the nearest neighbor terms of $\delta\mathcal{H}_1$ and $\delta\mathcal{H}_2$ read
\begin{eqnarray}
    &(\delta \mathcal{H}_1 + \delta\mathcal{H}_2)_{j,j+1} \ket{\downarrow \uparrow} = 0,\nn \\
    &(\delta\mathcal{H}_1 + \delta\mathcal{H}_2)_{j,j+1} \ket{\uparrow \downarrow} = V_1 \ket{\uparrow \downarrow},\nn \\
    &(\delta\mathcal{H}_1 + \delta\mathcal{H}_2)_{j,j+1} \ket{\uparrow \uparrow} = V_2 \ket{\uparrow \uparrow},\nn \\
    &(\delta\mathcal{H}_1 + \delta\mathcal{H}_2)_{j, j+1} \ket{\downarrow \downarrow} = V_2 \ket{\downarrow \downarrow}.\nn \\
\label{eq:elecact}
\end{eqnarray}
The above actions of the electrostatic terms are succinctly encoded in the Hamiltonian $\delta\mathcal{H} = \delta\mathcal{H}_1 + \delta\mathcal{H}_2$
\begin{align}
    \delta\mathcal{H} &= \sumal{j = 1}{N_b}{\left(\frac{V_2}{2}\left(1 + \sigma^z_j \sigma^z_{j+1}\right) + \frac{V_1}{4}\left(1 + \sigma^z_j\right)\left(1 - \sigma^z_{j+1}\right)\right)}\nn \\
    &= \sumal{j = 1}{N_b}{\left(\frac{V_1 + 2 V_2}{4} + \frac{2V_2 - V_1}{4}\sigma^z_j \sigma^z_{j+1}\right)} + \frac{V_1}{4}\sumal{j=1}{N_b}{\left(\sigma^z_j - \sigma^z_{j+1}\right)}. \nn \\
\label{eq:electrohamil}
\end{align}
where the unit cell index $j$ is defined modulo $N$ for PBC. 
Thus, Eq.~(\ref{eq:electrohamil}) reduces to
\begin{eqnarray}
    &\delta\mathcal{H} = \sumal{j = 1}{N_b}{\left(\frac{V_1 + 2 V_2}{4} + \frac{2V_2 - V_1}{4}\sigma^z_j \sigma^z_{j+1}\right)}\nn \\
    &+ \twopartdef{\frac{V_1}{4}\left(\sigma^z_1 - \sigma^z_N\right)}{OBC}{0}{PBC}.
\end{eqnarray}
Thus, the restriction of the total Hamiltonian---the pair-hopping Hamiltonian in addition to the electrostatic terms---to the spin Krylov subspace maps onto (for PBC and and infinite chain for OBC)
\begin{equation}
    H_T = \sumal{j}{}{\left(\frac{V_1 + 2V_2}{4} + \sigma^+_j \sigma^-_{j+1} + \sigma^-_j \sigma^+_{j+1} + \frac{2V_2 - V_1}{4}\sigma^z_j \sigma^z_{j+1} \right)}, 
\end{equation}
which is the translation invariant XXZ model and is thus Bethe Ansatz integrable.
In contrast, the integrable dipole subspaces discussed in Secs.~\ref{sec:dipolesubspace} and~\ref{sec:multidipolesubspace} become non-integrable upon the addition of electrostatic terms. 
To see this, consider the action of the nearest-neighbor terms of $\left(\delta \mathcal{H}_1 + \delta \mathcal{H}_2\right)$ on the dipole, which are given by (using Eq.~(\ref{eq:electroterms}))
\begin{eqnarray}
    \left(\delta \mathcal{H}_1 + \delta \mathcal{H}_2\right) \ket{\uparrow + -} &=& \left(V_1 + V_2\right)\ket{\uparrow + - }, \nn \\
    \left(\delta \mathcal{H}_1 + \delta \mathcal{H}_2\right)\ket{\downarrow + -} &=& V_2\ket{\downarrow + - }, \nn \\
    \left(\delta \mathcal{H}_1 + \delta \mathcal{H}_2\right)\ket{+ - \uparrow} &=& 0, \nn \\
    \left(\delta \mathcal{H}_1 + \delta \mathcal{H}_2\right)\ket{+ - \downarrow} &=& 0.
\label{eq:dipoleelectroaction}
\end{eqnarray}
When $V_1 \neq 0$, the actions encoded in Eq.~(\ref{eq:dipoleelectroaction}) break the symmetry between the configurations $\ket{+ - \ast}$ and $\ket{\ast + -}$, where $\ast =\ \uparrow, \downarrow$.
Thus the dipole \textit{cannot} be identified with an $\uparrow$, as is the case in the absence of electrostatic terms.   
We have verified that upon addition of electrostatic terms, the energy levels within any quantum number sector of the dipole subspace show GOE level statistics. 
The same is true for Krylov subspaces with $-+$ dipoles, for which the action of the electrostatic terms follows from Eq.~(\ref{eq:dipoleelectroaction}) upon the application of inversion symmetry. 
\subsection{Disorder}\label{sec:disorderaction}
Consider the disordered pair-hopping Hamiltonian, 
\begin{equation}
    H = \sumal{j = 1}{L_b}{H_j} = \sumal{j = 1}{L_b}{J_j \left(\cd_j \cd_{j+3} c_{j + 2} c_{j + 1} + h.c.\right)},
\label{eq:disoderedpairhopping}
\end{equation}
where $L_b = L - 3$ (resp. $L_b = L$) for OBC (resp. PBC), and $\{J_j\}$ are the disordered couplings. 
Assuming $L = 2N$,  we divide the Hamiltonian Eq.~(\ref{eq:disoderedpairhopping}) into two parts to preempt the mapping onto composite degrees of freedom, defined in Eq.~(\ref{eq:halffillingpart}):
\begin{eqnarray}
    &H = \sumal{j = 1}{N^{(o)}_b}{J_{2j - 1}\left(\cd_{2j -1 } \cd_{2j + 2} c_{2j + 1} c_{2j} + h.c.\right)} \nn \\
    &+\sumal{j = 1}{N^{(e)}_b}{J_{2j} \left(\cd_{2j} \cd_{2j+3} c_{2j + 2} c_{2j + 1} + h.c.\right)},
\label{eq:disoderedpairhopping2}
\end{eqnarray}
where $N^{(o)}_b = N - 1$ (resp. $N^{(o)}_b = N$) and $N^{(e)}_b = N -2$ (resp. $N^{(e)}_b = N$) for OBC (resp. PBC). 
Once the sites $(2j - 1)$ and $2j$ are grouped into one unit cell, the actions of the Hamiltonian terms are as follows (see Eqs.~(\ref{eq:spinscattering})-(\ref{eq:-fracton})):
\begin{eqnarray}
    \overset{2j\ 2j+1}{\ket{\ \fbox{0\ 1}\ \fbox{1\ 0}\ }} &\xleftrightarrow{J_{2j-1}}& \overset{2j\ 2j+1}{\ket{\ \fbox{1\ 0}\ \fbox{0\ 1}\ }}\nn \\
    \iff \ket{\uparrow \downarrow} &\xleftrightarrow{J_{2j-1}}& \ket{\downarrow\uparrow} \, , \label{eq:disspinscattering}\\
    \overset{2j\;\;\;\;\;\;\;\;\;2j+3}{\ket{\ \fbox{1\ 0}\ \fbox{1\ 1}\ \fbox{0\ 0}\ }} &\xleftrightarrow{J_{2j}}& \overset{2j\;\;\;\;\;\;\;\;\;2j+3}{\ket{\ \fbox{1\ 1}\ \fbox{0\ 0}\ \fbox{1\ 0}\ }} \nn \\
    \iff \ket{\downarrow + -} &\xleftrightarrow{J_{2j}}& \ket{+ - \downarrow} \, , \label{eq:disdipolescattering1} \\
    \overset{2j\;\;\;\;\;\;\;\;\;2j+3}{\ket{\ \fbox{0\ 0}\ \fbox{1\ 1}\ \fbox{0\ 1}\ }} &\xleftrightarrow{J_{2j}}& \overset{2j\;\;\;\;\;\;\;\;\;2j+3}{\ket{\ \fbox{0\ 1}\ \fbox{0\ 0}\ \fbox{1\ 1}\ }} \nn \\
    \iff \ket{- + \uparrow} &\xleftrightarrow{J_{2j}}& \ket{\uparrow - +} \, , \label{eq:disdipolescattering2} \\
    \overset{2j\;\;\;\;\;\;\;\;\;2j+3}{\ket{\ \fbox{1\ 0}\ \fbox{1\ 1}\ \fbox{0\ 1}\ }} &\xleftrightarrow{J_{2j}}& \overset{2j\;\;\;\;\;\;\;\;\;2j+3}{\ket{\ \fbox{1\ 1}\ \fbox{0\ 0}\ \fbox{1\ 1}\ }} \nn \\
    \iff \ket{\downarrow + \uparrow} &\xleftrightarrow{J_{2j}}& \ket{+ - +} \, , \label{eq:dis+fracton} \\
    \overset{2j\;\;\;\;\;\;\;\;\;2j+3}{\ket{\ \fbox{0\ 1}\ \fbox{0\ 0}\ \fbox{1\ 0}\ }} &\xleftrightarrow{J_{2j}}& \overset{2j\;\;\;\;\;\;\;\;\;2j+3}{\ket{\ \fbox{0\ 0}\ \fbox{1\ 1}\ \fbox{0\ 0}\ }} \nn \\
    \iff \ket{\uparrow - \downarrow} &\xleftrightarrow{J_{2j}}& \ket{- + -}, \label{eq:dis-fracton}
\end{eqnarray}
where $\ket{a} \xleftrightarrow{J} \ket{b}$ denotes that the action of a term of the Hamiltonian on the configuration $\ket{a}$ results in $\ket{b}$ with a coefficient $J$, and vice-versa. 
Since the spin Krylov subspace discussed in Sec.~\ref{sec:spinsubspace} is only sensitive to the action of the Hamiltonian on the spin degrees of freedom, according to Eq.~(\ref{eq:disspinscattering}) the Hamiltonian restricted to the Krylov subspace maps onto the disordered XX model:
\begin{equation}
    H = \sumal{j = 1}{N_b}{J_{2j - 1}\left(\sigma^+_j \sigma^-_{j + 1} + \sigma^-_j \sigma^+_{j+1}\right)}, 
\end{equation}
where $N_b = N - 1$ (resp. $N_b = N$) for OBC (resp. PBC). Thus, we expect that the spin Krylov subspace exhibits Anderson localization~\cite{anderson1958} upon the addition of disorder. 
We now analyze the effect of disorder on the single or multi-dipole Krylov subspaces. 
Recall that the Hamiltonian restricted to the dipole Krylov subspaces in Secs.~\ref{sec:dipolesubspace} and \ref{sec:multidipolesubspace} maps onto the XX model by identifying the $+ -$ (resp. $- +$) dipole with an $\uparrow$ (resp. $\downarrow$) and noting that Eq.~(\ref{eq:dipolescattering1}) (resp. Eq.~(\ref{eq:dipolescattering2})) is identical to Eq.~(\ref{eq:spinscattering}) upon this identification.  
However, if $J_{2j -1} \neq J_{2j}$, Eq.~(\ref{eq:disdipolescattering1}) (resp. Eq.~(\ref{eq:disdipolescattering2})) is \textit{no longer} identical to Eq.~(\ref{eq:disspinscattering}) when $+-$ (resp. $-+$) is identified with $\uparrow$ (resp. $\downarrow$).
Hence, the Hamiltonian restricted to dipole Krylov subspaces does not map onto the disordered XX model as one would naively expect. 
%
%
%
\section{Properties of the Fracton Krylov subspace}
\label{sec:fractonkrylov}
Here, we discuss some properties of the Fracton Krylov subspace discussed in Sec.~\ref{sec:nonintkrylov}.
To understand the effects of this constrained Krylov subspace, we focus on odd system sizes $N$ and on the root state consisting of a $+$ fracton on the center site $(N + 1)/2$ along with an equal number of $\uparrow$ and $\downarrow$ spins enveloping it, as shown in Eq.~(\ref{eq:fractonrootstate}).
Such a configuration has charge $Q = 1$, spin $S^z = 0$, and dipole moment $D= \exp\left(i \pi (N+1)/N\right)$ (see Eq.~(\ref{eq:dipoleoperator}) for the definition of dipole moment with PBC). 
We refer to the site containing the fracton in the root state as the \textit{middle site}. 

For purposes of illustration, we consider the root state $\ket{\psi_0} = \ket{\uparrow\uparrow\uparrow+\downarrow\downarrow\downarrow}$ with $N = 7$ and with PBC, which has charge $Q = 1$, spin $S = 0$, and dipole moment $D = e^{8 \pi i/7}$. 
Using the actions of Eqs.~(\ref{eq:spinscattering})-(\ref{eq:-fracton}), the product state configurations in $\mathcal{K}\left(H, \ket{\psi_0}\right)$ are then 

\begin{widetext}
\begin{eqnarray}
    &\ket{\uparrow \uparrow \uparrow + \downarrow \downarrow \downarrow}, \nn \\
    &\ket{\downarrow \uparrow \uparrow + \downarrow \downarrow \uparrow},\nn \\
    &\ket{\uparrow \downarrow \uparrow + \downarrow \downarrow \uparrow},\;\; \ket{\downarrow \uparrow \uparrow + \downarrow \uparrow \downarrow}, \nn \\
    &\ket{\uparrow \uparrow \downarrow + \downarrow \downarrow \uparrow},\;\; \ket{\uparrow \downarrow \uparrow + \downarrow \uparrow \downarrow},\;\; \ket{\downarrow \uparrow \uparrow + \uparrow \downarrow \downarrow}\nn \\
    &\ket{\uparrow \uparrow \downarrow + \downarrow \uparrow \downarrow},\;\; \ket{\uparrow \downarrow \uparrow + \uparrow \downarrow \downarrow},\;\;\ket{\downarrow\downarrow\uparrow+\downarrow\uparrow\uparrow} \nn \\
    &\ket{\uparrow \uparrow \downarrow + \uparrow \downarrow \downarrow},\;\; \ket{\downarrow \uparrow \downarrow + \uparrow \downarrow \uparrow},\;\;\ket{\downarrow \downarrow \uparrow + \uparrow \downarrow \uparrow},\;\;\ket{\downarrow\uparrow\downarrow+\downarrow\uparrow\uparrow} \nn \\
    &\ket{\uparrow \uparrow + - + \downarrow \downarrow}, \;\; \ket{\downarrow \uparrow + -+ \downarrow \uparrow},\;\;\ket{\uparrow\downarrow\downarrow+\downarrow\uparrow\uparrow},\;\;\ket{\uparrow \downarrow \downarrow + \uparrow \downarrow \uparrow},\;\;\ket{\downarrow \downarrow \uparrow + \uparrow \uparrow \downarrow},\;\;\ket{\downarrow \uparrow \downarrow + \uparrow \uparrow \downarrow}\nn \\
    &\ket{\uparrow \downarrow + - + \downarrow \uparrow},\;\;\ket{\downarrow \uparrow + - + \uparrow \downarrow},\;\;\ket{\uparrow \downarrow \downarrow + \uparrow \uparrow \downarrow} \nn \\
    &\ket{\uparrow + - \downarrow + \downarrow \uparrow}, \;\; \ket{\uparrow \downarrow + - + \uparrow \downarrow},\;\; \ket{\downarrow \uparrow + \uparrow - + \downarrow},\;\;\ket{\downarrow \downarrow \downarrow + \uparrow \uparrow \uparrow} \nn \\
    &\ket{\downarrow \downarrow  + - + \uparrow \uparrow},\;\;\ket{\uparrow + - \downarrow + \uparrow \downarrow}, \;\;\ket{\uparrow \downarrow + \uparrow - + \downarrow} \nn \\
    &\ket{\uparrow + - + - + \downarrow}, \;\; \ket{\downarrow \downarrow + \uparrow - + \uparrow},\;\;\ket{\downarrow + - \downarrow + \uparrow \uparrow} \nn \\
    &\ket{\downarrow + - + - + \uparrow}, \;\; \ket{+ - \downarrow\downarrow + \uparrow\uparrow},\;\;\ket{\downarrow \downarrow + \uparrow \uparrow - +},\;\;\ket{\uparrow + \uparrow - \downarrow + \downarrow} \nn \\
    &\ket{+ - \downarrow + - + \uparrow}, \;\; \ket{\downarrow + - + \uparrow - +},\;\;\ket{\downarrow + \uparrow - \downarrow + \uparrow},\nn \\
    &\ket{+ - + - \downarrow + \uparrow}, \;\; \ket{\downarrow + \uparrow - + - +},\;\;\ket{+ - \downarrow + \uparrow - +}\nn \\
    &\ket{+ \uparrow - \downarrow \downarrow + \uparrow},\;\;\ket{+ - + - + - +},\;\;\ket{\downarrow  + \uparrow \uparrow - \downarrow +} \nn \\
    &\ket{+ - + \uparrow - \downarrow +},\;\;\ket{+ \uparrow - \downarrow + - +},\;\;\ket{+ \uparrow - + - \downarrow +},\;\;\ket{+ \uparrow\uparrow - \downarrow \downarrow +},
\label{eq:fractonkrylovexample}
\end{eqnarray}
\end{widetext}
where configurations on the $n$-th row are product configurations belonging to $\textrm{span}\left\{\ket{\psi_0}, H\ket{\psi_0}, \cdots, H^{n-1}\ket{\psi_0}\right\}$ but not to $\textrm{span}\left\{\ket{\psi_0}, H\ket{\psi_0}, \cdots, H^{n-2}\ket{\psi_0}\right\}$. That is, they are the new product configurations obtained on the $(n - 1)$-th action of the Hamiltonian $H$ on $\ket{\psi_0}$.

As discussed in Sec.~\ref{sec:nonintkrylov} (see Eq.~(\ref{eq:quasilocalization})), in order to obtain the infinite temperature expectation value of the charge on the middle site \textit{within} the Krylov subspace, we must compute the Hilbert space dimension $\mathcal{D}_N$ of the Krylov subspace as well as $\mathcal{Q}_N$, the difference between the number of product states with a $+$ fracton and with a $-$ fracton on the middle site.
These quantities can be enumerated numerically for various system sizes, and are tabulated in the following:
\begin{equation}
    \begin{tabular}{c|c|c}
        $N$ & $\mathcal{D}_N$ & $\mathcal{Q}_N$ \\
        \hline 
        3 &  3 & 1\\
        5 & 12 & 1\\
        7 & 50 & 14\\
        9 & 210 & 15 \\
        11 & 882 & 56 \\
        13 & 3696 & 210 \\
        15 & 15444 & 792 \\
        17 & 64350 & 3003
    \end{tabular}
\label{eq:dimchargetable}
\end{equation}
For example, the total number of configurations in Eq.~(\ref{eq:fractonkrylovexample}) ($N = 7$) is $50$, and it can be explicitly verified that the total middle site charge summed over all configurations equals $14$. 
We now compute the infinite temperature expectation value of the middle site charge. 
We find that the Hilbert space dimension $\mathcal{D}_N$ for odd system sizes, tabulated in Eq.~(\ref{eq:dimchargetable}) for $N \leq 17$, corresponds to the integer sequence OEIS A092443~\cite{oeis}, which takes the standard form
\begin{equation}
    \mathcal{D}_{N = 2n + 1} = \frac{n+2}{2}\binom{2n}{n}.
\label{eq:krylovdimexact}
\end{equation}
Similarly we find that $\mathcal{Q}_N$, tabulated in Eq.~(\ref{eq:dimchargetable}) for $N \leq 17$, corresponds to the integer sequence OEIS A051924~\cite{oeis}, which has the closed form 
\begin{equation}
    \mathcal{Q}_{N = 2n + 1} = \frac{3n - 2}{n}\binom{2(n-1)}{n-1}.
\label{eq:krylovchargeexact}
\end{equation}

Although we do not attempt to prove this rigorously here, we posit that Eqs.~(\ref{eq:krylovdimexact}) and (\ref{eq:krylovchargeexact}) accurately represent $\mathcal{D}_{N = 2n + 1}$ and $\mathcal{Q}_{N = 2n + 1}$ for all values of $n$.  
With these expressions in hand, we can then analytically obtain the infinite temperature charge density from the ratio $\mathcal{Q}_N/\mathcal{D}_N$.  
To find the asymptotic behavior for large $N$, we use Stirling's approximation
\begin{equation}
    n! \approx \sqrt{2\pi n}\left(\frac{n}{e}\right)^n \,. 
\label{eq:stirling}
\end{equation}
The asymptotic behaviour of $\mathcal{D}_N$ of Eq.~(\ref{eq:krylovdimexact}) is then 
\begin{eqnarray}
\mathcal{D}_{N = 2n + 1} &=& \frac{n + 2}{2}\frac{(2n)!}{(n!)^2} \approx  \frac{n + 2}{2\sqrt{\pi n}}\frac{\left(\frac{2n}{e}\right)^{2n}}{\left(\frac{n}{e}\right)^{2n}} \nn \\
&\sim& \sqrt{\frac{n}{4\pi}}2^{2n} \sim \sqrt{\frac{N}{32\pi}}2^{N}, 
\end{eqnarray}
whereas $\mathcal{Q}_N$ asymptotes to
\begin{eqnarray}
    \mathcal{Q}_{N = 2n + 1} &=& \frac{3n - 2}{n}\frac{\left(2(n-1)\right)!}{\left(\left(n-1\right)!\right)^2} \nn \\
    &\approx& \frac{\left(3n - 2\right)}{n\sqrt{\pi\left(n - 1\right)}}\frac{\left(\frac{2(n-1)}{e}\right)^{2(n-1)}}{\left(\frac{n-1}{e}\right)^{2(n-1)}} \nn \\
    &\sim& \frac{3}{\sqrt{16 \pi n}}2^{2n} \sim \frac{3}{\sqrt{32 \pi N}}2^{N}.
\end{eqnarray}
Hence, for large N, the infinite temperature expectation value of the charge on the middle site is given by
\begin{equation}
    \frac{Q_{N}}{\mathcal{D}_{N}} \sim  \frac{3}{N}.
\label{eq:quasilocapp}
\end{equation}
\section{Schrieffer-Wolff Transformation for the Bloch MBL Hamiltonian}\label{sec:blochSW}
In this Appendix, we explicitly derive the pair-hopping Hamiltonian Eq.~(\ref{eq:pairhopping}) in the large $E/t$ limit of the Bloch MBL Hamiltonian Eq.~(\ref{eq:starkhamil}):
\begin{align}
    H_{\textrm{Bloch}} =\ & t \sumal{j}{}{\left(\cd_j c_{j+1} + h.c.\right)} + E\,\sumal{j}{}{j \, \hat{n}_j} \nonumber \\
    & + V_0 \sumal{j}{}{\hat{n}_j} + V_1 \sumal{j}{}{w_j \hat{n}_j \hat{n}_{j+1}},
\label{eq:blochmblhamil}
\end{align}
where we have omitted the limits on the sums since we consider a chain of infinite length.
Furthermore, we treat $t$, $V_0$, $V_1$ perturbatively and hence rescale Eq.~(\ref{eq:starkhamil}) by $E$, such that the Hamiltonian is recast as
\begin{equation}
    H \equiv \frac{H_{\textrm{Bloch}}}{E} =  \hC + \lambda\left(\hT_+ + \hT_- + \hV \right),
\end{equation}
where $\hC$ is the CoM operator (for OBC)
\beq
\hC = \sum_j  j n_j \, ,
\eeq
$\hT = \hT_+ + \hT_-$ and $\hV = \hV_0 + \hV_1$, with
\begin{eqnarray}
\label{eq:defns}
    &\hT_+ = \sumal{j}{}{\cd_{j+1} c_j}, \quad &\hT_- = \sumal{j}{}{\cd_j c_{j+1}} = \hT_+^\dagger, \nn \\
    &\hV_0 = \alpha_0 \sumal{j}{}{w_j \hn_j},\quad &\hV_1 = \alpha_1\sumal{j}{}{\hn_j \hn_{j+1}} \, .
\end{eqnarray}
Here, the parameters are defined as
\beq
\lambda = \frac{t}{E}, \quad \alpha_\nu = \frac{V_\nu}{t},
\eeq
for $\nu \in \{0,1\}$ and where we work in the regime where $\alpha_\sigma \sim \mathcal{O}\left( 1 \right)$.

As is clear from Eq.~(\ref{eq:defns}), $\hT_+$ and $\hT_-$ correspond to hopping processes that increase and decrease energies by one unit with respect to the CoM term $\hC$.
That is,
\begin{equation}
    \hC \ket{\mu} = \mE_\mu\ket{\mu} \implies \hC \left(\hT_\pm\ket{\mu}\right) = \left(\mE_{\mu} \pm 1\right) \hT_\pm \ket{\mu}. 
\end{equation}
Following the standard Schrieffer-Wolff procedure~\cite{bravyi2011schrieffer}, we divide the Hilbert space into ``blocks", which are subspaces degenerate under the leading order term $\hC$.  
Terms of the Hamiltonian which only have non-vanishing matrix elements within the same block are called ``block diagonal" whereas terms which only have non-vanishing matrix elements between different blocks are called ``block off-diagonal".   
For the Hamiltonian $H$, the ``block diagonal" and ``block off-diagonal" parts $H_d$ and $H_{od}$ respectively read 
\begin{equation}
    H = \underbrace{\hC + \lambda \hV}_{H_d} + \underbrace{\lambda\hT}_{H_{od}}.
\label{eq:Hsplit}
\end{equation}
Next, we wish to perturbatively find a unitary transformation such that the resultant Hamiltonian has no ``block off-diagonal" parts:
\begin{equation}
    \heff = e^{\lambda S} H e^{-\lambda S},
\label{eq:effhamil}
\end{equation}
where $S$ is anti-Hermitian.
Here, each block diagonal subspace of $\heff$ corresponds to a subspace degenerate under $\hC$---since $\hC$ is the center-of-mass operator with OBC (see Eq.~(\ref{eq:comoperator})), different block diagonal parts of $\heff$ correspond to subspaces labelled by distinct center-of-mass quantum numbers. 
%
The effective Hamiltonian can be expressed as
\begin{equation}
    \heff = e^{\lambda S} H e^{-\lambda S} = \sumal{n = 0}{\infty}{\lambda^n \heff^{(n)}},
\end{equation}
where $\heff^{(n)}$ is the effective Hamiltonian in $n$-th order perturbation theory.
In what follows, we show that the pair-hopping term arises in $\heff^{(3)}$ i.e., in the effective Hamiltonian restricted to one center-of-mass sector of the Bloch MBL Hamiltonian.
We now derive the expression for $\heff$ up to third order in perturbation theory. 
Expanding $\heff$, defined in Eq.~(\ref{eq:effhamil}), in powers of $\lambda$, we obtain
\begin{align}
    \heff =\ &H + \lambda \left[S, H\right] + \frac{\lambda^2}{2}\left[S, \left[S, H\right]\right] \nn \\
    &+ \frac{\lambda^3}{6}\left[S, \left[S, \left[S, H\right]\right]\right] +  \mathcal{O}\left(\lambda^4\right).
\label{eq:heffexpansion}
\end{align}
We also expand $S$ in powers of $\lambda$ as
\begin{equation}
    S = S_0 + \lambda S_1 + \lambda^2 S_2 +  \mathcal{O}\left(\lambda^3\right).
\end{equation}
Using Eqs.~(\ref{eq:Hsplit}) and (\ref{eq:heffexpansion}), we obtain
\begin{widetext}
\begin{eqnarray}
    &\heff = \hC + \lambda \left\{\hV + \hT + \left[S_0, \hC\right]\right\} + \lambda^2\left\{\frac{1}{2}\left[S_0, \left[S_0,\ \hC\right]\right] + \left[S_0,\  \hV + \hT\right] + \left[S_1, \hC\right]\right\} \nn \\
    &+ \lambda^3\left\{\frac{1}{6}\left[S_0, \left[S_0, \left[S_0, \hC\right]\right]\right] + \frac{1}{2}\left(\left[S_1, \left[S_0, \hC\right]\right] + \left[S_0, \left[S_1, \hC\right]\right] + \left[S_0, \left[S_0, \hV + \hT\right]\right]\right)+ \left[S_1, \hV + \hT\right]  + \left[S_2, \hC\right]\right\} + \mathcal{O}\left(\lambda^4\right).\nn \\
\label{eq:Heffexact}
\end{eqnarray}
\end{widetext}
Since $\hV$ is diagonal, to cancel the block off-diagonal component $\hT$ at $\mathcal{O}\left(\lambda\right)$ in Eq.~(\ref{eq:Heffexact}) we require that $S_0$ satisfies
\begin{equation}
    \left[S_0, \hC\right] = -\hT. 
\label{eq:SWcondition1}
\end{equation}
Simplifying Eq.~(\ref{eq:Heffexact}) using Eq.~(\ref{eq:SWcondition1}), we obtain
\begin{widetext}
\begin{eqnarray}
    &\heff = \hC + \lambda \hV + \lambda^2\left\{\frac{1}{2}\left[S_0, \hT\right] + \left[S_0,\  \hV\right] + \left[S_1, \hC\right]\right\} \nn \\
    &+ \lambda^3\left\{\frac{1}{3}\left[S_0, \left[S_0, \hT\right]\right] + \frac{1}{2}\left(\left[S_1, \hT\right] + \left[S_0, \left[S_1, \hC\right]\right] + \left[S_0, \left[S_0, \hV\right]\right]\right)+ \left[S_1, \hV\right] + \left[S_2, \hC\right]\right\} + \mathcal{O}\left(\lambda^4\right).
\label{eq:Heffexactfirst}
\end{eqnarray}
\end{widetext}
To determine the block off-diagonal terms at $\mathcal{O}\left(\lambda^2\right)$ in Eq.~(\ref{eq:Heffexactfirst}), we note that since $\hC$ and $\hT$ are block diagonal and block off-diagonal respectively, we can always choose $S_0$ in Eq.~(\ref{eq:SWcondition1}) to be block off-diagonal. 
Thus, $\left[S_0, \hT\right]$ can have block diagonal terms, whereas $\left[S_0, \hV\right]$ is completely block off-diagonal.
To cancel the block off-diagonal terms at $\mathcal{O}\left(\lambda^2\right)$ in Eq.~(\ref{eq:Heffexactfirst}), we hence require that $S_1$ satisfies
\begin{equation}
\label{eq:SWcondition2}
    \left[S_1, \hC\right] = -\left[S_0, \hV\right] - \frac{1}{2}\left(\left[S_0, \hT\right] - \mathcal{P}\left[S_0, \hT\right]\mathcal{P}\right),
\end{equation}
where $\mathcal{P}$ is a projector that kills block off-diagonal components. That is, if a matrix $X$ has both block diagonal and block off-diagonal components, $\mathcal{P} X \mathcal{P}$ (resp. $\left(X - \mathcal{P} X \mathcal{P}\right)$) is completely block diagonal (resp. off-diagonal). 
Simplifying the expression for $\heff$ in Eq.~(\ref{eq:Heffexactfirst}) using Eq.~(\ref{eq:SWcondition2}), we obtain
\begin{widetext}
\begin{equation}
    \heff = \hC + \lambda \hV + \frac{\lambda^2}{2}\mathcal{P}\left[S_0, \hT\right]\mathcal{P} + \lambda^3\left\{\frac{1}{3}\left[S_0, \left[S_0, \hT\right]\right] + \frac{1}{2}\left[S_1, \hT\right] + \left[S_1, \hV\right] + \left[S_2, \hC\right]\right\} + \mathcal{O}\left(\lambda^4\right).
\label{eq:Heffexactsecond}
\end{equation}
\end{widetext}
%
%
In Eq.~(\ref{eq:SWcondition2}), since the RHS is block off-diagonal, and $\hC$ is block diagonal, $S_1$ can be chosen to be block off-diagonal.
Since $S_0$ and $S_1$ are block off-diagonal, $\left[S_1, \hT\right]$ and $\left[S_0, \left[S_0, \hT\right]\right]$ can have block diagonal components whereas $\left[S_1, \hV\right]$ is completely block off-diagonal. 
Moreover, in the Schrieffer-Wolff procedure, $S_2$ is chosen such that the term $\left[S_2, \hC\right]$ cancels block off-diagonal terms at $\mathcal{O}\left(\lambda^3\right)$. 
That is,
\begin{eqnarray}
    &\left[S_2, \hC\right] = -\left[S_1, \hV\right] - \frac{1}{2}\left(\left[S_1, \hT\right] - \mathcal{P}\left[S_1, \hT\right]\mathcal{P}\right) \nn \\
    &- \frac{1}{3}\left(\left[S_0, \left[S_0, \hT\right]\right] - \mathcal{P}\left[S_0, \left[S_0, \hT\right]\right]\mathcal{P}\right).
\end{eqnarray}
Thus, $\heff$ reads
\begin{widetext}
\begin{equation}
    \heff = \hC + \lambda \hV + \frac{\lambda^2}{2}\mathcal{P}\left[S_0, \hT\right]\mathcal{P} + \lambda^3\mathcal{P}\left(\frac{1}{2}\left[S_1, \hT\right] + \frac{1}{3}\left[S_0, \left[S_0, \hT\right]\right]\right)\mathcal{P} + \mathcal{O}\left(\lambda^4\right).
\label{eq:Heffproj}
\end{equation}
\end{widetext}
Thus, we find that $\heff^{(2)}$ and $\heff^{(3)}$ are given by
\begin{align}
    &\heff^{(2)} = \frac{1}{2}\mP \left[S_0, \hT\right] \mP \nn \\
    &\heff^{(3)} = \mP \left(\frac{1}{2}\left[S_1, \hT\right] + \frac{1}{3}\left[S_0, \left[S_0, \hT\right]\right]\right) \mP. 
\label{eq:Heffnexpr}
\end{align}
We now compute $S_0$ and $S_1$ in order to obtain the effective Hamiltonians $\heff^{(2)}$ and $\heff^{(3)}$.
According to Eq.~(\ref{eq:SWcondition1}), $S_0$ is determined by
\begin{equation}
    \left[S_0, \hC\right] = -\hT = - \left(\hT_+ + \hT_-\right). 
\end{equation}
We first compute some useful commutators:
\begin{equation}
    \left[\hT_+, \hT_-\right] = 0,\;\;\; \left[\hT_+, \hC\right] = -\hT_+,\;\;\;\left[\hT_-, \hC\right] = \hT_-,
\label{eq:Tcomms}
\end{equation}
Thus, Eq.~(\ref{eq:SWcondition1}) is satisfied by choosing
\begin{equation}
    S_0 = \hT_+ - \hT_-. 
\label{eq:Schoice}
\end{equation}
Note that $S_0$ in Eq.~(\ref{eq:Schoice}) is block off-diagonal and anti-Hermitian.
Using Eqs.~(\ref{eq:Schoice}) and (\ref{eq:Tcomms}), we obtain
\begin{equation}
    \left[S_0, \hT \right] = 0,\;\;\; \heff^{(2)} = 0.
\label{eq:Heff2}
\end{equation}
$S_1$ is computed using Eq.~(\ref{eq:SWcondition2}), and the relevant commutators read
\begin{align}
    \left[\hT_+, \hV_1\right] &= \alpha_1\sumal{j}{}{\left(\hn_{j-1} \cd_{j+1} c_j - \cd_j c_{j-1} \hn_{j+1}\right)} \nn \\
    &\equiv \alpha_1\left(\hO_{+-} - \hO_{++}\right) \nn \\
    \left[\hT_-, \hV_1\right] &= \alpha_1\sumal{j}{}{\left(- \hn_{j-1} \cd_{j} c_{j+1} + \cd_{j-1} c_{j} \hn_{j+1}\right)} \nn \\
    &\equiv \alpha_1\left(\hO_{-+} - \hO_{--}\right), \nn \\
    \left[\hT_+, \hV_0\right] &=
    \alpha_0\sumal{j}{}{\left(w_j -w_{j+1}\right) \cd_{j+1} c_{j}} \nn \\
    &\equiv \alpha_0\left(-\hF_+ + \hB_+\right) \nn \\
    \left[\hT_-, \hV_0\right] &=
    \alpha_0\sumal{j}{}{\left(w_{j+1} -w_j\right) \cd_{j} c_{j+1}} \nn \\
    &\equiv \alpha_0\left(\hF_- - \hB_-\right) \nn \\
    \implies \left[S_0, \hV\right] &= \alpha_1\left(\hO_{+-} + \hO_{--} - \hO_{-+} - \hO_{++}\right) \nn \\
    &+ \alpha_0\left(-\hF_+ - \hF_- + \hB_+ + \hB_-\right).
\label{eq:SVcommutator}
\end{align}
where we have defined the operators
\begin{eqnarray}
    &\hO_{++} = \sumal{j}{}{\cd_j c_{j-1} \hn_{j+1}},\;\;
    \hO_{-+} = \hO_{++}^\dagger = \sumal{j}{}{\cd_{j-1} c_{j} \hn_{j+1}} \nn \\
    &\hO_{+-} = \sumal{j}{}{\cd_{j+1} c_{j} \hn_{j-1}},\;\;\hO_{--} = \hO_{+-}^\dagger = \sumal{j}{}{\cd_j c_{j+1} \hn_{j-1}},\nn \\
    &\hF_+ = \sumal{j}{}{w_{j+1} \cd_{j+1} c_j},\;\;\hF_- = \hF_+^\dagger = \sumal{j}{}{w_{j+1} \cd_{j} c_{j+1}},\nn \\
    &\hB_+ = \sumal{j}{}{w_j \cd_{j+1} c_j},\;\;\hB_- = \hB_+^\dagger = \sumal{j}{}{w_j \cd_{j} c_{j+1}}.
\label{eq:usefulops}
\end{eqnarray}
Thus, according to Eq.~(\ref{eq:SWcondition2}), $S_1$ should satify
\begin{align}
    \left[S_1, \hC\right] = -\left[S_0, \hV\right] &= \alpha_1\left(-\hO_{+-} - \hO_{--} + \hO_{-+} + \hO_{++}\right)\nn \\
    &+ \alpha_0\left(\hF_+ + \hF_- - \hB_+ - \hB_-\right).
\label{eq:SVO}
\end{align}
We now show that $S_1$ can be chosen to be a linear superposition of $\hO_{\mu\nu}$, $\hF_\mu$, and $\hB_\mu$, where $\mu, \nu \in \{+, -\}$. 
The commutators $\left[\hO_{\mu\nu}, \hC\right]$, $\left[\hF_\mu, \hC\right]$, and $\left[\hB_\mu, \hC\right]$ read
\begin{eqnarray}
    &\left[\hO_{++}, \hC\right] = -\hO_{++},\;\;\;\left[\hO_{-+}, \hC\right] = \hO_{-+},\nn \\
    &\left[\hO_{+-}, \hC\right] = -\hO_{+-},\;\;\;\left[\hO_{--}, \hC\right] = \hO_{--}, \nn \\
    &\left[\hF_+, \hC\right] = -\hF_+,\;\;\;\left[\hF_-, \hC\right] = \hF_-,\nn \\
    &\left[\hB_+, \hC\right] = -\hB_+,\;\;\;\left[\hB_-, \hC\right] = \hB_-.
\label{eq:comrels}
\end{eqnarray}
Thus, Eq.~(\ref{eq:SVO}) is satisfied by choosing 
\begin{align}
    S_1 &= \alpha_1\left(\hO_{+-}  - \hO_{--}  +\hO_{-+} -\hO_{++} \right) \nn \\
    &+ \alpha_0\left(-\hF_+ + \hF_- + \hB_+ - \hB_-\right).
\label{eq:S1expr}
\end{align}
Noting that $\left[S_0, \hT\right] = 0$, $\heff^{(3)}$ in Eq.~(\ref{eq:Heffnexpr}) reads
\begin{align}
    \heff^{(3)} &= \frac{\lambda^3}{2}\mathcal{P}\left[S_1, \hT_+ + \hT_-\right]\mathcal{P}\nn \\
    &= -\frac{\alpha_1 \lambda^3}{2}\mathcal{P}\left[\hT_+ + \hT_-,\ \hO_{+-} - \hO_{--} + \hO_{-+} - \hO_{++}\right]\mathcal{P} \nn \\
    &- \frac{\alpha_0 \lambda^3}{2}\mathcal{P}\left[\hT_+ + \hT_-,-\hF_+ + \hF_- + \hB_+ - \hB_-\right]\mathcal{P}.
\label{eq:effHamill3}
\end{align}
We obtain the following commutators
\begin{widetext}
\begin{eqnarray*}
    &\left[\hT_+, \hO_{+-}\right] = \sumal{j}{}{\left(-\hn_{j}\cd_{j+1} c_{j-1} + \cd_{j+2} c_j \hn_{j-1} - \cd_{j+1} c_j \cd_{j-1} c_{j-2}\right)}\nn \\
    &\left[\hT_+, \hO_{-+}\right] = \sumal{j}{}{\left(\left(\hn_j - \hn_{j-1}\right) \hn_{j+1} + \cd_{j-1} \cd_{j+2} c_{j+1} c_{j}\right)} \nn \\
    &\left[\hT_+, \hO_{++}\right] = \sumal{j}{}{\left(-\cd_j \hn_{j+1} c_{j-2} + \cd_{j+1} c_{j-1} n_j + \cd_j c_{j-1} \cd_{j+2} c_{j+1}\right)} \nn \\
\end{eqnarray*}
\begin{eqnarray}
    &\left[\hT_-, \hO_{+-}\right] = \left[\hT_+^\dagger, \hO_{--}^\dagger\right] = -\left[\hT_+, \hO_{--}\right]^\dagger\;\;
    \left[\hT_-, \hO_{++}\right] = \left[\hT_+^\dagger, \hO_{-+}^\dagger\right] = -\left[\hT_+, \hO_{-+}\right]^\dagger,\nn \\
    &\left[\hT_-, \hO_{-+}\right] = \left[\hT_+^\dagger, \hO_{++}^\dagger\right] = -\left[\hT_+, \hO_{++}\right]^\dagger,\;\;\left[\hT_-, \hO_{--}\right] = \left[\hT_+^\dagger, \hO_{+-}^\dagger\right] =  -\left[\hT_+, \hO_{+-}\right]^\dagger \nn \\
    &\left[\hT_+, \hF_+\right] = \sumal{j}{}{\left(w_{j} - w_{j+1}\right) \cd_{j+1} c_{j-1}},\;\;\left[\hT_+, \hF_-\right] = \sumal{j}{}{\left(w_j - w_{j+1}\right)\hn_j},\nn \\
    &\left[\hT_+, \hB_+\right] = \sumal{j}{}{\left(w_{j-1} - w_j\right) \cd_{j+1} c_{j-1}},\;\;\left[\hT_+, \hB_-\right] = \sumal{j}{}{\left(w_{j-1} - w_{j}\right)\hn_j} \nn \\
    &\left[\hT_-, \hF_-\right] = \left[\hT_+^\dagger, \hF_{+}^\dagger\right] =  -\left[\hT_+, \hF_{+}\right]^\dagger,\;\;\left[\hT_-, \hF_+\right] = \left[\hT_+^\dagger, \hF_{-}^\dagger\right] =  -\left[\hT_+, \hF_{-}\right]^\dagger \nn \\
    &\left[\hT_-, \hB_-\right] = \left[\hT_+^\dagger, \hB_{+}^\dagger\right] =  -\left[\hT_+, \hB_{+}\right]^\dagger,\;\; \left[\hT_-, \hB_+\right] = \left[\hT_+^\dagger, \hB_{-}^\dagger\right] =  -\left[\hT_+, \hB_{-}\right]^\dagger. 
\label{eq:allcommutators}
\end{eqnarray}
Note that
\begin{align}
    &\mathcal{P} \left[\hT_+, \hO_{+-}\right] \mathcal{P} = \mathcal{P} \left[\hT_-, \hO_{--}\right] \mathcal{P} = 0,\;\;\;\mathcal{P} \left[\hT_+, \hO_{++}\right] \mathcal{P} = \mathcal{P} \left[\hT_-, \hO_{-+}\right] \mathcal{P} = 0\nn \\
    &\mathcal{P} \left[\hT_+, \hF_{+}\right] \mathcal{P} = \mathcal{P} \left[\hT_-, \hF_{-}\right] \mathcal{P} = 0,\;\;\;\mathcal{P} \left[\hT_+, \hB_{+}\right] \mathcal{P} = \mathcal{P} \left[\hT_-, \hB_{-}\right] \mathcal{P} = 0,\nn \\
\label{eq:commvanish}
\end{align}
since according to Eq.~(\ref{eq:allcommutators}), these terms change the energy of eigenstates of $\hC$, and are hence block off-diagonal.
Simplifying Eq.~(\ref{eq:effHamill3}) using Eqs.~(\ref{eq:allcommutators}) and (\ref{eq:commvanish}), we obtain
\begin{eqnarray}
    &\heff^{(3)} = -\frac{\alpha_1\lambda^3}{2}\mP\left\{\left(-\left[\hT_+, \hO_{--}\right] + \left[\hT_+, \hO_{-+}\right]\right) + h.c.\right\}\mP -\frac{\alpha_0\lambda^3}{2}\mP\left\{\left(\left[\hT_+, \hF_{-}\right] - \left[\hT_+, \hB_{-}\right]\right) + h.c.\right\}\mP \nn \\
    &= -\alpha_1\lambda^3\sumal{j}{}{\left\{\left(\cd_j \cd_{j+3} c_{j+2} c_{j+1} + h.c.\right) + 2\left(\hn_j\hn_{j+1} - \hn_j \hn_{j+2}\right)\right\}} -\alpha_0\lambda^3\sumal{j}{}{\left(2 w_j - w_{j-1} - w_{j+1}\right) \hn_j}\nn \\
\end{eqnarray}
Finally, re-introducing the overall factor of $E$, the full effective Hamiltonian restricted to one center-of-mass sector is:
\begin{eqnarray}
    &\heff = V_0\sumal{j}{}{\widetilde{w}_j\hn_j} + \widetilde{V}_1\sumal{j}{}{\hn_j \hn_{j+1}} + \widetilde{V}_2 \sumal{j}{}{\hn_j \hn_{j+2}} - \frac{t^2 V_1}{E^2}\sumal{j}{}{\left(\cd_j \cd_{j+3} c_{j+2} c_{j+1} + h.c.\right)} + \mathcal{O}\left(\frac{t^3}{E^3}\right),\nn \\
\label{eq:hefffinal}
\end{eqnarray}
where we have omitted the term $E\sumal{j}{}{j \hn_j}$ since it is a symmetry of the effective Hamiltonian, and we have defined
\begin{equation}
    \widetilde{w}_j \equiv \left(1 - \frac{2 t^2}{E^2}\right) w_j + \frac{t^2}{E^2}\left(w_{j-1} + w_{j+1}\right),\;\;\widetilde{V}_1 \equiv V_1\left(1 - \frac{2 t^2}{E^2}\right),\;\;\widetilde{V}_2 \equiv \frac{2 t^2 V_1}{E^2}.
\label{eq:tildedefns}
\end{equation}
\end{widetext}
%
%
\newpage
\bibliography{fracton}
\end{document}